\documentclass[11pt,a4paper]{scrartcl}

\usepackage{ILD}

\usepackage{xpatch}
\makeatletter
\xpatchcmd\@HepConStyle
 {\edef\@upcode{\updefault}}
 {\ifdefined\shapedefault\edef\@upcode{\shapedefault}\else\edef\@upcode{\updefault}\fi}
 {}{}
\makeatother

\usepackage[symbol]{footmisc}
\usepackage{feynmf}
\usepackage{multirow}
\usepackage{textpos}
\usepackage{heppennames2}
\usepackage{LCVision_definitions}

\usepackage{siunitx}
\usepackage{lineno}

\usepackage{booktabs,multirow,array}
\usepackage{amssymb}

\title{Update of the Higgs Self-coupling Projections from Di-Higgs Production in Detailed Simulation of the ILD Concept}

\ildphys{2025}{001}

\date{\today}

\addauthor{Mikael Berggren}{\institute{1}}
\addauthor{Bryan Bliewert}{\institute{1}\institute{2}}
\addauthor{Jenny List}{\institute{1}}
\addauthor{Dimitris Ntounis}{\institute{3}}
\addauthor{Taikan Suehara}{\institute{4}}
\addauthor{Junping Tian}{\institute{4}}
\addauthor{Julie Munch Torndal}{\institute{1}\institute{2}}
\addauthor{Caterina Vernieri}{\institute{3}}
\addinstitute{1}{Deutsches Elektronen\-/Synchrotron DESY, Germany}
\addinstitute{2}{Department of Physics, Universit\"at Hamburg, Germany}
\addinstitute{3}{SLAC National Accelerator Laboratory, United States}
\addinstitute{4}{International Center for Elementary Particle Physics (ICEPP), The University of Tokyo, Japan}

\abstract{This contribution summarizes the update of the projections for the determination of the tri-linear Higgs self-coupling from di-Higgs production at future $\Pep\Pem$ colliders. In particular, we will present an update of the analysis of $\PZ\PH\PH$ production at \SI{500}{\giga\electronvolt} in detailed simulations of the ILD concept, covering the $\PH\PH \rightarrow \PQb\PAQb \PQb\PAQb$ and $\PZ \rightarrow \PQq\PAQq / \Pep\Pem / \PGmp\PGmm / \PGn\PAGn$ channels. Based on the experience of previous analyses, we will extrapolate these to contain some of the remaining decay modes, e.g.\ $\PH\PH \rightarrow \PQb\PAQb \PW\PW^*$ or $\PZ \rightarrow \PGtp\PGtm$, as well as the contribution from the $\PW\PW$ fusion production mode. We will study the dependency of the results on the centre-of-mass energy (in particular discussing \SI{550}{\giga\electronvolt} and \SI{1}{\tera\electronvolt}) as well as on the value of the trilinear coupling realised in nature.
}

\DeclareTOCStyleEntry[beforeskip=.2cm]{section}{section}

\graphicspath{ {./logos/}{./figures/} }

\begin{document}

\titlepage
\pagenumbering{arabic}\setcounter{page}{1}

\tableofcontents
\clearpage

\section{Introduction}
\label{sec:intro}
Di-Higgs production is the main class of processes that can access the trilinear coupling(s) of the Higgs boson(s) at the leading order. Detecting the simultaneous production of two 125-GeV Higgs bosons is a key objective of the LHC, and of future $\Pp\Pp$, $\PGmp\PGmm$ and $\Pep\Pem$ colliders. In particular, $\Pep\Pem$ colliders with centre-of-mass energies $\sqrt{s} \geq$ \SI{450}{\giga\electronvolt} will be able to observe both di-Higgs production from $\PW\PW$ fusion, $\PGn\PAGn\PH\PH$, dominant at centre-of-mass energies above \SI{1}{\tera\electronvolt}, and double Higgs-strahlung, $\PZ\PH\PH$, dominant at lower centre-of-mass energies~\cite{Durig:2016jrs, Tian:2013}. Both processes have been studied about $10$ years ago in detailed, Geant4-based simulation of the ILD concept, considering the $\PH\PH$ $ \rightarrow $ $4\PQb$~\cite{Durig:2016jrs, Tian:2013} and $\PH\PH \rightarrow \PQb\PAQb\PW\PW^*$~\cite{Kurata:2013} final states, and of the CLICdp detector concept~\cite{Roloff:2019crr}.
It should be noted that in particular for the $\PZ\PH\PH$ process, it has been shown explicitly that at an $\Pep\Pem$ collider, all other parameters entering the interpretation of the $\PZ\PH\PH$ cross-section in dim-6 SMEFT will be determined precisely enough that their impact on the extraction of $\uplambda$ is negligible~\cite{Barklow:2017awn}.

Since the above-mentioned studies have been performed, both the machine planning and physics projections for future colliders have evolved significantly. In particular, ongoing analyses of LHC Run 3 data and the development of machine-learning-based tools have significantly improved the prospects for measuring di-Higgs production at the HL-LHC~\cite{ATL-PHYS-PUB-2025-018}. The SM prediction for di-Higgs production is expected to be observed at the level of $7\,\sigma$ assuming 3 ab$^{-1}$ and combining both ATLAS and CMS. This would translate into a precision of $30\%$ on $\uplambda$, assuming all the other Higgs boson couplings to be set at their SM value. For the first time, HL-LHC projections have been presented as a function of $\uplambda$, showing a reduced sensitivity for  values of $\uplambda$ larger than the SM prediction, as we will discuss in Sec.~\ref{sec:results}.

With regard to $\Pep\Pem$ colliders, running scenarios have been revised. Where earlier studies for the ILC~\cite{TDR1,TDR2,TDR31,TDR32,TDR4} assumed a total integrated luminosity of \SI{4}{\abinv} at a centre-of-mass energy of \SI{500}{GeV}, with beam polarisations of 80\% for the electron and 30\% for the positron beam ($|$P($\Pem,\Pep$)$|$=($80\%$, $30\%$)), C$^3$~\cite{Vernieri:2022fae} adopted a centre-of-mass energy of \SI{550}{GeV}, and the recently proposed Linear Collider Facility (LCF) at CERN~\cite{LinearCollider:2025lya} targets \SI{8}{\abinv} at \SI{550}{GeV} with $|$P($\Pem,\Pep$)$|$=($80\%$, $60\%$). Originally motivated by a drastic increase in $\PQt\PAQt\PH$ production, the increase in centre-of-mass energy significantly enhances the sensitivity to the $\PW\PW$-fusion production of di-Higgs events. Machine learning (ML) has led to significant advances in event reconstruction and analysis algorithms, as evidenced by LHC experiments' results and recent HL-LHC projections. Advancing event reconstruction with ML has also been explored at $\Pep\Pem$ colliders~\cite{Altmann:2025feg}. Modern tools now exploit detector granularity much more effectively and derive substantial additional benefits from ML-based techniques. All these updates call for a full update of the more than 10-year-old di-Higgs projections for $\Pep\Pem$ colliders. The purpose of this note is to summarize the status of this update for the Briefing Book of the current update of the European Strategy for Particle Physics.

This note is structured as follows: In Sec.~\ref{sec:ILDMC}, we will briefly introduce the ILD concept and its detailed simulation and reconstruction tools as well as the event samples used. 
After recalling the level of complexity, main findings and limitations of the 2014 ILD analysis of di-Higgs production in Sec.~\ref{sec:summary_claude}, and discussing the recent developments in high-level reconstruction tools in Sec.~\ref{sec:hlreco_extrapol}, we will describe the status of the new analysis in Sec.~\ref{sec:ana550}.
In Sec.~\ref{sec:results}, we will derive updated projection on the di-Higgs production cross-sections as well as on the trilinear Higgs coupling $\uplambda$ as a function of $\uplambda$ as well as of the centre-of-mass energy and compare them to projections for other current and future colliders. A summary and an outlook will be given in Sec.~\ref{sec:concl}.

\section{The ILD Concept, MC Samples and Standard Reconstruction}
\label{sec:ILDMC}
ILD~\cite{ILDIDR, ILDESU, TDR4, abramowicz2025ilddetectorversatiledetector} is one of the proposed detector concepts for future $\Pep\Pem$ Higgs factories.
It is a highly-hermetic multi-purpose detector optimised for particle flow reconstruction~\cite{PFA}. ILD consists of a high precision vertex detector, a time projection chamber (TPC), silicon tracking detectors, a highly granular calorimeter system and a forward detector system, all placed inside of a solenoid providing, in the larger version of the detector~\cite{ILDIDR} used in this analysis, a magnetic field of $3.5$\,T, surrounded by an iron yoke instrumented for muon detection. The TPC of ILD provides also a dE/dx measurement of better than 5\%, which, however, is not yet leveraged in the analyses presented in this note, thus offers room for further improvements.

The particle-flow approach allows to reconstruct nearly each detector-stable particle individually, either as charged or neutral particle-flow object (PFO), allowing to measure the four-momentum  -- since recently along with the corresponding covariance matrix -- of isolated particles as well as of jets. In addition to the vertex information, modern jet-flavour tagging tools use all PFOs inside a jet. The recent improvements will be discussed in more detail in Sec.~\ref{sec:hlreco_extrapol}.

This section will introduce the relevant physics processes and corresponding event samples, the  full, Geant4-based simulation and reconstruction of ILD as well as the fast ILD simulation based on SGV.   

\subsection{Event Generation}
\label{sec:ILDMC:evtgen}
The event generation for the di-Higgs analyses primarily follows the standard procedure of the Generator Group of the  Linear Collider Collaboration (LCC)~\cite{Berggren:2021sju}. This includes generating events for 100\% beam polarisation, for all sign combinations with a non-zero cross-section (up to four). These can be reweighted in the analysis to any desired actual beam polarisation setting. For the majority of the events, the generator Whizard~\cite{Kilian:2007gr} is used to generate events for various processes, defined in terms of parton-level final-states, based on the leading-order matrix elements. Processes are classified by the number of fermions (or anti-fermions) in the final-state, usually up to 6. Parton shower and hadronisation are performed with Pythia 6.4~\cite{Sjostrand:2006za}, $\PGt$-lepton decays are handled by Tauola~\cite{Was:2000st}. 
An exemption are processes containing a Higgs boson: in these cases, the Higgs boson is part of the matrix-element-level final-state, and its decay is handled by Pythia, using the branching fractions provided by the LHC Higgs cross-section working group~\cite{LHCHiggsCrossSectionWorkingGroup:2016ypw}. For the di-Higgs analysis, the most important 8-fermion processes are included as well, in particular $\PQt\PAQt\PH$ and $\PQt\PAQt\PZ$. As Whizard does not yet implement the $\PQt\PAQt$ threshold enhancement for these processes, these were generated with PhysSim~\cite{physsimweb}.

The 2014 analysis~\cite{Durig:2016jrs} used the standard ILC500 beam conditions. The beam energy spectrum has been calculated for ILC500 with GuineaPig~\cite{Schulte:1998au} and is passed to Whizard~1.95. 
A list of processes, cross-sections and available MC statistics for the 2014 analysis can be found in Table~5.3 of Ref.~\cite{Durig:2016jrs}.  The total number of MC events used at the time corresponds to about 30 millions. 

For \SI{550}{GeV}, the same spectrum from GuineaPig was
parameterised to serve as input to CIRCE2~\cite{Ohl:1996fi},
and was subsequently scaled to the slightly higher centre-of-mass energy, and new event samples have been generated with Whizard~{3.1.5}. So far, the total number of events amounts to about 170 millions, focusing on 4-, 6- and 8-fermion events. Table~\ref{tab:MC550} summarises the cross-sections and event numbers of the processes so far included in the 550-GeV analysis.
\begin{table}[htb]
\begin{center}
\begin{tabular}{llcccc}
Category & Final state & Polarisation & Cross-section [fb] & N$_{gen}$ \\
\hline
$\Pep\Pem \to \PZ\PH\PH$ & $\Pl\Pl\PH\PH$ & LR & \num{3.85e-2} & 3.00\,M \\
& & RL & \num{2.45e-2} & 3.00\,M \\
& & LL & \num{5.12e-4} & 1.00\,M \\
& & RR & \num{5.12e-4} & 1.00\,M \\
& $\PGn\PGn\PH\PH$ & LR & \num{9.08e-2} & 2.00\,M \\
& & RL & \num{4.80e-2} & 2.00\,M \\
& $\PQq\PQq\PH\PH$ & LR & \num{2.56e-1} & 1.00\,M \\
& & RL & \num{1.64e-1} & 1.00\,M \\
\hline
$\Pep\Pem \to 4f \, \PH$ & $\Pl\Pl\PQq\PQq\PH$ & LR & \num{2.56e-1} & 3.00\,M \\
& & RL & \num{1.19e-1} & 3.00\,M \\
& & LL & \num{5.16e-3} & 1.00\,M \\
& & RR & \num{5.17e-3} & 1.00\,M \\
& $\PGn\PGn\PQq\PQq\PH$ & LR & \num{7.05e-1} & 2.00\,M \\
& & RL & \num{1.71e-1} & 2.00\,M \\
& $\PQq\PQq\PQq\PQq\PH$ & LR & 9.03 & 1.00\,M  \\
&  & RL & \num{5.72e-1} & 1.00\,M \\
\hline
$\Pep\Pem \to 6f$ & $\Pl\Pl\PQq\PQq\PQq\PQq$ & LR & 21.9 & 187\,k  \\
& & RL & 2.15 & 120\,k   \\
& & LL & 4.50 & 62.3\,k \\
& & RR & 4.48 & 62.2\,k   \\
& $\PGn\PGn\PQq\PQq\PQq\PQq$ & LR & 19.9 & 133\,k  \\
& & RL & \num{4.1e-1} & 60.0\,k \\
& $\PQq\PQq\PQq\PQq\PQq\PQq$ & LR & 585 & 2.82\,M \\
&  & RL & 214 & 1.16\,M \\
& $\Pl\PGn\PQb\PQb\PQq\PQq$ & LR & 566 & 1.88\,M \\
&  & RL & 220 & 1.07\,M \\
\hline
$\Pep\Pem \to 4f$ & $\Pl\Pl\PQq\PQq$ & LR & \num{2.33e3} & 23.2\,M  \\
& & RL & \num{1.95e3} & 19.4\,M  \\
& & LL & \num{1.86e3} & 3.67\,M \\
& & RR & \num{1.86e3} & 3.7\,M \\
& $\Pl\PGn\PQq\PQq$ & LR & \num{7.72e3} & 77.2\,M \\
& & RL & 55.8 & 435\,k \\
& & LL & 857 & 1.61\,M \\
& & RR & 858 & 1.70\,M \\

\end{tabular}
\end{center} 
\caption{Cross-sections and numbers of generated events for all considered processes at $\sqrt{s}=550$\,GeV. For the cross-sections listed here as well as for the event generation, 100\% polarisation is used of $P(\Pem,\Pep)$. Where LL and RR cross sections are not given, the SM cross section is zero (or negligible).}
\label{tab:MC550}
\end{table}

\subsection{Detector Simulation}
The ILD is implemented in a detailed, Geant4-based ``full'' simulation, including not only the subdetectors, but also detailed modelling on mechanics and services~\cite{TDR4}. The full simulation is based on DD4HEP, and is complemented with a digitisation step emulating the read-out electronics. The performance of the various subdetectors has been gauged against test-beam data of prototypes.

The 2014 analysis was based on this full simulation. Due to resource limitations, it was not possible to produce sufficient statistics of full simulation events to exclusively base the current update upon. Instead, only the $\PZ\PH\PH$ events and a subset of the 6f samples corresponding to the most signal-like background from $\PZ\PZ\PH$ have been processed in full simulation.

\begin{figure}[htb!]
    \centering
    \includegraphics[width=0.84\linewidth]{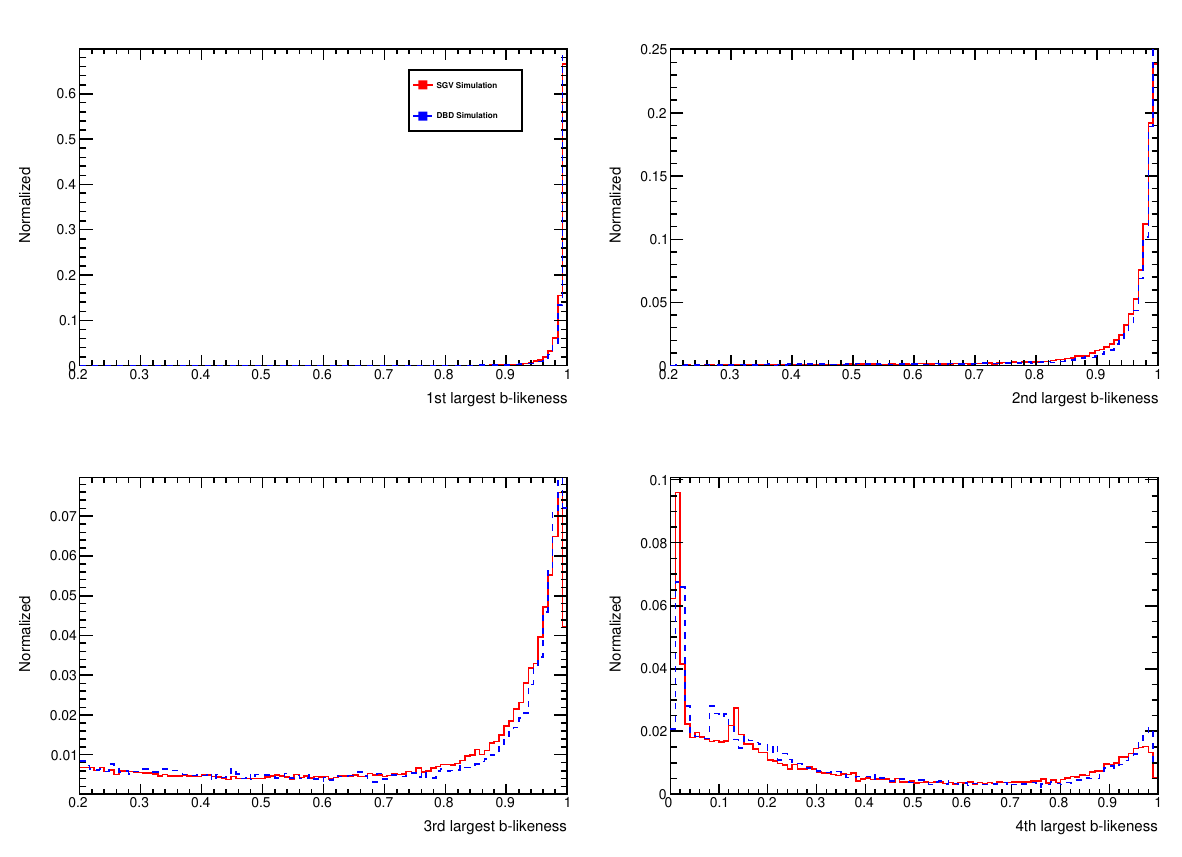}
    \caption{$\PQb$-jet likeliness by LCFIPlus on $\PZ\PH\PH$ events at 1\,TeV with four true $\PQb$-jets, ordered by highest found $\PQb$-jet likeliness, simulated and reconstructed either by SGV (red solid) or full simulation and
      reconstruction (blue dashed), from the 2014 analysis.
    \label{fig:comp_btag}}
    \end{figure}

For large-scale event processing, the implementation of ILD in the detailed fast simulation program SGV~\cite{Berggren:2012ar} has been used instead. SGV models the performance of the tracking system of any collider detector from first principles (i.e.\ positions, materials and point resolutions of tracker layers), while the calorimeter response is parameterised. The level of agreement between SGV and full simulation of ILD is excellent, see~\cite{Berggren:2012ar} for examples from tracking and calorimetry.
An excellent agreement between fast and full simulation was already observed in the previous analysis as shown in Fig.~\ref{fig:comp_btag} with the comparison of the distribution of the b-tag values for the first to fourth highest values for di-Higgs events in ILD at 1\,TeV
using LCFIPlus.
The result when LCFIPlus is applied to data processed by full simulation and
reconstruction is shown in dashed blue, while the results when it is applied to same data processed by SGV is shown in solid red.

In the context of this study, a new validation of the current SGV vs the current full simulation with a special focus on the new flavour tag and kinematic reconstruction tools was performed, as will be discussed further in Sec.~\ref{sec:hlreco_extrapol}.

\subsection{Standard Reconstruction}
The standard reconstruction of ILD is implemented in Marlin~\cite{Marlin}. After the digitisation step, pattern recognition in the tracker, track fitting, vertex finding and fitting, clustering in the calorimeters and overall particle flow based on PandoraPFA~\cite{PFA} are performed, delivering a unique set of particle-flow objects (PFOs) with their four-momenta and a set of vertices. Since the 2014 analysis, the standard reconstruction has been augmented by algorithms which determine a covariance matrix in four-momentum space for each PFO, capturing its individual measurement uncertainties. The output of SGV corresponds to that of the standard reconstruction, called DST.

The standard reconstruction is usually followed by analysis-dependent high-level reconstruction step, which uses the same algorithms on full simulation and SGV. This comprises the identification of isolated electrons, muons and photons, jet clustering, tau identification, kinematic fitting and jet flavour tagging. 

The isolated electron and muon tagging was developed for the 2014 analysis and is, with small re-optimisations, still used today (IsolatedLeptonTaggingProcessor in Marlin). Jet clustering can be done with FastJet~\cite{Cacciari:2006sm}, usually using the Durham~\cite{Durham} algorithm. Alternatively, the Durham algorithm can be used inside the flavour tagging package LCFIPlus~\cite{Suehara:2015ura}, which ensures that PFOs from the same secondary vertex are not torn apart in jet clustering. The kinematic fitting is implemented via MarlinKinFit~\cite{Beckmann:2010ib}, which features a dedicated technique to model photon radiation escaping down the beam pipe. This was already used in the 2014 analysis. Among the short-comings at the time was the well-known feature of kinematic fits to pull the background towards the signal. Since then, this effect has been very significantly reduced by the combination of a drastically improved jet error estimation and a correction for missing momentum from neutrinos in semi-leptonic b-decays. Both will be discussed in more detail in Sec.~\ref{sec:hlreco_extrapol}.


\section{Conclusions of the 2014 ZHH Analysis in full ILD simulation at 500\,GeV}
\label{sec:summary_claude}

In this section we give a brief summary of the ILD di-Higgs analyses performed around 2014 at a centre-of-mass energy of 500\,GeV, assuming 2\,\abinv for each of the two opposite-sign polarisation configurations. At the time, the $\PH\PH \to 4\PQb$ and $\PH\PH \to \PQb\PQb\PW\PW^*$ decay modes were analysed. In particular for the $\PH\PH \to 4\PQb$, potential improvements of various parts of the analysis were propagated through the analysis to estimate their impact on the final result. 

\subsection{Analysis of the 4b decay mode}
The analysis of the $\PH\PH \to 4\PQb$ decay mode is documented in the PhD thesis of Claude D\"urig~\cite{Durig:2016jrs}. At the time, the $\Pe\Pe\PH\PH$, $\PGm\PGm\PH\PH$, $\PGn\PAGn\PH\PH$ and $\PQq\PQq\PH\PH$ channels were considered, with the latter split into $\PQq = \PQb$ and $\PQq \neq \PQb$. 
The analysis considered all relevant 4f and 6f (and in some channels 8f) backgrounds with at least two $\PQb$ quarks in the final state in addition to the ``channel defining'' fermion-antifermion pair, processed in full simulation and passed through ILD standard reconstruction, c.f.\ Sec.~\ref{sec:ILDMC}. For each of the channels, a cut-based preselection was performed, targeting a well-reconstructed $\PZ$ candidate. After the preselection, BDTs exploiting kinematic observables were trained against each of the major remaining background classes, and in a final step $\PQb$-tagging information was exploited. The analysis was done separately for the two data sets with opposite-sign polarisations, assumed to have an integrated luminosity of 2\,\abinv each, and for 0\%, 30\% and 60\% positron polarisation. 
The hadronic channel was split further into the cases $\PZ \to \PQb\PQb$ and $\PZ \to \PQq\PQq$ (non-$\PQb\PQb$). 

For all channels, an inclusive approach was taken in order to maximise the sensitivity to the Higgs self-coupling from this analysis alone. This means that while the analysis was targeting $\PH\PH \to 4\PQb$, any surviving $\PH\PH$ event was included in the final signal count. The size of the non-4$\PQb$ contribution is strongly channel-dependent: The lepton channels with their clear kinematic signature require only a very soft cut on the $\PQb$-tag output, and thus nearly half of the final signal count originates from non-4$\PQb$ contributions. In the neutrino and hadron channels, strong $\PQb$-tagging requirements were applied, leading to a cleaner $\PH\PH \to 4\PQb$ signal. This approach cannot be followed one-to-one in the new analysis for two reasons: first of all since the modern, much more powerful $\PQb$-tagging tools strongly reduce the cross-talk between channels, and also the future ambition should be to design dedicated analyses for the most important decay modes next to $\PH\PH \to 4\PQb$, and combine them properly without overlap.

For the purpose of the extrapolation (c.f.\ Sec.~\ref{sec:extrapol}) and for comparison with the new analysis (c.f.\ Sec.~\ref{sec:ana550}), we summarize  the final signal counts and significances\footnote{In the 2014 analysis, the measurement and excess significances were calculated with the likelihood-ratio method. For the purpose of comparisons and extrapolations, we will use here a simplified $S/\sqrt{S+B}$ approach.} for the 2014 analysis in total and for the $\PH\PH \to 4\PQb$ contribution alone in Table~\ref{tab:Nevt_Claude}, based on Table~9.1 of~\cite{Durig:2016jrs}.

\begin{table}[htb]
\begin{center}
\begin{tabular}{llc|cccc}
P($\Pem$), P($\Pep$) & channel & decay mode &  S & B & $\sigma_{m}$ & S/$\sqrt{S+B}$\\
\hline
$-80\%$, $+30\%$ & $\Pe\Pe\PH\PH$ & inclusive & 3.9 & 7.0 & 1.07 & 1.14\\
 & & $\PH\PH \to 4\PQb$ &  2.6  &   &  & 0.84 \\
 & $\PGm\PGm\PH\PH$ & inclusive & 5.1 & 8.9 & 1.26 & 1.36\\
 & & $\PH\PH \to 4\PQb$ &  2.8  &   &  & 0.82\\
 & $\PGn\PGn\PH\PH$ & inclusive & 5.6 & 6.5 & 1.50 & 1.58\\
 & & $\PH\PH \to 4\PQb$ &  5.5  &   &  & 1.56\\
 & $\PQb\PQb\PH\PH$ & inclusive & 8.5 & 21.9 & 1.57 & 1.54\\
 & & $\PH\PH \to 4\PQb$ &  8.0  &   &  & 1.46\\
 & $\PQq\PQq\PH\PH$ & inclusive & 12.6 & 55.0 & 1.55 & 1.53\\
 & & $\PH\PH \to 4\PQb$ &  10.9  &   &  & 1.34\\
\hline
$-80\%$, $+30\%$ & combined & inclusive &  &  &   & 3.2 \\
 & & $\PH\PH \to 4\PQb$ &     &   &  & 2.8 \\
\hline
\hline
$+80\%$, $-30\%$ &  $\Pe\Pe\PH\PH$ & inclusive & 2.9& 4.2 & 0.92 & 1.09\\
 & & $\PH\PH \to 4\PQb$ & 1.9 & &  & 0.77\\
 & $\PGm\PGm\PH\PH$ & inclusive & 3.8 & 5.3 & 1.10 & 1.26\\
 & & $\PH\PH \to 4\PQb$ &  2.0  &   &  & 0.74\\
 & $\PGn\PGn\PH\PH$ & inclusive & 3.6 & 1.1 & 1.54 & 1.66\\
 & & $\PH\PH \to 4\PQb$ &  3.5  &   &  & 1.63\\
 & $\PQb\PQb\PH\PH$ & inclusive & 5.9 & 7.0 & 1.58 & 1.64\\
 & & $\PH\PH \to 4\PQb$ &  5.6  &   &  & 1.58\\
 & $\PQq\PQq\PH\PH$ & inclusive & 8.3 & 16.0 & 1.64 & 1.68\\
 & & $\PH\PH \to 4\PQb$ &  7.8  &   &  & 1.60\\
\hline
$+80\%$, $-30\%$ & combined & inclusive &  &  &   & 3.3 \\
 & & $\PH\PH \to 4\PQb$ &     &   &  & 3.0  \\
\hline
\hline
combined & combined & inclusive &  &  &   & 4.6 \\
 & & $\PH\PH \to 4\PQb$ &     &   &  & 4.1  \\
\hline
\hline

\end{tabular}
\end{center} 
\caption{Final signal and background counts and measurement significances $\sigma_{m}$ from the 2014 $\PH\PH\to 4\PQb$ analysis, based on Table~9.1 of~\cite{Durig:2016jrs}. The $\PZ\PH\PH$ cross-section uncertainty was given as 30.3\% (29.4\%) for the P($\Pem,\Pep$) = ($-80\%$, $+30\%$) (P($\Pem,\Pep$) = ($+80\%$, $-30\%$)) data sets from a likelihood-ratio combination. A naive combination from the inverse sum of the squared significances as will be used in the extrapolation in Sec.~\ref{sec:results} gives with 30.9\% (30.1\%) very similar, if at all slightly conservative, results.}
\label{tab:Nevt_Claude}
\end{table}

Table~9.1 of~\cite{Durig:2016jrs} also gives the expected $\PZ\PH\PH$ cross-section precisions for both polarisation sign configurations and three different absolute values of the positron polarisation, which we do not repeat here. For the standard ILC assumption ( P($\Pem,\Pep$)=($\pm80\%$, $\mp 30\% $) ) and 2\,\abinv for each opposite-sign configuration, the total $\PZ\PH\PH$ cross-section precision was estimated to 21.1\% for the SM case from a likelihood-ratio combination. A naive combination from the inverse sum of the squared significances as will be used in the extrapolation in Sec.~\ref{sec:results} gives with 21.5\% a very similar, if at all slightly conservative, result.

Another important result of the 2014 analysis was an assessment of the limitations and thus the improvement potential of the analysis, discussed in Sec.~9.3 of~\cite{Durig:2016jrs}. In particular, a 5\% higher b-tagging efficiency at the same purity was evaluated to lead to an 11\% improvement on the cross-section precision. Another important limitation at the time was the quality of the kinematic reconstruction. An improvement here was emulated by cheating the jet clustering. The resulting improvement in reconstructed di-jet masses and other kinematic observables gave a 40\% improvement on the $\PZ\PH\PH$ cross-section precision. The inclusion of the $\PZ \to \PGt\PGt$ channel was estimated to improve the result by 8\%. 

\subsection{Combination with bbWW}
\label{sec:masakazu}

The analysis of the $\PH\PH \to \PQb\PAQb\PW\PW^*$ decay mode is documented in the ILD note~\cite{Kurata:2013} prepared for the Detailed Baseline Design (DBD) study~\cite{TDR4}. Depending on how $\PZ$ and $\PW$ decay, four signal channels were taken into account in the analysis: (i) fully hadronic, $\PZ\to\PQb\PAQb/\PQc\PAQc$, $\PH\PH \to \PQb\PAQb\PW\PW^*\to\PQb\PAQb+$4-jet, labeled as $8\HepParticle{j}{}{}\Xspace$ mode; (ii) one charged lepton (electron or muon) together with four jets and one missing neutrino, $\PZ\to\PQb\PAQb/\PQc\PAQc$, $\PH\PH \to \PQb\PAQb\PW\PW^*\to\PQb\PAQb+\Pl+\PGn+$2-jet, labeled as $\Pl\PGn 6\HepParticle{j}{}{}\Xspace$ mode; (iii) two charged leptons and four jets, $\PZ\to \Pl\Pl$, $\PH\PH \to \PQb\PAQb\PW\PW^*\to\PQb\PAQb$+4-jet, labeled as $2\Pl6\HepParticle{j}{}{}\Xspace$ mode; (iv) three charged leptons together with two jets and one missing neutrino, $\PZ\to \Pl\Pl$, $\PH\PH \to \PQb\PAQb\PW\PW^*\to\PQb\PAQb+\Pl+\PGn+$2-jet, labeled as $3\Pl\PGn 4\HepParticle{j}{}{}\Xspace$ mode. The general strategy for selecting signal events in each mode is rather similar to the $\PH\PH\to 4\PQb$ channel: first looking for required isolated charged lepton candidates; then clustering the remaining particles into required number of jets using Durham jet algorithm; after that the charged leptons, missing neutrino (reconstructed assuming 4-momentum conservation) and jets are paired into one on-shell $\PZ$, one on-shell $\PW$, one off-shell $\PW^*$, and one on-shell $\PH$ (from two $\PQb$-tagged jets, the other $\PH$ is reconstructed from $\PW\PW^*$). After those initial event selections, multi-variate analysis based on BDT was carried out to optimize the discrimination with background events which were dominated by $\PQt\PAQt$, $\PW\PW\PZ$ and $\PZ\PZ\PH$. In the first two modes where $\PZ$ decays hadronically, only $\PZ\to\PQb\PAQb/\PQc\PAQc$ were investigated because that otherwise it was very hard to suppress $\PQt\PAQt$ background efficiently. 

\begin{table}[htb]
\begin{center}
\begin{tabular}{lc|ccc}
 channel & decay mode &  $S$ & $B$ &  $S/\sqrt{S+B}$\\
\hline
 $\PQq\PQq\PH\PH$ & $8\HepParticle{j}{}{}\Xspace$ & 15 & 88 & 1.5\\
 $\PQq\PQq\PH\PH$ & $\Pl\PGn6\HepParticle{j}{}{}\Xspace$ & 1.6 & 18 & 0.38 \\
 $\Pl\Pl\PH\PH$ & $2\Pl6\HepParticle{j}{}{}\Xspace$ & 2.2 & 8.4 & 0.69 \\
 $\Pl\Pl\PH\PH$ & $3\Pl\PGn4\HepParticle{j}{}{}\Xspace$ & 1.0 & 2.6 & 0.55 \\
 
\hline
combined &  &  &  & 1.9 \\
\hline
\hline

\end{tabular}
\end{center} 
\caption{Final signal and background counts and significances in the major signal categories from the 2013 $\PH\PH\to \PQb\PAQb\PW\PW^*$ analysis~\cite{Kurata:2013}. }
\label{tab:Nevt_Kurata}
\end{table}

Table~\ref{tab:Nevt_Kurata} summarizes the event counts and significances of the $\PH\PH \to \PQb\PAQb\PW\PW^*$ analysis for 2\,\abinv at 500\,GeV with P($\Pem,\Pep$)=($-80\%$, $+30\%$). The combined significance of 1.9 is obtained from including only the $\PZ \to \Pep\Pem / \PGmp\PGmm / \PQb\PAQb / \PQc\PAQc$ modes. Adding the 1.9 in quadrature to the significance of the $\PH\PH \to 4\PQb$ channel  of 2.8 (which includes all decays of the $\PZ$ apart from the $\PZ \to \PGtp\PGtm$) would correspond already to a 20\% improvement.

For the 2014 result, no actual event-level combination of the $\PQb\PAQb\PW\PW^*$ and 4$\PQb$ analyses was performed, but instead a 20\% relative improvement expected from such a combination with the $\PH\PH \to \PQb\PAQb\PW\PW^*$ channel was included, leading to a cross-section uncertainty of 16.8\%. In a single-parameter extraction\footnote{It has been shown in~\cite{Barklow:2017awn} that in case of $\PZ\PH\PH$, this is equivalent to the precision expected in a full dim6-SMEFT interpretation, when assuming all other Higgs couplings to be known with typical Higgs-Factory-like precisions.} from event counts weighted in $m_{\PH\PH}$, this corresponds to the well-known result of 26.6\% on $\lambda_{\mathrm{SM}}$~\cite{Durig:2016jrs}. 


\section{Recent Developments in High-Level Reconstruction and Expected Impact on the ZHH Analysis}
\label{sec:hlreco_extrapol}

In this section, we summarize the most important developments in high-level reconstruction and discuss their impact on the di-Higgs analysis.

\subsection{Kinematic Reconstruction}
\begin{figure}[htbp]
    \centering
    \begin{subfigure}{.5\textwidth}
    \centering
        \includegraphics[width=0.95\textwidth]{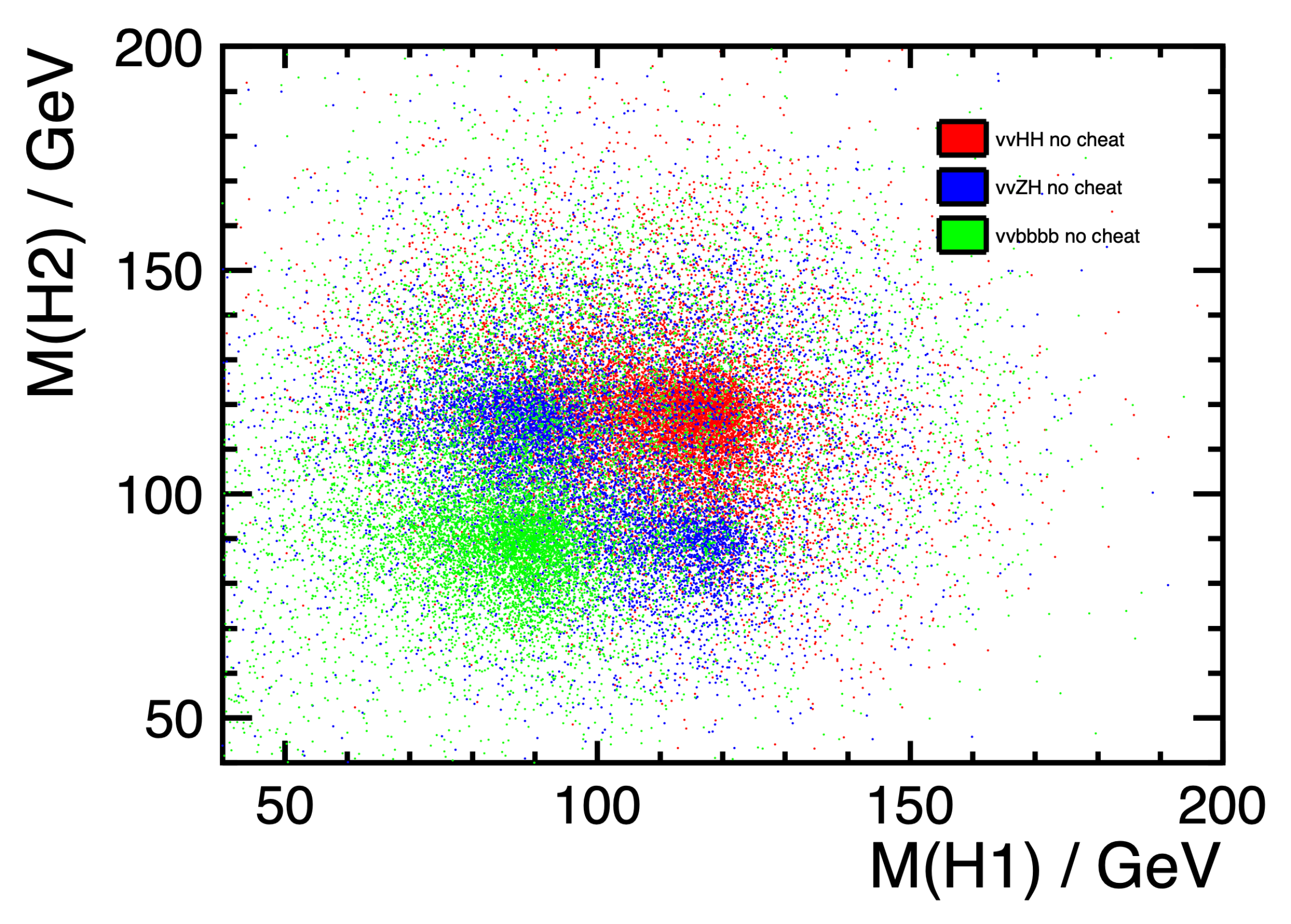}
        \caption{}
        \label{fig:dijet_junping:reco}    
    \end{subfigure}\hfill%
    \begin{subfigure}{.5\textwidth}
        \centering
        \includegraphics[width=0.95\textwidth]{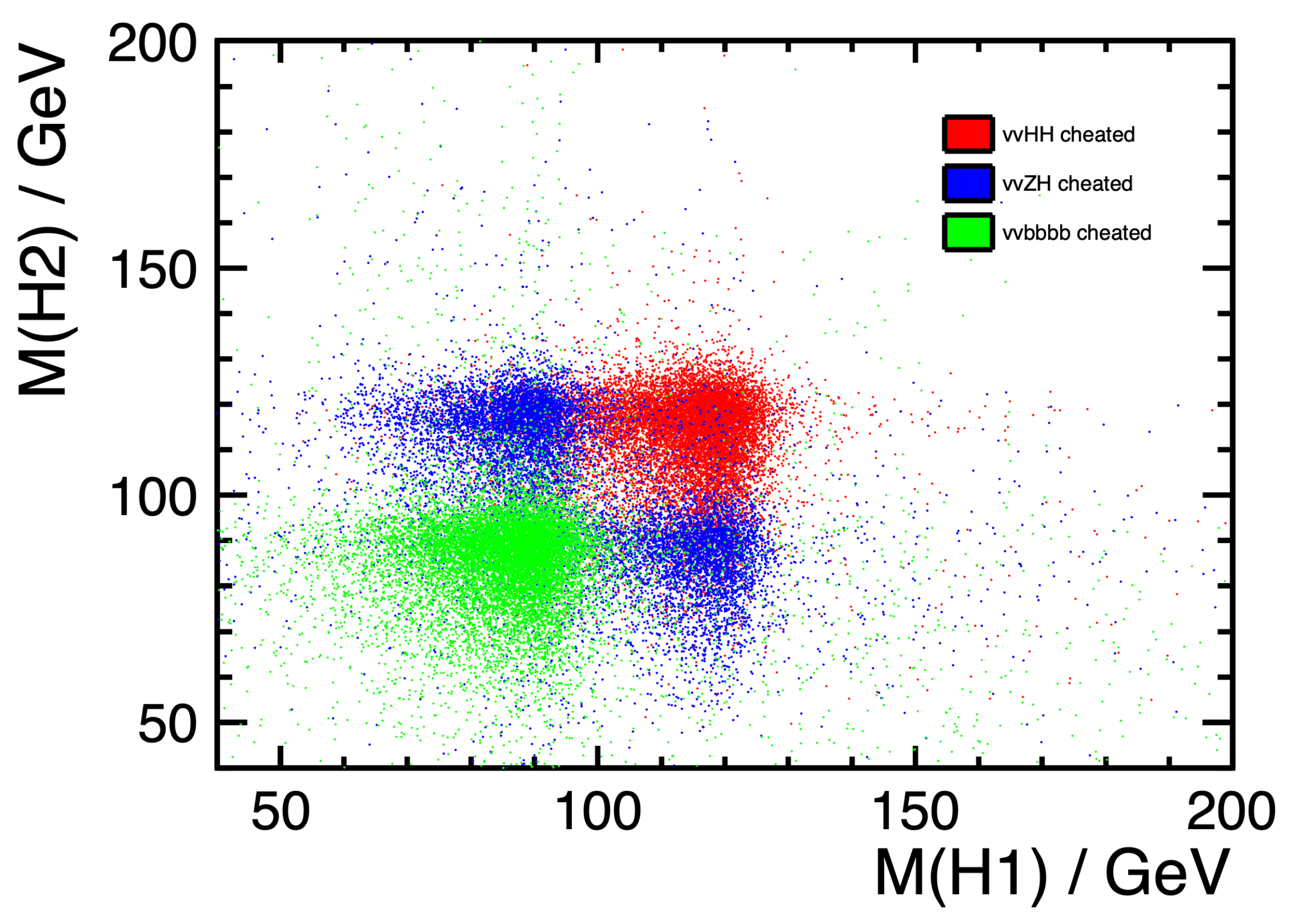}
        \caption{}
        \label{fig:dijet_junping:cheated}    
    \end{subfigure}%
    \caption{Invariant di-$\PQb$-jet masses in the lepton channel of the 2014 analysis, (a) from particle flow reconstruction at the time followed by Durham jet clustering and (b) from particle flow reconstruction at the time followed by cheated clustering and pairing. If (b) could be achieved in reality, the $\PZ\PH\PH$ cross-section precision would improve by 40\%. The normalisation is arbitrary.}
    \label{fig:dijet_junping}
\end{figure}

The precise reconstruction of invariant di-jet masses (and related kinematic observables) for all jet flavours, including $\PQb$-jets, has been recognised already in  the 2014 as one of the main limitations, offering the potential to improve the precision of the $\PZ\PH\PH$ cross-section measurement by up to 40\%. This large potential for improvement originates from the central role of the kinematics in separating $\PZ\PH\PH$ from $\PZ\PZ\PH$ events, in particular in case the second $\PZ$ decays into $\PQb$-jets. Figure~\ref{fig:dijet_junping} illustrates this effect with the two di-jet masses plotted against each other for $\PZ\PH\PH$, $\PZ\PZ\PH$ and $\PZ\PZ\PZ$, for the state of the reconstruction in 2014 and for cheated jet clustering.

While there is ongoing work in this direction, there is up to now no new (e.g.\ ML-based) jet clustering algorithm performing much better than Durham. However, a number of other aspects of the kinematic reconstruction have been improved considerably, leading to improvements of the invariant di-jet mass reconstruction. These comprise track refitting with improved mass hypotheses after charged hadron identification and using matrix element probabilities as part of the selection. A first-principle estimate of the covariance matrix of the four-momentum of each jet also leads to a significantly improved performance of the kinematic fit, as well as a first-principle correction for missing four-momentum from neutrinos in semi-leptonic $\PQb$-decays. We estimate conservatively that these four changes lead to an improvement equivalent to about one quarter of the effect shown in Fig.~\ref{fig:dijet_junping}, i.e.\ a 10\% improvement on the $\PZ\PH\PH$ cross-section measurement. The details of this estimate will be given below in Sec.~\ref{sec:errorflow} and Sec.~\ref{sec:sldcorr}.

\subsubsection{Particle-ID and ErrorFlow for Kinematic Fitting}
\label{sec:errorflow}
Kinematic fitting plays a central role for the accurate reconstruction of $\PZ\PH\PH$ events and offers several use-cases also benefiting the background suppression. Measurements are fitted to a model expressed in the form of conditions or constraints, with a procedure relying on the principle of least squares and Lagrange multipliers. For instance, this fitting procedure can be used to improve mass resolutions, or for hypothesis testing to suppress background processes, as well as the neutrino correction (c.f.\ Sec.~\ref{sec:sldcorr}) and jet pairing. The kinematic fitting is particularly powerful at $\Pep\Pem$ colliders as it benefits from the well-known initial state and has been widely used already at LEP. Its application at higher-energy $\Pep\Pem$ colliders requires a dedicated treatment of collinear photon radiation escaping undetected through the beam-pipe, which was already in place at the time of the previous analysis~\cite{Beckmann:2010ib}. The other crucial ingredient are well-estimated measurement resolutions. While the highly-granular, particle flow optimised detector concept of ILD already offers excellent resolutions at PFO-level, not all the information offered is fully exploited in the reconstruction, in particular for heavy-flavour jets. Several advancements in this regard have been achieved, e.g.\ in~\cite{Dudar:2024quz} and~\cite{Radkhorrami:2021cuy, Radkhorrami:2021fbp, Radkhorrami:2024phd}, and will be summarised here.  

\begin{figure}[htbp]
    \centering
     \includegraphics[width=0.95\textwidth]{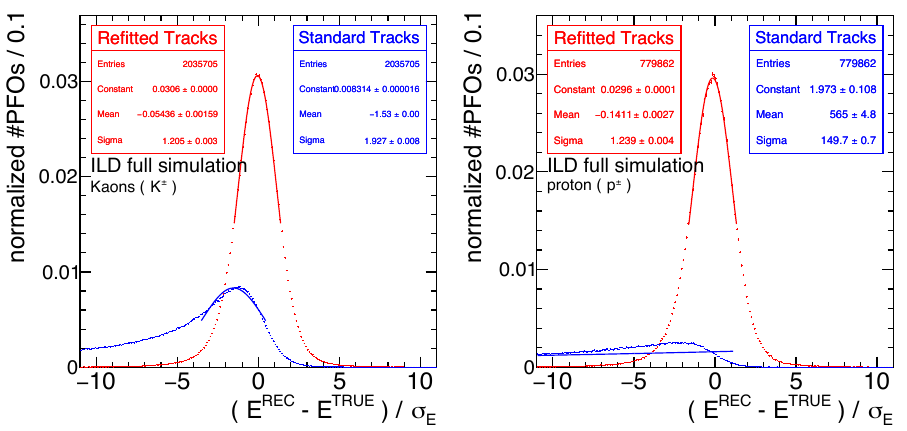}
    \caption{Normalised residual of the reconstructed energy of kaon and proton PFOs in $\Pep\Pem\to\PZ\PH$ events at $\sqrt{s}=250$\,GeV in standard reconstruction and when accounting for the correct mass~\cite{Radkhorrami:2024phd}.}
    \label{fig:E_proton_kaon_refit}
\end{figure}

For charged particles, the standard reconstruction assumes the pion mass when interpreting the four-momenta and covariance matrices from the track of each PFO. However, at low momenta with $p<$\SI{5}{\giga\electronvolt}, the differences in mass introduces discrepancies for the kaons and protons which can be resolved by refitting the tracks using the mass provided by the PID information~\cite{Dudar:2024quz}. Neutral PFOs with two tracks, e.g.\ V$^0$ decays, where an unobserved neutral strange particle decay into two observed charged daughter particles, also benefit from being refitted with the mass provided by the PID information. The dominating effect of the kaon and proton identification, however, is the consideration of the correct mass when forming the four-momentum vector of the PFO and its covariance matrix, as shown in Fig.~\ref{fig:E_proton_kaon_refit}, where the fit only converges for refitted tracks.

Neutral PFOs without tracks have their four-momentum interpreted from cluster properties where the deposited energy is considered to be the total energy of the PFO, and the momentum component is energy weighted spatial distribution of the cluster. This interpretation works well for massless particles such as photons but leads to inconsistent four-momenta of neutral hadrons resulting in incorrect covariance matrices. By introducing the mass for neutral hadrons (neutrons in full simulation and $\PKzL$ in fast simulation) to four-momentum interpretation of the cluster, the error estimation of the momentum-components can be improved  by about 15\%~\cite{Radkhorrami:2024phd}. The reinterpretation of the neutral hadron four-momenta is only adopted for recalculating the covariance matrices and do not affect the angular uncertainties which are underestimated and need to be corrected by factors of 1.3 and 1.8 for photons and neutral hadrons, respectively~\cite{Radkhorrami:2024phd}. 

\begin{figure}[htbp]
    \centering
     \includegraphics[width=0.45\textwidth]{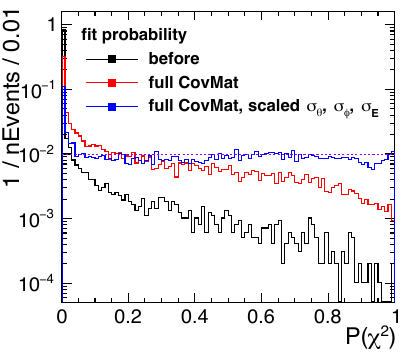}
    \caption{$\chi^2$ probability of a 4C kinematic fit on $\Pep\Pem\to\PZ\PH\to\PGmp\PGmm\PQb\PAQb$ events at $\sqrt{s}=250$\,GeV with the standard reconstruction, the ErrorFlow and ErrorFlow with scaled uncertainties for neutral PFOs~\cite{Radkhorrami:2024phd}.}
    \label{fig:ErrorFlow}
\end{figure}

The PFO-level covariance matrices can be leveraged to estimate the jet-level covariance matrix for each jet  based on its individual composition. In analogy to ``ParticleFlow'', this technique has been dubbed ``ErrorFlow''. The ErrorFlow algorithm adds up the covariance matrices of all PFOs in the jet, and estimates the effect of ParticleFlow confusion based on the density of the jet~\cite{Ebrahimi:2017ouw}. Figure~\ref{fig:ErrorFlow} shows the overall effect of the ErrorFlow-based jet covariance matrices on the fit $\chi^2$ probability of a kinematic fit with four constraints (for total energy and momentum conservation) on $\Pep\Pem\to\PZ\PH\to\PGmp\PGmm\PQb\PAQb$ events at $\sqrt{s}=250$\,GeV, where the flat $\chi^2$ probability distribution confirms the proper estimate of measurement uncertainties and correlations in the fit. While the ErrorFlow already improves this flatness a lot when relying on the raw PFO-level covariance matrices as obtained from the ParticleFlow, the final version improves the uncertainty modelling further by tuning the uncertainties for the neutral PFOs~\cite{Radkhorrami:2024phd}, as illustrated in Fig.~\ref{fig:ErrorFlow}.

\subsubsection{Neutrino Correction}
\label{sec:sldcorr}
\begin{figure}[htbp]
    \centering
    \begin{subfigure}{.5\textwidth}
    \centering
        \includegraphics[width=0.8\textwidth]{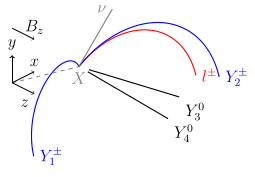}
        \caption{}
        \label{fig:sld:stable}    
    \end{subfigure}\hfill%
    \begin{subfigure}{.5\textwidth}
        \centering
        \includegraphics[width=0.95\textwidth]{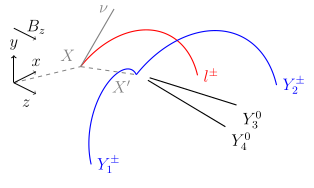}
        \caption{}
        \label{fig:sld:unstable}    
    \end{subfigure}%
    \caption{Schematic view of semi-leptonic decays where the missing energy carried away by the neutrino can be corrected, where in (a) the decay includes all stable particles, and in (b) the decay includes an unstable particle travelling a small distance before decaying further~\cite{Radkhorrami:2024phd}.}
    \label{fig:sld}
\end{figure}

Neutrinos present in semi-leptonic decays of heavy ($\PQb$ and $\PQc$) jets degrade the jet energy scale and resolution as energy is carried away by the undetectable neutrinos, and thus limit the accurate reconstruction of di-jet systems like $\PH \to \PQb\PQb$. The missing energy from the neutrinos can be corrected on an event-by-event basis from kinematics by identifying the partner lepton within the jet and their associated semi-leptonic decay. There are two cases where this is possible; when the lepton is associated to a secondary vertex, or when the lepton is a single track but another vertex exists in the jet. Figure~\ref{fig:sld} shows a schematic view of the semi-leptonic decay for each of the two cases. In Fig.~\ref{fig:sld:stable} the heavy hadron, X, decays to detector-stable particles, and in Fig.~\ref{fig:sld:unstable} the decay of the heavy hadron includes an unstable particle, X$^{\prime}$, that flies a small distance before decaying to stable particles that can be detected. The neutrino correction requires the mass and flight direction of the heavy hadron as well as the four-momentum of all the visible decay products. From that the energy of the neutrino can be derived down to an ambiguity resulting from the unknown sign of the momentum component~\cite{Radkhorrami:2024phd}, which can be solved using a kinematic fit only imposing four-momentum conservation on the event. Currently only semi-leptonic decays of $\PB$-hadrons are corrected but in principle the method also applies to charmed hadrons.

The flight direction of the heavy hadron is used as a reference frame and therefore needs to be well reconstructed. When all the decay particles of the heavy hadron are stable, the flight direction can be taken from the straight line between the primary and secondary vertex, but if there is an unstable particle separating the tertiary vertex from the lepton, the flight direction is found from their intersection point~\cite{Radkhorrami:2024phd}. For both cases, a set of charged particles, denoted Y$^{\pm}$ in Fig.~\ref{fig:sld}, will be associated with the vertex but there might be additional charged particles as well as neutral particles, Y$^0$ in Fig.~\ref{fig:sld}, that should also be associated with the vertex belonging to the semi-leptonic decay. These particles are found using a cone technique using a critical angle cut as well as multiple mass cuts depending on the type of semi-leptonic decay and whether an associated particle is charged or neutral. More information can be found in~\cite{Radkhorrami:2024phd}. 

\begin{figure}[htbp]
    \centering
     \includegraphics[width=0.85\textwidth]{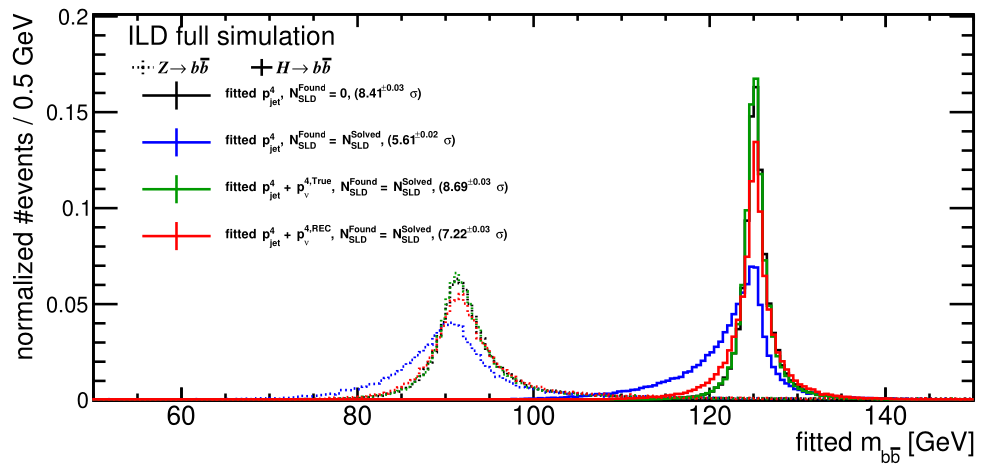}
    \caption{Fitted $\PH/\PZ\rightarrow \PQb\PAQb$ mass reconstruction in $\PZ\PH/\PZ\PZ\rightarrow \PGmp\PGmm \PQb\PAQb$ events for events with no semi-leptonic decays and in black compared to events with correctable semi-leptonic decays with no neutrino correction added to the fit in blue, with the true neutrino energy added to fit in green, and with the neutrino correction reconstructed from kinematics added to the fit in red~\cite{Radkhorrami:2024phd}.}
    \label{fig:nucorr}
\end{figure}

Solving the sign ambiguity of the neutrino correction requires additional information about the event. By imposing four-momentum conservation on the event each of the $\pm$ neutrino corrections can be compared against each other as well as against a zero solution where no neutrino correction is added to a jet. As multiple jets can contain a semi-leptonic decay, each combination of neutrino corrections has to be tested. For example, for two $\PQb$-jets, each with a correctable semi-leptonic decay, there would be 9 combinations to fit. In addition to the four-momentum neutrino correction, an uncertainty on the correction must also be added to the jet covariance matrix akin to ErrorFlow. However, this is not an event-specific error estimation, but is instead parametrised by comparing the neutrino correction energy and direction to truth level~\cite{Radkhorrami:2024phd}. Figure~\ref{fig:nucorr} shows how the neutrino correction improves the invariant $\PQb\PAQb$ mass reconstruction in $\PZ\PH\rightarrow \PGmp\PGmm \PQb\PAQb$ events and $\PZ\PZ\rightarrow \PGmp\PGmm \PQb\PAQb$ events, compensating the missing energy carried away. In black are event with no semi-leptonic decays compared to events with correctable semi-leptonic decays in blue, green, and red. In blue only the visible jet energy and momentum is fitted, showing that the fit cannot compensate for the missing neutrino energy on its own. In green the true neutrino four-momentum has been added to the reconstructed jet four-momentum, showing a more than perfect compensation, which can be explained by the fact that in case of jets without semi-leptonic decays, the full jet energy is from reconstructed information, thus comes with its measurement uncertainty. Finally, in red histogram, the neutrino correction is used, showing a compensation that comes close to the reconstruction quality of jets without semi-leptonic decays. 

\subsection{Flavour Tagging}
\begin{figure}[htbp]
    \centering
    \begin{subfigure}{.45\textwidth}
        \centering
        \includegraphics[width=0.95\textwidth]{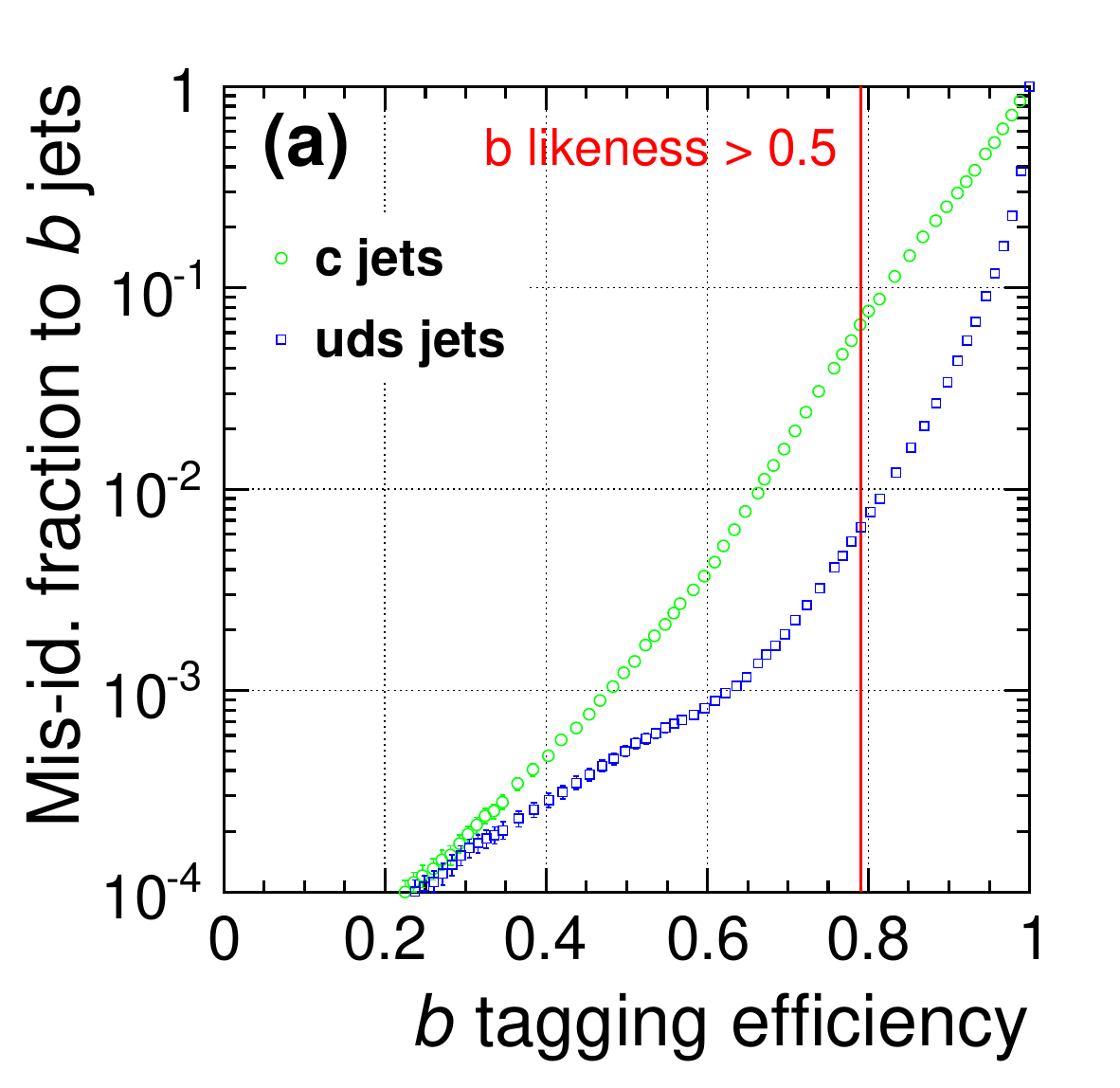}
        \caption{}
        \label{fig:complcfi:plot1}
    \end{subfigure}\hfill%
    \begin{subfigure}{.45\textwidth}
        \centering
        \includegraphics[width=0.95\textwidth]{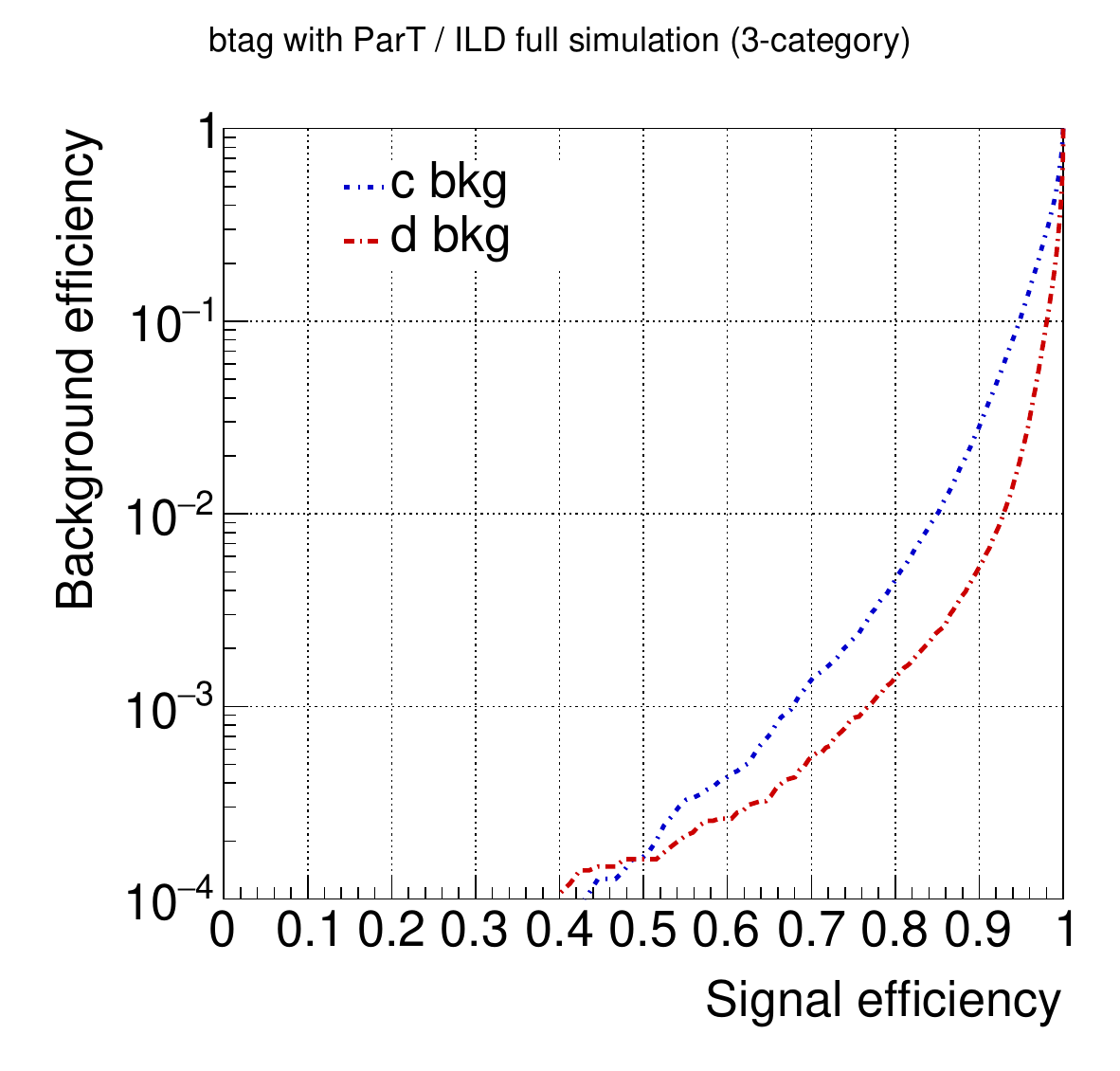}
        \caption{}
        \label{fig:complcfi:plot2}    
    \end{subfigure}%
    \caption{$\PQb$-tagging performance in ILD full simulation in terms of mis-ID fraction vs efficiency for (a) LCFIPlus on $\Pep\Pem \to \PQq\PAQq$ events at $\sqrt{s}=91$\,GeV produced at the time on the 2014 analysis (DBD simulation) and (b) ParT (3 cat.) on recently produced (mc2020) $\Pep\Pem \to \PGn\PAGn\PQq\PAQq$ events at $\sqrt{s}=250$\,GeV.}
    \label{fig:complcfi}
\end{figure}

Highly efficient identification of jet flavour, especially $\PQb$-jet tagging, is essential for the di-Higgs analysis, especially in the $\PH\PH \to 4\PQb$ mode, since many backgrounds have -- always or often -- at most two $\PQb$-jets.
The previous jet flavour tagging algorithm included in the LCFIPlus package~\cite{Suehara:2015ura} utilizes high-level features of jets such as
properties of secondary vertices, jet mass, etc.\ as inputs to Boosted Decision Trees (BDT) to obtain a classification into $\PQb$, $\PQc$ and light jets.
The algorithm was used in the 2014 $\PZ\PH\PH$ analysis as well as various ILD-related analyses until recently.
It has classically been trained on $\Pep\Pem \to \PQq\PQq$ events at $\sqrt{s}=91$\,GeV. Its performance is shown in Fig.~\ref{fig:complcfi:plot1}.

In response to the recent rapid developments on neural network technologies such as Transformers, showing already tremendous impact on jet flavour tagging at LHC and $\Pep\Pem$ Higgs Factories (for a recent review see e.g.~\cite{Altmann:2025feg}),
we are replacing the algorithm with one based on the Particle Transformer (ParT) architecture~\cite{Qu:2022mxj, Tagami:2024gtc}. In contrast to LCFIPlus, the ParT-based algorithm
utilizes features of individual reconstructed tracks as well as neutral particles inside the jets. The transformer architecture used to learn relationships between two objects (particles in our case) by the self-attention mechanism enables to efficiently derive characteristics of each jet from individual particles. In addition to the standard transformer, ParT uses additional kinematic relations between two particles to bias attention weights of the particles. We implemented separate embedding layers for charged and neutral particles, which have different
types of information. 

The training of the ParT-based algorithm has been performed on $\Pep\Pem \to \PGn\PAGn \PH \to \PGn\PAGn\PQq\PAQq$ (with \PQq = \PQb, \PQc, \PQs, \PQu, \PQd, \Pg)
final states at a centre-of-mass energy of 250\,GeV, pre-clustered into a 2-jet configuration with the Durham~\cite{Durham} algorithm. The available number of jets is 10$^6$ with ILD full detector
simulation and reconstruction, while 10$^7$ jets are available from SGV simulation. 80\% of the full statistics are used for training with the remaining used for validation and performance evaluation. The performance of the 3-class ($\PQb$, $\PQc$ and light jets) training is shown in Fig.~\ref{fig:complcfi:plot2}.

Though the event characteristics is different between the plots, the comparison of Fig.~\ref{fig:complcfi:plot1} and Fig.~\ref{fig:complcfi:plot2} already gives a good impression of the performance difference since the performance of LCFIPlus is not heavily dependent on the jet energy~\cite{Suehara:2015ura}. The comparison shows a reduction of the surviving background by about a factor of 10 at a $\PQb$-tagging efficiency of 80\%. 

In addition to the 3-category tagging of $\PQb$, $\PQc$ and light jets, the ParT-based algorithm can also be configured for 6-category
($\PQb$, $\PQc$, $\PQs$, $\PQu$, $\PQd$, $\Pg$) or 11-category (adding separation of quark and anti-quark) classification.
For strange tagging, we have added features related to Particle ID derived from Comprehensive PID (CPID) algorithm~\cite{Einhaus:2023oqy} implemented in the ILD reconstruction chain
to identify kaons, which is essential to separate strange jets. CPID is a BDT-based multivariate algorithm
utilizing dE/dx and time-of-flight of tracks as features. The time detection is assumed to be performed in the first 10 layers of the electromagnetic calorimeter
with 100\,ps timing resolution. For a detailed discussion of timing information in ILD we refer to~\cite{Dudar:2024quz}. The PID information is not yet fully implemented in SGV, therefore the strange-jet tagging is not expected to reach the same performance as in full simulation. This, however, is not expected to influence the analysis at hand, which only uses $\PQb$-jet identification.

\begin{figure}[htbp]
    \centering
    \begin{subfigure}{.3\textwidth}
        \centering
        \includegraphics[width=0.95\textwidth]{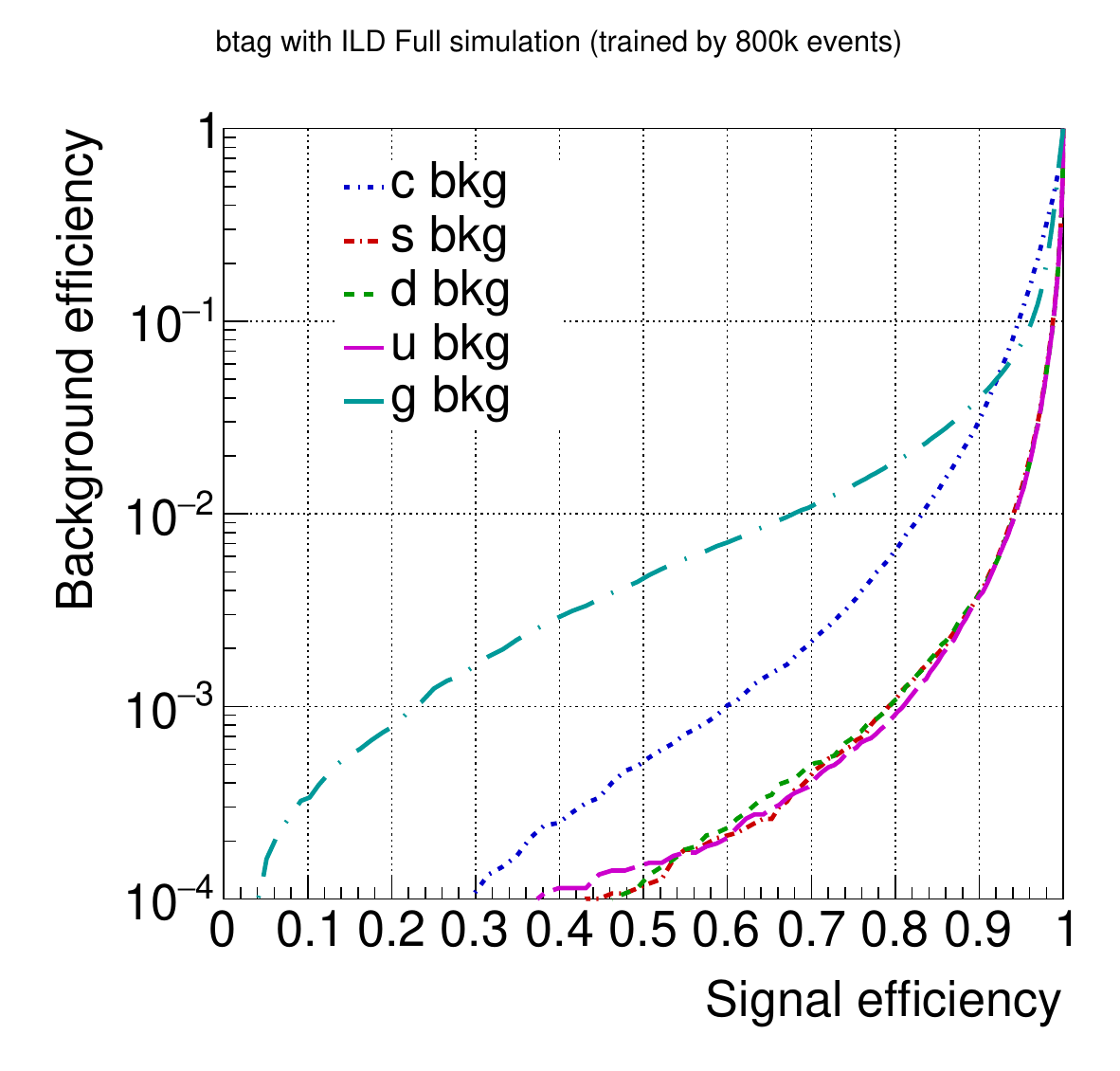}
        \caption{}
        \label{fig:compfullsgv:plot1}
    \end{subfigure}\hfill%
    \begin{subfigure}{.3\textwidth}
        \centering
        \includegraphics[width=0.95\textwidth]{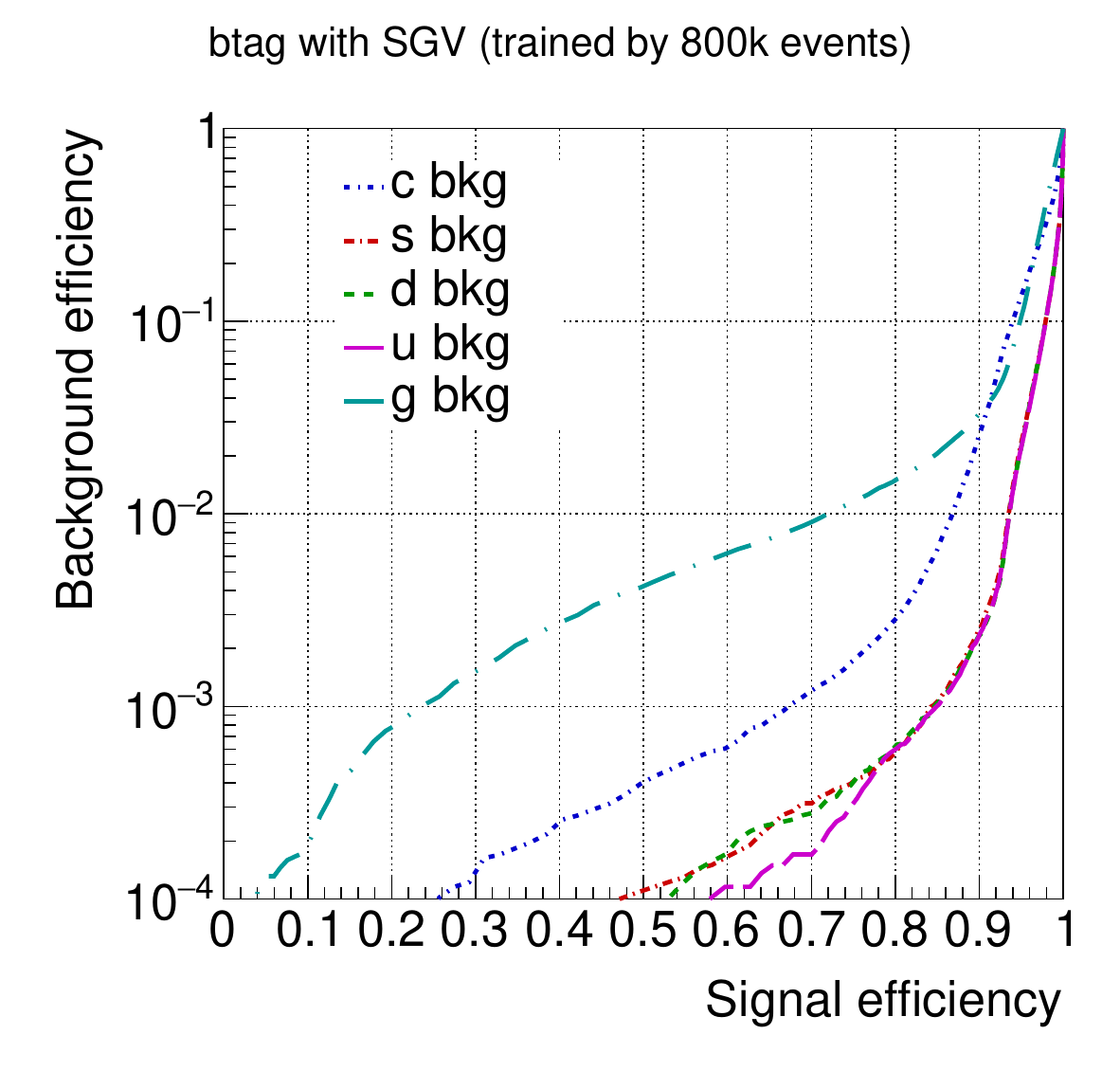}
        \caption{}
        \label{fig:compfullsgv:plot2}
    \end{subfigure}\hfill%
    \begin{subfigure}{.3\textwidth}
        \centering
        \includegraphics[width=0.95\textwidth]{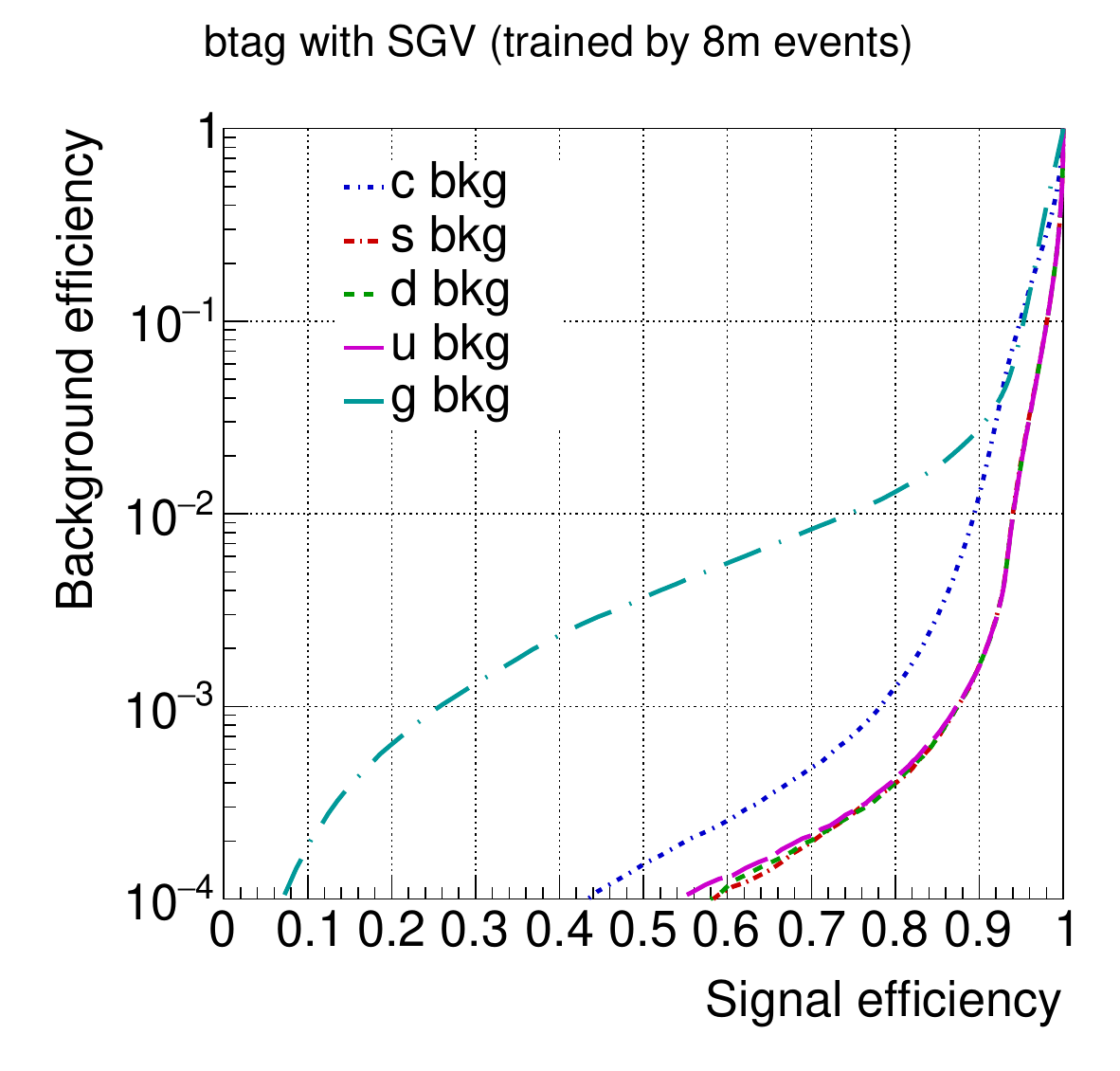}
        \caption{}
        \label{fig:compfullsgv:plot3}
    \end{subfigure}\hfill%
    \caption{Jet-flavour tagging performance of ParT (11 cat.) on $\Pep\Pem \to \PGn\PAGn\PQq\PAQq$ events at $\sqrt{s}=250$\,GeV (a) on 800k-jets training sample in ILD full simulation (mc2020) (b) on 800k-jets training sample in ILD SGV simulation (c) 8-million-jets training sample in ILD SGV  simulation. Quark and antiquark probabilities are added to derive the efficiencies.}
    \label{fig:compfullsgv}
\end{figure}

\begin{figure}[htbp]
    \centering
    \begin{subfigure}{.45\textwidth}
        \centering
        \includegraphics[width=0.95\textwidth]{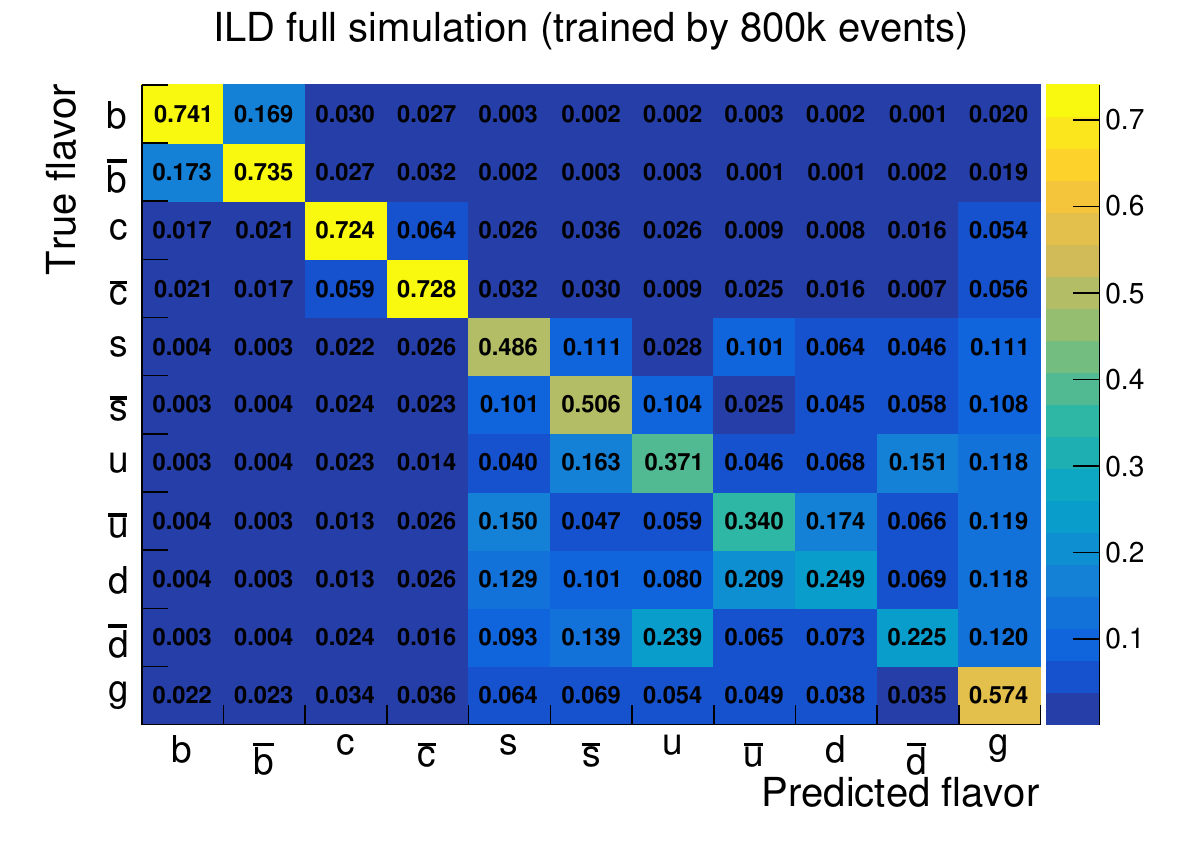}
        \caption{}
        \label{fig:matrix:plot1}
    \end{subfigure}\hfill%
    \begin{subfigure}{.45\textwidth}
        \centering
        \includegraphics[width=0.95\textwidth]{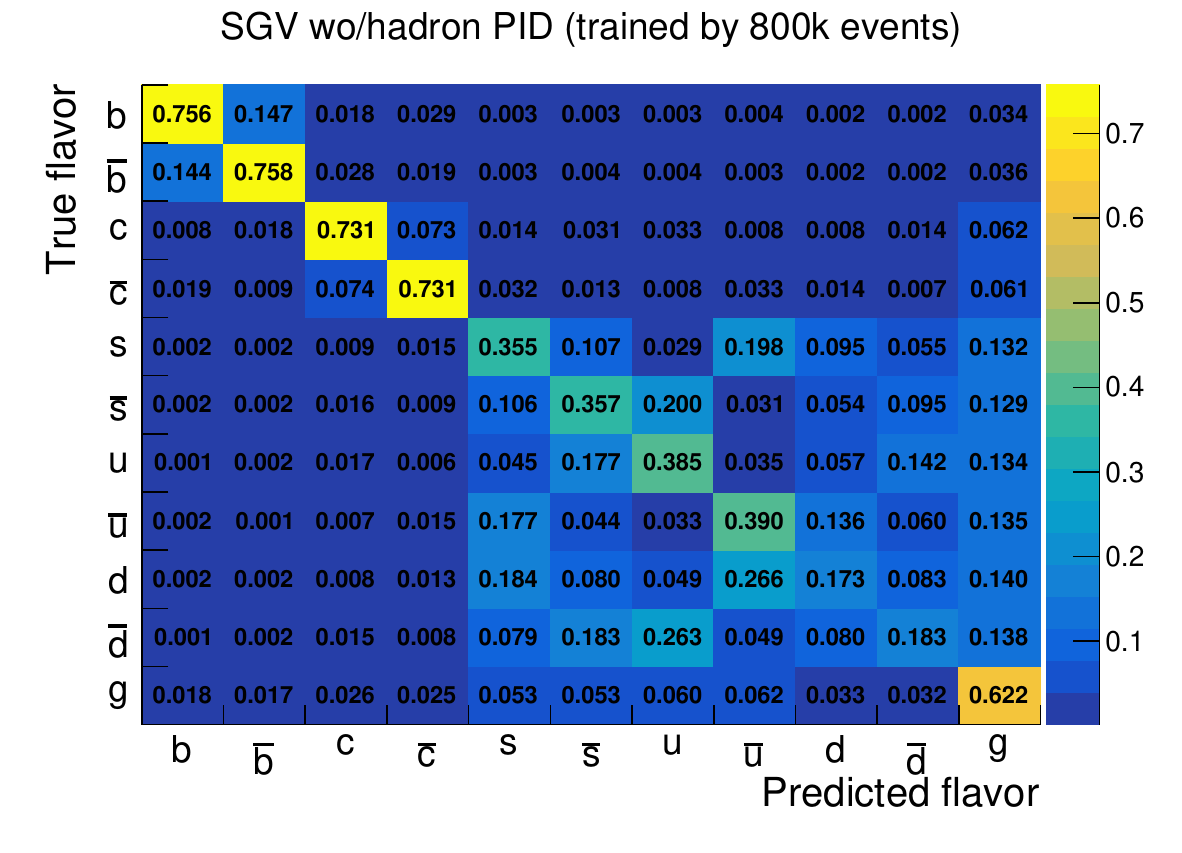}
        \caption{}
        \label{fig:matrix:plot2}    
    \end{subfigure}%
    \caption{Jet-flavour tagging performance of ParT (11 cat.) on $\Pep\Pem \to \PGn\PAGn\PQq\PAQq$ events at $\sqrt{s}=250$\,GeV (a) on 800k-jets training sample in ILD full simulation (mc2020) (b) on 800k-jets training sample in ILD SGV simulation. The numbers in the bins correspond to purities, i.e.\ add up to unity per row.}
    \label{fig:matrix}
\end{figure}

Figure~\ref{fig:compfullsgv} gives b-tag performance of 11-category tagging based on ParT-based algorithm on full ILD simulation as well as on SGV fast simulation for two different sample sizes. 
Comparing full simulation and SGV trained on the same number of events shows a very good agreement between the two, in particular in view of the simpler PID information available in SGV. This is also confirmed by the matrix classification shown in Fig.~\ref{fig:matrix}, apart from SGV -- as expected -- giving worse performance for strange tagging due to the lack of PID-related variables. Figure~\ref{fig:compfullsgv} also shows that a ten times larger training sample, currently only available in fast simulation, improves the performance. This indicates that further improvement with full simulation sample should also be possible. A large-scale full detector simulation is under preparation to confirm this.

Table~\ref{tab:flavtag} shows differences between LCFIPlus and ParT on full simulation and SGV in terms of the $\PQb$-tag efficiency at fixed $\PQc$-jet mis-ID rates of 10\%, 3\% and 1\%. For all three mis-ID rates, all the ParT configurations give a significantly higher $\PQb$-jet efficiency than LCFIPlus by about 15\%, 20\% and 25\%, relative to LCFIPlus. For mis-ID rates of 10\% and 3\%, representative for the $\PZ\PH\PH$ analysis as we will discuss in more detail in Sec.~\ref{sec:extrapol:flav}, there is excellent agreement between full ILD simulation and SGV. Only at the highest purities, ParT achieves a few percent higher efficiency on SGV than on full simulation, and also profits most from increased training statistics.

\begin{table}[htb]
\centering
\begin{tabular}{lrrr }
\hline
$\PQc$-jet mis-ID rate & 10\% & 3\% & 1\% \\ 
\hline
LCFIPlus, 91 GeV $\PQq\PAQq$ and fullsim                               & 82\%  & 74\% & 67\% \\ 
ParT, 250 GeV $\PGn\PAGn\PQq\PAQq$ and fullsim, 800k jets, 3-category    & 95\%  & 90\% & 85\% \\ 
ParT, 250 GeV $\PGn\PAGn\PQq\PAQq$ and fullsim, 800k jets, 11-category   & 95\%  & 90\% & 83\% \\ 
ParT, 250 GeV $\PGn\PAGn\PQq\PAQq$ and SGV, 800k jets, 11-category       & 94\%  & 90\% & 87\% \\ 
ParT, 250 GeV $\PGn\PAGn\PQq\PAQq$ and SGV, 8 million jets, 11-category  & 95\%  & 92\% & 89\% \\ 
\hline
\end{tabular}%
    \caption{b-tag efficiency in fixed c-background rejection with various configuration. }
    \label{tab:flavtag}
\end{table}

\subsection{Expected Impact on ZHH Analysis}
\label{sec:extrapol}
In this section, we comment on the performance of the new high-level reconstruction tools introduced in the previous two sections on $\PZ\PH\PH$ events and some of the main background processes in the $\PZ\PH\PH$ analysis.

\subsubsection{Kinematic Reconstruction}
\label{sec:extrapol:kin}
\begin{figure}[htbp]
    \centering
    \begin{subfigure}{.5\textwidth}
    \centering
        \includegraphics[width=0.95\textwidth]{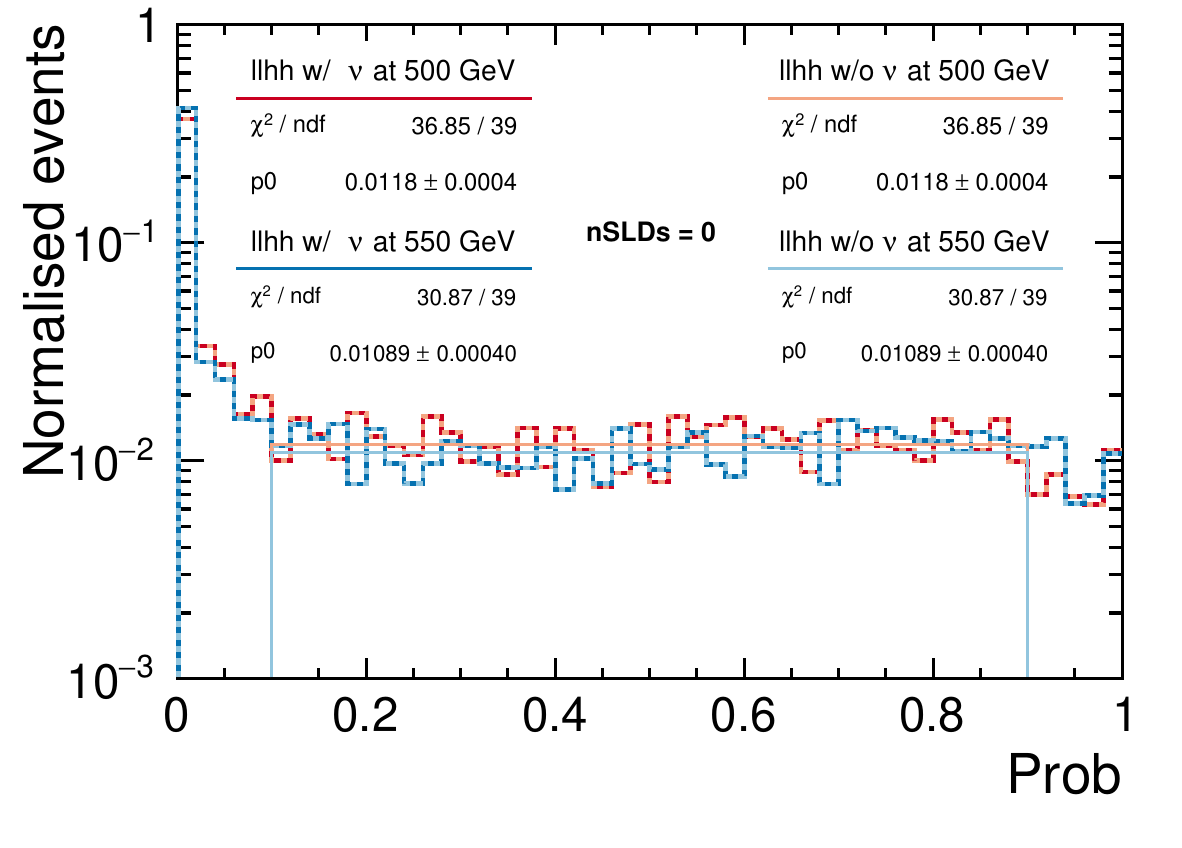}
        \caption{}
        \label{fig:4Cprob:wosld}    
    \end{subfigure}\hfill%
    \begin{subfigure}{.5\textwidth}
        \centering
        \includegraphics[width=0.95\textwidth]{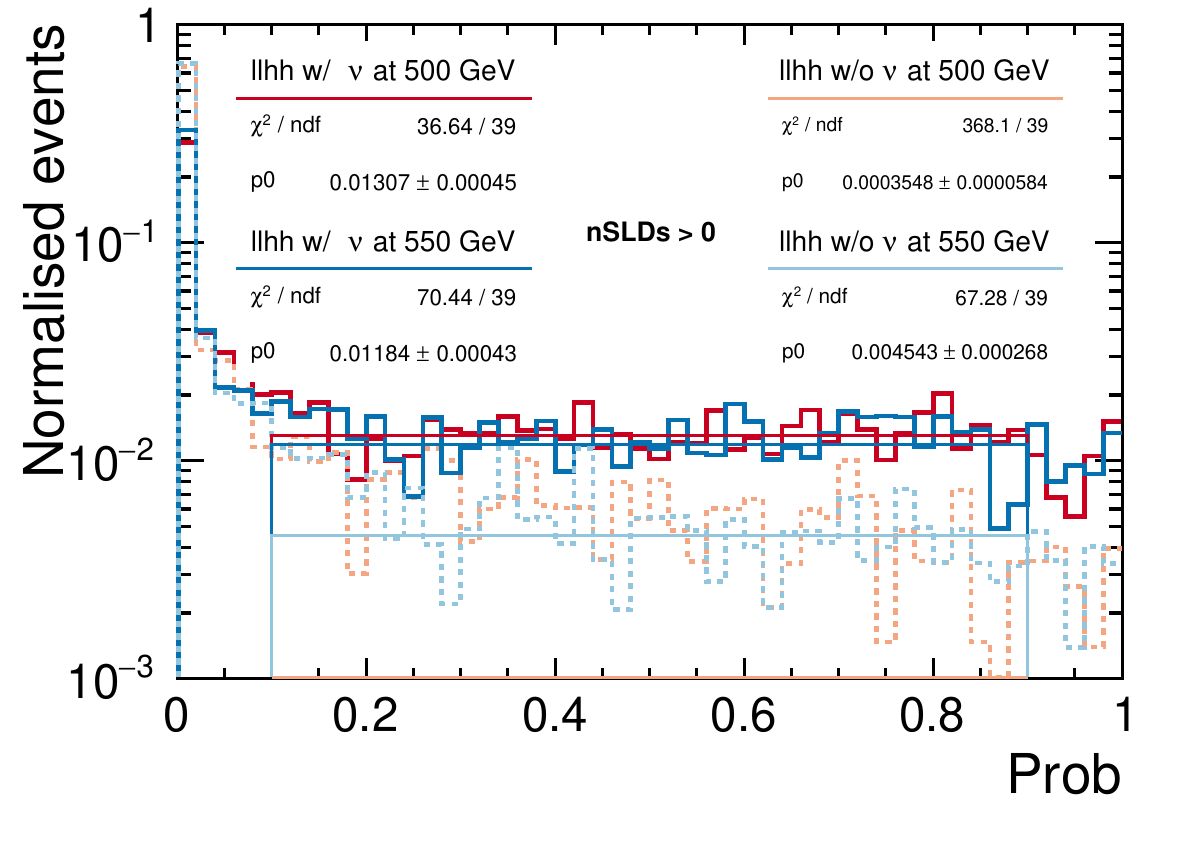}
        \caption{}
        \label{fig:4Cprob:wsld}    
    \end{subfigure}%
    \caption{Impact of including the neutrino correction on the fit probability of a 4C kinematic fit on $\PZ\PH\PH$ events without (a) and with (b) semi-leptonic decays present in the fitted events, also comparing centre-of-mass energies of 500 GeV to 550 GeV.}
    \label{fig:4Cprob}
\end{figure}

In the context of the di-Higgs analysis, the algorithms described in Sec.~\ref{sec:errorflow} and Sec.~\ref{sec:sldcorr} and developed on $\PZ\PH$ events, have been successfully ported to the more complex case of $\PZ\PH\PH$ events. Here, at least one semi-leptonic decay exists in about 2/3 of the events and for half of the events the neutrino correction can be applied. As described in Sec.~\ref{sec:sldcorr}, the ambiguities in the neutrino correction are resolved with a 4C kinematic fits where 4 constraints from four-momentum conservation are imposed on the event. The fit probabilities are shown for events without and with semi-leptonic decays present in Figs.~\ref{fig:4Cprob:wosld} and~\ref{fig:4Cprob:wsld}, respectively, exhibiting the improvement of including the neutrino correction. For a well-functioning fit, well-estimated errors are crucial, and a uniform distribution is expected which is evident of Fig.~\ref{fig:4Cprob:wosld}, testifying to the success of ErrorFlow and the improvements described in Sec.~\ref{sec:errorflow}. For events where no semi-leptonic decays are present there is of course nothing to correct, hence equal distributions, but the fit probability clearly improves when the missing neutrinos are corrected, and the effect on the fit probability appears completely compensated by the neutrino correction when comparing the uniformity between the distributions in Fig.~\ref{fig:4Cprob:wosld} and Fig.~\ref{fig:4Cprob:wsld}.     

\begin{figure}[htbp]
    \centering
    \begin{subfigure}{.5\textwidth}
    \centering
        \includegraphics[width=0.95\textwidth]{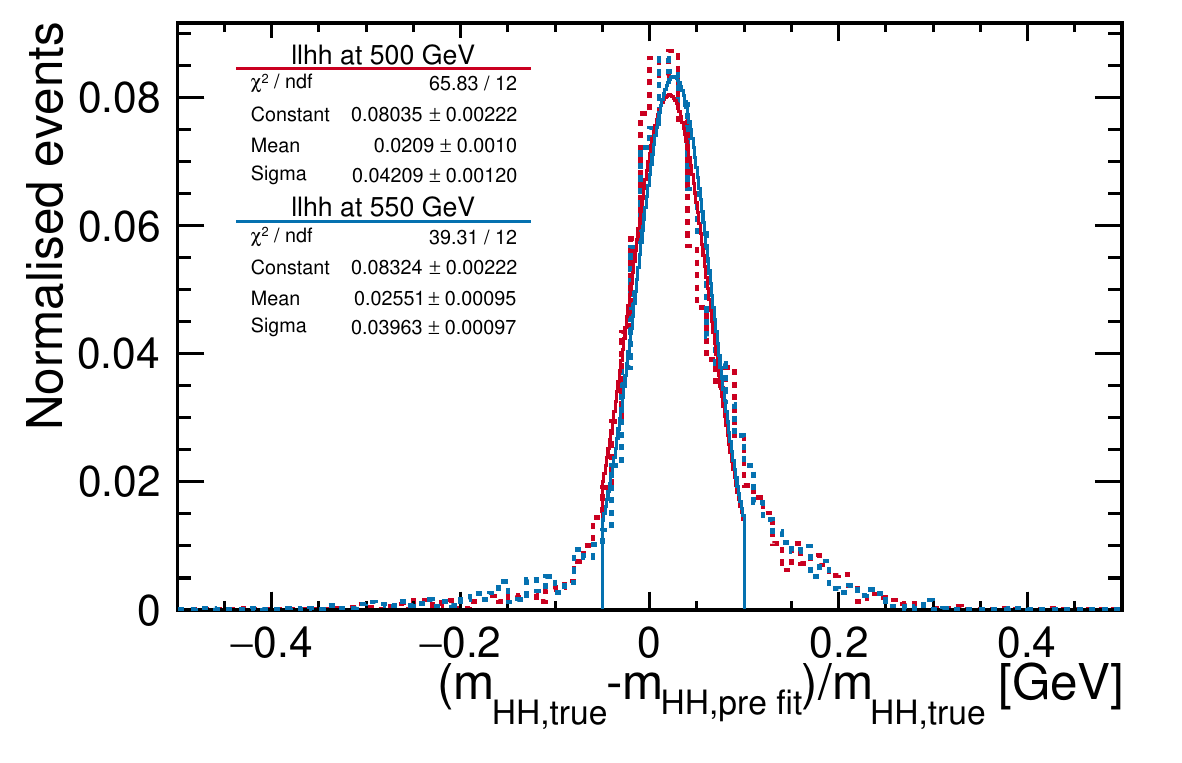}
        \caption{}
        \label{fig:mhh:chi2}    
    \end{subfigure}\hfill%
    \begin{subfigure}{.5\textwidth}
        \centering
        \includegraphics[width=0.95\textwidth]{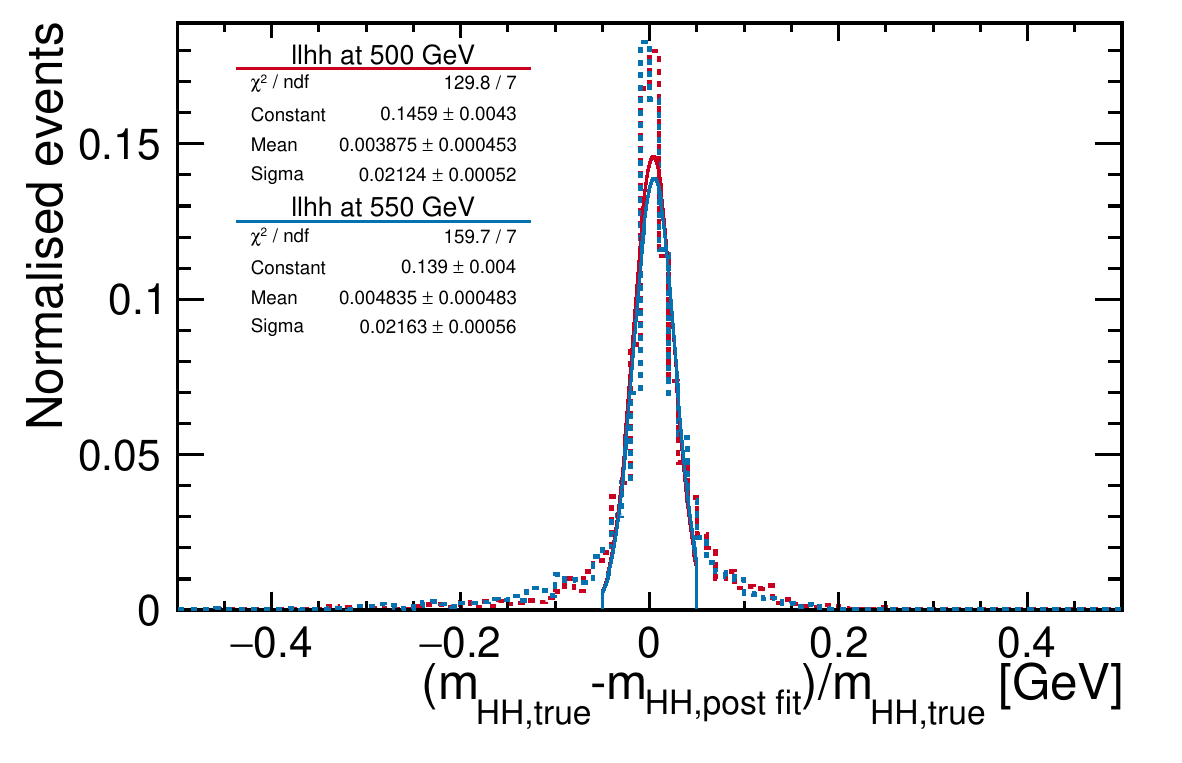}
        \caption{}
        \label{fig:mhh:kinfit}    
    \end{subfigure}%
    \caption{Impact of the full new kinematic fit including the neutrino correction on the reconstruction of $m_{\PH\PH}$, here showing the resolution of $m_{\PH\PH}$ prior to the 4C fit in (a) and posterior to the 4C fit in (b) improving by a factor 2, to the level of 2\%, also comparing centre-of-mass energies of 500 GeV to 550 GeV.
    }
    \label{fig:mhh}
\end{figure}

The 4C fit also improves the reconstruction of the di-Higgs invariant mass, in addition to resolving the neutrino correction. Figure~\ref{fig:mhh} shows how the di-Higgs mass resolution in the lepton channel, while already good at a few percent before the 4C fit, improves by a factor 2, to the level of 2\%, after applying it. In the future, the di-Higgs mass will be very useful in two ways: it is a quite powerful discriminating observable against background processes -- and it is known to depend on the value of the self-coupling, i.e.\ it could be used to improve the extraction of the self-coupling beyond the total cross-section method and could provide decisive clues with respect to the presence of BSM Higgs bosons~\cite{Arco:2025pgx}.  In the results presented in Sec.~\ref{sec:ana550} and Sec.~\ref{sec:results:self}, the improved $m_{\PH\PH}$ resolution is not yet taken into account.

Both the 2014 analysis and the work presented in Sec.~\ref{sec:ana550:lepton} make use of further kinematic fits which apply mass constraints in addition to four-momentum conservation. These mass constrained fits are used for jet pairing (needed for the reconstruction of di-jet masses) as well as hypothesis testing. One of the main irreducible backgrounds of $\PZ\PH\PH$ are $\PZ\PZ\PH$ events, hence mass reconstruction also at di-jet  level is crucial for background suppression in the MVA selection described in Sec.~\ref{sec:ana550:lepton}. In the 6C fits, six constraints are imposed on the events, the same four constraints from energy and momentum conservation with two additional mass-constraints on the di-jets. In the $\PZ\PH\PH$ hypothesis, the hard mass constraints imposed are $m_{j_1j_2}=m_{j_3j_4}=\SI{125}{\giga\electronvolt}$ with 3 possible permutations of the jets, and in the $\PZ\PZ\PH$ hypothesis they are $m_{j_1j_2}=\SI{125}{\giga\electronvolt}$ and $m_{j_3j_4}=\SI{91.2}{\giga\electronvolt}$ with 6 possible permutations of the jets. The pairing is chosen from the fit of the 9 permutations with the best $\chi^2$. As input to the fit, the neutrino correction chosen by the 4C fit is added to the pre-fitted jets and the corrected pre-fit jets and the pre-fit leptons are (re)-fitted. 

\begin{figure}[htbp]
    \centering
    \begin{subfigure}{.46\textwidth}
    \centering
        \includegraphics[width=0.95\textwidth]{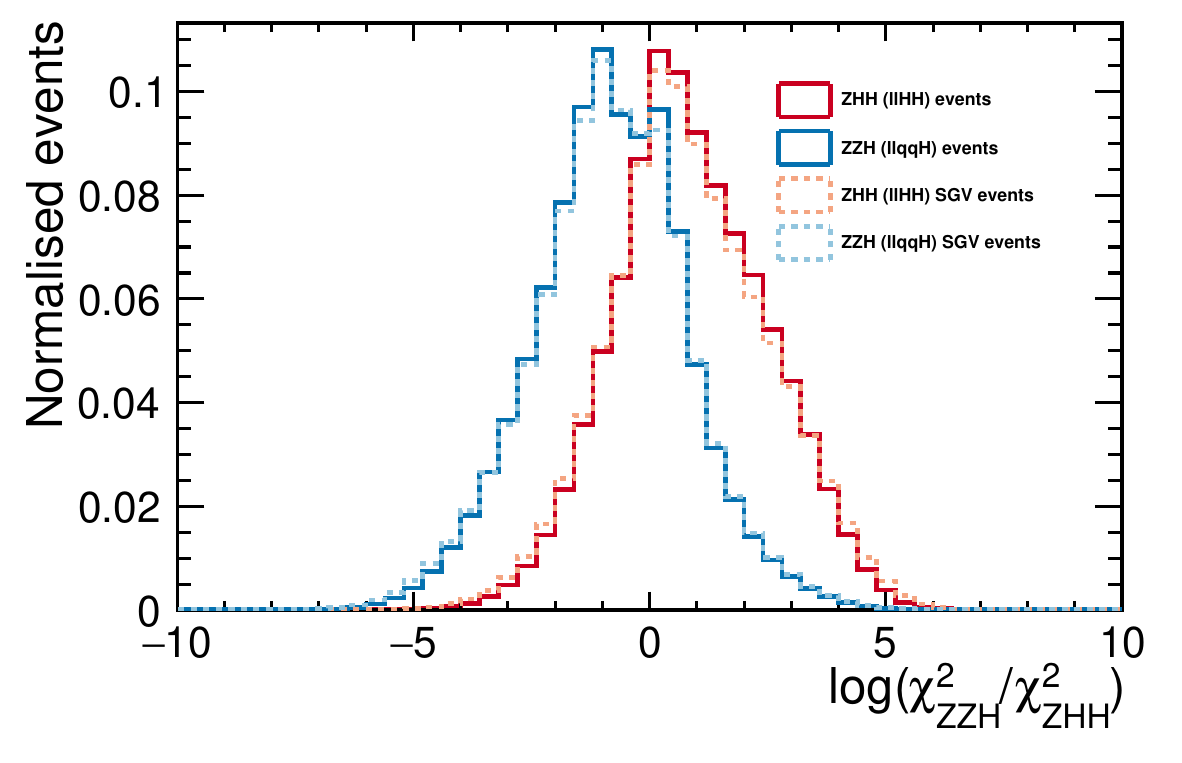}
        \caption{}
        \label{fig:chi2sgv:ratio}    
    \end{subfigure}\hfill%
    \begin{subfigure}{.54\textwidth}
        \centering
        \includegraphics[width=0.95\textwidth]{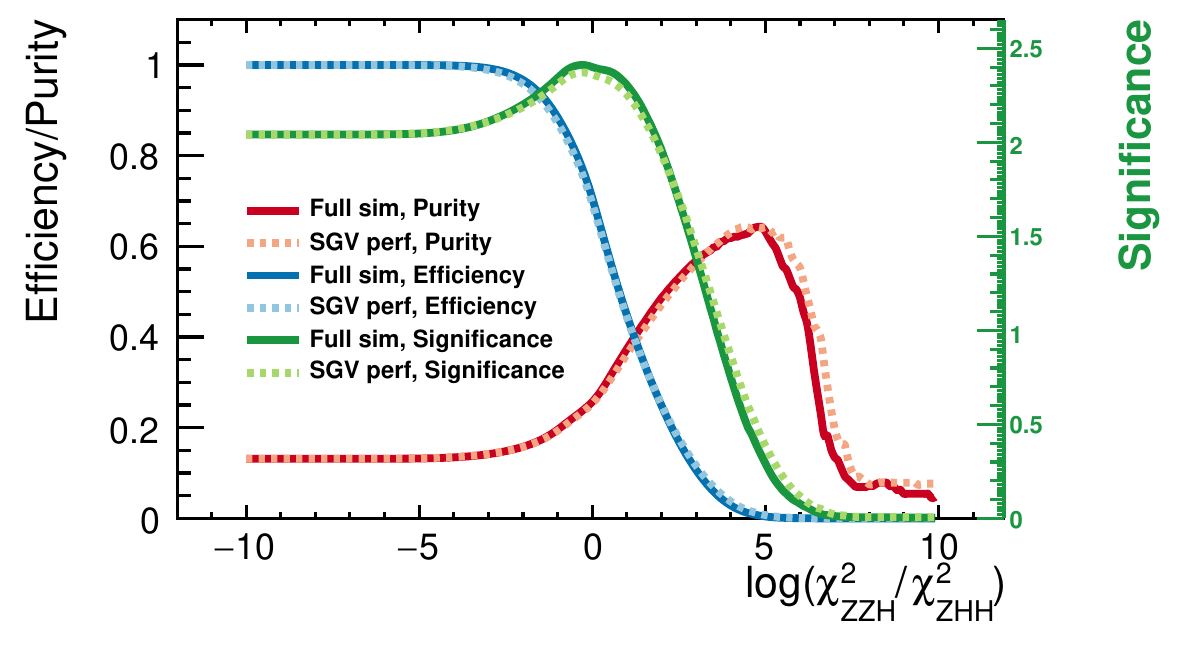}
        \caption{}
        \label{fig:chi2sgv:significance}    
    \end{subfigure}%
    \caption{Kinematic fit performance in SGV and full ILD simulation: (a) $\chi^2$ ratios of the $\PZ\PH\PH$ and $\PZ\PZ\PH$ hypotheses on both $\PZ\PH\PH$ and $\PZ\PZ\PH$ events, (b) efficiency, purity and significance obtained from a simple cut on the $\chi^2$ ratio.}
    \label{fig:chi2sgv}
\end{figure}

The behaviour of the kinematic fit on full ILD simulation and SGV has been cross-checked in detail, and the modelling of the ErrorFlow performance in SGV has been adjusted such that the $\chi^2$ of the 6C fits give equivalent performance for hypothesis testing. Figure~\ref{fig:chi2sgv:ratio} shows an excellent agreement of the $\chi^2$ ratios of the $\PZ\PH\PH$ and $\PZ\PZ\PH$ hypotheses on both $\PZ\PH\PH$ and $\PZ\PZ\PH$ events. While in a real analysis, these observables serve as input to an MVA, their distributions in Fig.~\ref{fig:chi2sgv:significance} illustrate the agreement in terms of the significance obtained from a simple cut on the $\chi^2$ ratio.

\begin{figure}[htbp]
    \centering
    \begin{subfigure}{.5\textwidth}
        \centering
        \includegraphics[width=0.95\textwidth,page=1]{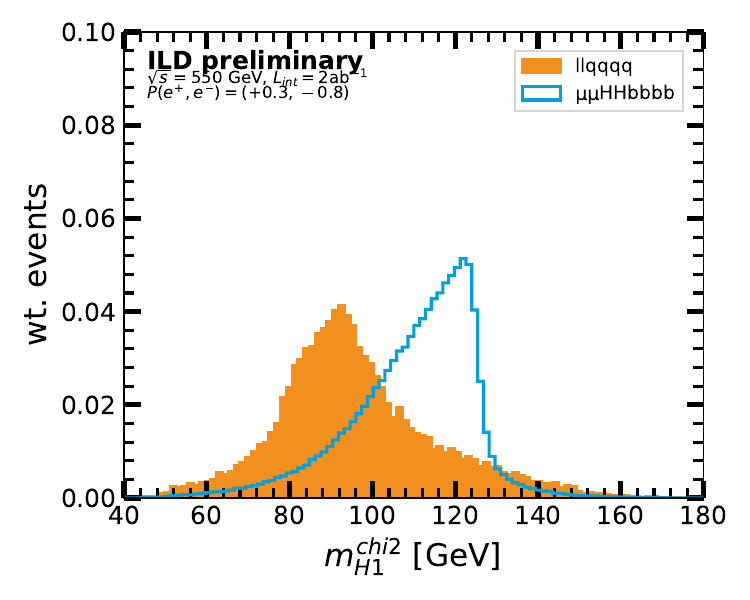}
        \caption{}
        \label{fig:mhllqqqq:mh1chi2}    
    \end{subfigure}%
    \begin{subfigure}{.5\textwidth}
        \centering
        \includegraphics[width=0.95\textwidth,page=2]{figures/Bryan_mHi_llqqqq_mumubbbb_250830.pdf}
        \caption{}
        \label{fig:mhllqqqq:mh1kinfit}    
    \end{subfigure}\hfill%
    \begin{subfigure}{.5\textwidth}
        \centering
        \includegraphics[width=0.95\textwidth,page=3]{figures/Bryan_mHi_llqqqq_mumubbbb_250830.pdf}
        \caption{}
        \label{fig:mhllqqqq:mh2chi2}
    \end{subfigure}%
    \begin{subfigure}{.5\textwidth}
    \centering
        \includegraphics[width=0.95\textwidth,page=4]{figures/Bryan_mHi_llqqqq_mumubbbb_250830.pdf}
        \caption{}
        \label{fig:mhllqqqq:mh2kinfit}    
    \end{subfigure}\hfill%
    \caption{Invariant masses of the two di-jet Higgs candidates in the $\PZ \to \Pl\Pl$ channel as used in the old (top row) and new (bottom row) analysis for signal events and $\Pl\Pl\PQq\PQq\PQq\PQq$ background: (a) lower and (b) higher of the two di-jet masses using pre-fit momenta and a simple $\chi^2$ jet pairing, and (c) lower and (d) higher of the two di-jet masses using the 4C-fit momenta and 6C-fit jet pairing. } 
    \label{fig:mhllqqqq}
\end{figure}

\begin{figure}[htbp]
    \centering
    \begin{subfigure}{.33\textwidth}
    \centering
        \includegraphics[width=0.95\textwidth,page=1]{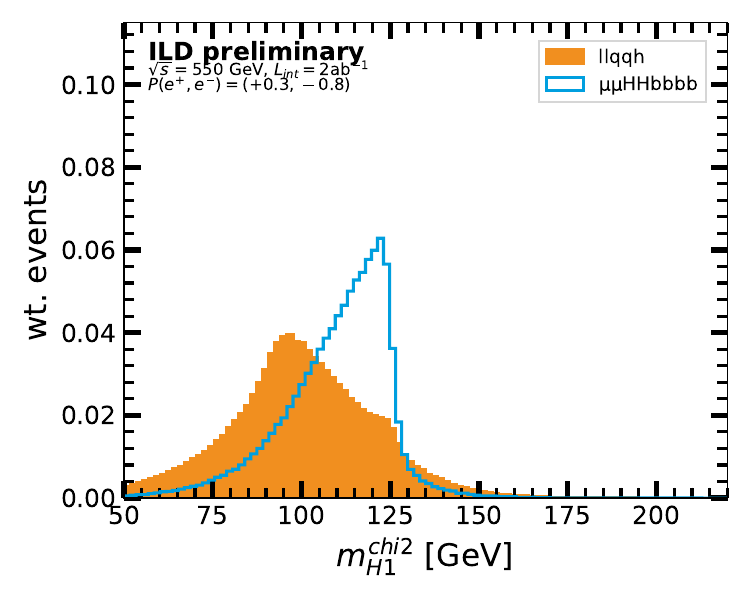}
        \caption{}
        \label{fig:mhllqqH:mh1chi2}    
    \end{subfigure}\hfill%
    \begin{subfigure}{.33\textwidth}
        \centering
        \includegraphics[width=0.95\textwidth,page=3]{figures/Bryan_mHi_llqqh_mumubbbb_250830.pdf}
        \caption{}
        \label{fig:mhllqqH:mh1kinfit}    
    \end{subfigure}%
    \begin{subfigure}{.33\textwidth}
    \centering
        \includegraphics[width=0.95\textwidth]{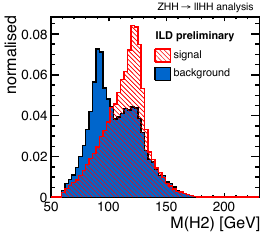}
        \caption{}
        \label{fig:mhllqqH:mh2claude}    
    \end{subfigure}\hfill%
    \begin{subfigure}{.33\textwidth}
        \centering
        \includegraphics[width=0.95\textwidth,page=2]{figures/Bryan_mHi_llqqh_mumubbbb_250830.pdf}
        \caption{}
        \label{fig:mhllqqH:mh2chi2}    
    \end{subfigure}%
    \begin{subfigure}{.33\textwidth}
    \centering
        \includegraphics[width=0.95\textwidth,page=4]{figures/Bryan_mHi_llqqh_mumubbbb_250830.pdf}
        \caption{}
        \label{fig:mhllqqH:mh2kinfit}    
    \end{subfigure}\hfill%
    \begin{subfigure}{.33\textwidth}
        \centering
        \includegraphics[width=0.95\textwidth]{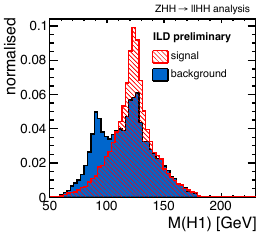}
        \caption{}
        \label{fig:mhllqqH:mh1claude}    
    \end{subfigure}%
    \caption{Invariant masses of the two di-jet Higgs candidates in the $\PZ \to \Plp\Plm$ channel as used in the old (right column) and new (middle column) analysis for signal events and $\Pl\Pl\PQq\PQq\PH$ background (for the $\Plp\Plm\PQq\PAQq\PQq\PAQq$ case see Fig.~\ref{fig:mhllqqqq}) (a) lower of the two di-jet masses using pre-fit momenta and a simple $\chi^2$ jet pairing (b) lower of the two di-jet masses using the 4C-fit momenta and the 6C-fit jet pairing (c) corresponding distribution from the 2014 analysis (d) higher of the two di-jet masses using pre-fit momenta and a simple $\chi^2$ jet pairing (e) higher of the two di-jet masses using the 4C-fit momenta and the 6C-fit jet pairing (f) corresponding distribution from the 2014 analysis. (c) and (f) are taken from Fig.~8.8 of~\cite{Durig:2016jrs}, the background is not only  $\Plp\Plm\PQq\PAQq\PH$, but includes also the $\Plp\Plm\PQb\PAQb\PQb\PAQb$ component of $\Plp\Plm\PQq\PAQq\PQq\PAQq$, which gives rise to the $\PZ$ peak in (f). Note that the definition of $M_{\PH 1}$ and $M_{\PH 2}$ is swapped with respect to the current analysis and that the bin width is 1\,GeV in (a), (b), (d), (e) while it is about 3\,GeV in (c) and (f).}
    \label{fig:mhllqqH}
\end{figure}

The impact of the improved kinematic reconstruction on the di-jet masses is illustrated in Figs.~\ref{fig:mhllqqqq} and~\ref{fig:mhllqqH}. Figure~\ref{fig:mhllqqqq} shows the di-jet Higgs candidate masses, split into the higher and lower di-jet mass, without and with kinematic fit for $\Plp\Plm\PH\PH$ signal and $\Plp\Plm\PQq\PAQq\PQq\PAQq$ background. The improvement due to the 4C fit with the neutrino correction as well as the success of choosing the jet pairing via the best of all 6C fits from both the $\PZ\PH\PH$ and the $\PZ\PZ\PH$ hypotheses can clearly be seen. The small shoulder around 90\,GeV in the signal histogram in Fig.~\ref{fig:mhllqqqq:mh1kinfit} originates from the few events for which the $\PZ\PZ\PH$ hypothesis gives a higher fit probability than the best fit under the $\PZ\PH\PH$ hypothesis and thus the a wrong jet pairing is chosen. More importantly, the background is not pulled significantly towards the signal peak, as it was the case in the 2014 analysis, c.f.\ Figs.~7.29 and~7.30 of~\cite{Durig:2016jrs}, since at the time neither the full ErrorFlow nor a first-principle neutrino correction were available.

Figure~\ref{fig:mhllqqH} shows equivalent plots, but now for $\Plp\Plm\PH\PH$ signal and $\Plp\Plm\PQq\PAQq\PH$ background, and compares them to the corresponding mass distributions of the observables used as input to training an MVA discriminator against the $\Plp\Plm\PQq\PAQq\PH$ and $\Plp\Plm\PQb\PAQb\PQb\PAQb$ background, both contained in the ``background'' category in Figs.~\ref{fig:mhllqqH:mh2claude} and~\ref{fig:mhllqqH:mh1claude}.


\subsubsection{Flavour Tagging}
\label{sec:extrapol:flav}
In the $\PZ\PH\PH$ analysis, the essential function of the flavour tagging is to separate the signal signature with 4$\PQb$ jets in the final state from backgrounds with 2$\PQb$ jets final state. For this task, the quality of the third best $\PQb$-tag among the jets is decisive, and thus the improvements observed for single-jet tagging e.g.\ in Table~\ref{tab:flavtag} enter with a factor between up to 3. A rejection of $10^2$ to $10^3$ of major backgrounds via $\PQb$-tagging allows a 3-10\% background acceptance per jet, for which Table~\ref{tab:flavtag} showed an increase of signal yield by a relative 15-20\% per jet. Thus by replacing LCFIPlus by ParT, and assuming that the improvement enters with a factor 2 since some backgrounds will have more than two actual $\PQb$ jets, we expect an increase of $\PH\PH \to \PQb\PAQb\PQb\PAQb$ signal yield in the analysis by a relative 30-40\%. The extrapolation in Sec.~\ref{sec:results} assumes a 30\% improvement.

\begin{figure}[htbp]
    \centering
     \includegraphics[width=0.5\textwidth]{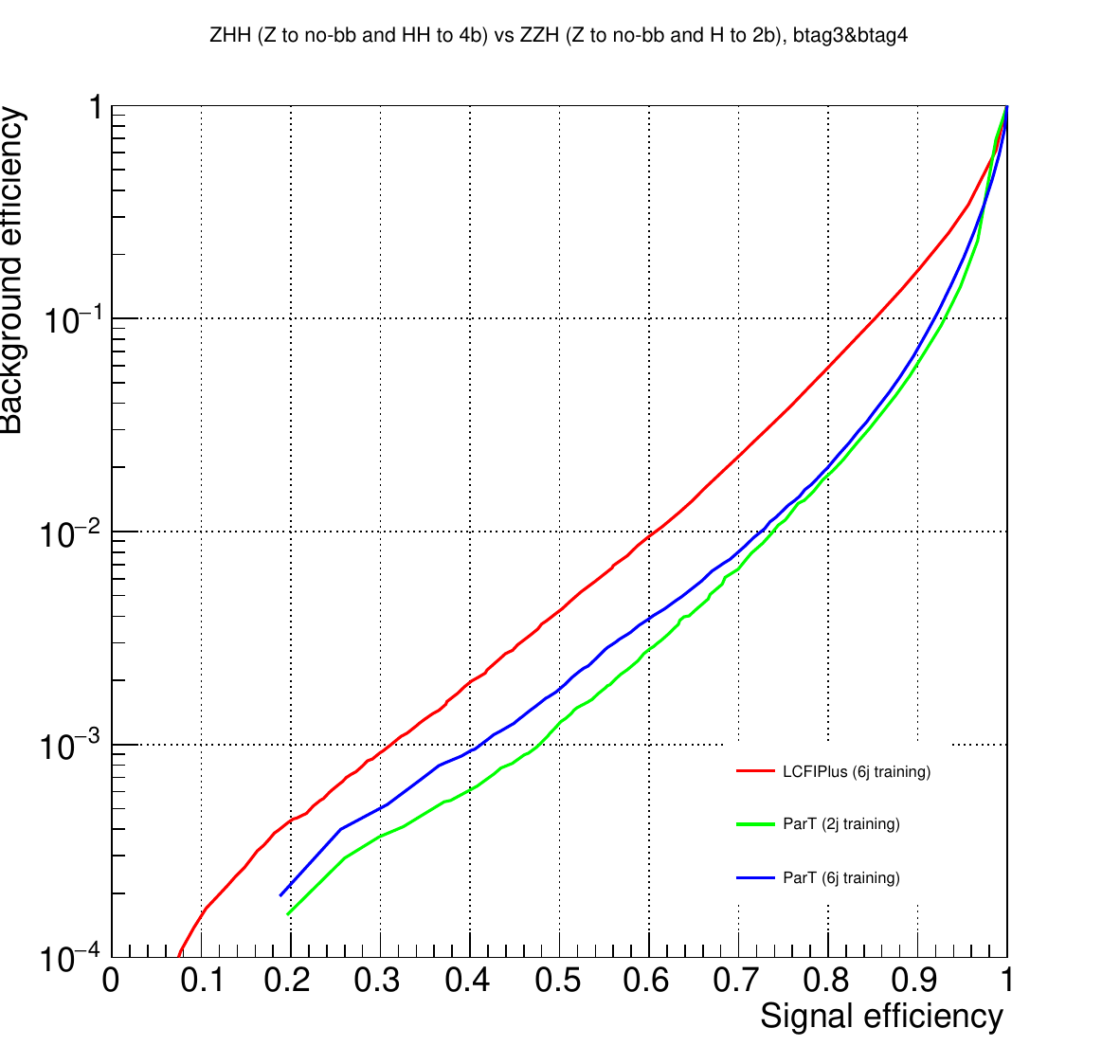}
    \caption{Acceptance of $\PZ\PZ\PH \to \PQq\PAQq \PQq\PAQq\PQb\PAQb$ background vs efficiency of  $\PZ\PH\PH \to \PQq\PAQq \PQb\PAQb\PQb\PAQb$ selection based purely on the third and forth highest $\PQb$-tag value.}
    \label{fig:zhhvszzh}
\end{figure}

This has been further examined in terms of separating $\PZ\PH\PH \to \PQq\PAQq \PQb\PAQb\PQb\PAQb$ and $\PZ\PZ\PH \to \PQq\PAQq \PQq\PAQq\PQb\PAQb $ events, with the Higgs and $\PZ$ boson decays being fixed to $\PH \to \PQb\PAQb$ and $\PZ \to \PQq\PAQq$ (non-$\PQb\PAQb$) for better illustration. Cuts on the third and fourth largest $\PQb$-tagging value have been applied in a 2-dimensional scan. Figure~\ref{fig:zhhvszzh} gives the resulting selection efficiency for $\PZ\PH\PH \to \PQq\PAQq \PQb\PAQb\PQb\PAQb$ versus the $\PZ\PZ\PH \to  \PQq\PAQq\PQq\PAQq\PQb\PAQb$ acceptance. If we allow 1\% background acceptance, the efficiency increases from 61\% to 74\% (21\% relative increase) by ParT, and if we allow only 0.1\% background acceptance, the efficiency increases from 31\% to 48\% (55\% relative increase). Though the relative increase depends on the working point, the observed relative increase roughly agrees with the observation from single-jet $\PQb$-tagging discussed previously.  

\begin{figure}[htbp]
    \centering

        \includegraphics[width=0.5\textwidth,page=3]{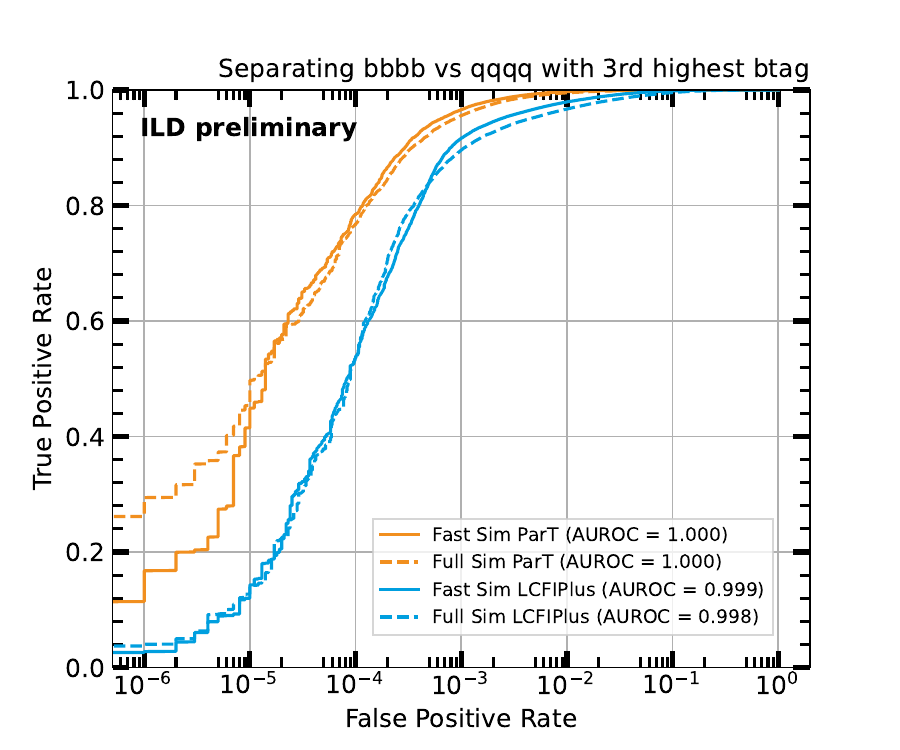}
    \caption{Efficiency vs mis-ID rate for the separation of $\PQb\PAQb\PQb\PAQb$ events from non-$\PQb$ $\PQq\PAQq\PQq\PAQq$ events with the third highest $\PQb$-tag from either LCFIPlus or the current ParT model in full and SGV simulation of the ILD concept. }
    \label{fig:flav4b4q}
\end{figure}


Figure \ref{fig:flav4b4q} compares the performance of flavour tagging between fast and full simulations and the previous baseline algorithm with the ParT implementation. The scenario here considers the separation of $\PQb\PAQb\PQb\PAQb$ and $\PQq\PAQq\PQq\PAQq$ events using the third highest $\PQb$-tag value. As shown previously in Fig.~\ref{fig:comp_btag}, the consistency of LCFIPlus $\PQb$-jet tagging (blue curve) between fast and full simulations is remarkable. This is also true for the ParT implementation used in the updated analysis (orange curve), and again underlines the appropriateness of using SGV fast simulation for the updated study.

To conclude, the developments in the field of jet flavour tagging over the last years through the use of machine learning are shown to be significant. The assumptions in the extrapolation of the full analysis result, originally derived when only performance evaluations on di-jet samples were available, have now been confirmed on multi-jet physics samples including the key processes $\PZ\PH\PH$ and $\PZ\PZ\PH$. Also for the new ParT-based algorithms, the agreement between full ILD simulation and SGV has been found to be excellent.


\section{Towards a new ZHH Analysis at 550\,GeV}
\label{sec:ana550}

As in the previous analysis, the analysis is split according to the decay mode of the $\PZ$ boson into a charged and neutral lepton channel as well as the hadronic channels. First, event-shape variables such as thrust are calculated. Then, the BDT-based models of the IsolatedLeptonTagging processor tag and identify isolated leptons among the particle flow objects (PFOs). At the moment, only muons and electrons are considered in this step. Afterwards, the events are clustered into jets using the Durham algorithm~\cite{Durham} as implemented in LCFIPlus~\cite{Suehara:2015ura} (i.e.\ respecting any found secondary and tertiary vertices), ErrorFlow as well as the neutrino correction are applied via the 4C fit based on MarlinKinfit~\cite{Beckmann:2010ib}, and the flavour tagging with the 11-category version of ParT~\cite{Tagami:2024gtc} is performed. For more information on these high-level reconstruction algorithms see Sec.~\ref{sec:hlreco_extrapol}. 

In the following, we summarize the current status of the analysis in the muon and electron channels, both giving very similar results. The re-analysis of the neutrino channel has started, and we comment here in particular on the studies on separating the Higgs-strahlung and $\PW$ boson fusion contributions. We have made extensive use of the parallelization framework law \cite{Rieger:2024} for straightforward bookkeeping and efficient job pipelining. The re-analysis of the hadronic channels has not yet started.

\subsection{Muon and Electron Channels}
\label{sec:ana550:lepton}

If at least two isolated leptons of the same flavour and opposite charge are found ($IsoMuons \ge 2$ or $IsoElectrons \ge 2$), the event is considered for the muon or electron channel. Among all possible opposite-charge-same-flavour pairs, the one with invariant mass closest to the $\PZ$ boson mass is chosen. In this process, final state radiation and bremsstrahlung photons are recovered by an angular cut around the lepton direction, using an algorithm developed originally for the Higgs recoil analysis~\cite{Yan:2016xyx}. The type of this lepton pair ($13$ for $\PGm$, $11$ for $\Pe$) is noted as $Paired Lepton Type$ and used in the preselection. After identifying the $\PZ\to\Plp\Plm$ candidate, the remaining PFOs are clustered into four jets\footnote{In addition, also  jet clustering into two jets is performed which allows to construct some specific observables against $\PZ \rightarrow \Pl\Pl\PQb\PQb$, which enter the main selection}. The jet and lepton four-momenta of the kinematic fit including only four momentum conservation (4C-fit) are used for the mass of the di-lepton system and -- together with the 6C-fit jet pairing -- for the masses of the  di-jet Higgs candidates. Further cut variables are the missing transverse momentum $p_t^{miss}$ and the thrust. 

All processes listed in Table~\ref{tab:MC550} have been included in the analysis, i.e.\ a much more complete set of backgrounds than in the 2014 analysis has been considered\footnote{For the $\Pl\PGn\PQq\PAQq$ samples, only a part of the available statistics has been used for now, though.}. In the cutflow tables, we list explicitly only the categories used in the 2014 analysis for ease of comparison. The contributions of all additionally considered backgrounds are summed and given in the ``Other bkgs'' column.

In comparison to the previous analysis, the preselection cuts have been loosened to increase the signal efficiency, still assuming that the final selection will be accurate enough such that the overall sensitivity will be improved. An overview of the cuts is given in Table~\ref{tab:mumuPreselCuts} for the muon channel and in Table \ref{tab:eePreselCuts} for the electron channel, both for an integrated luminosity  of 2\,\abinv with P($\Pem,\Pep$)=($-80\%$, $+30\%$) to ease the comparison with the previous analysis. 

In case of the muon channel, about 6\% of the signal events are lost because less than two isolated muon candidates are reconstructed in the very first place, while all other preselection cuts are now more than 99\% efficient. In case of the electron channel, in about 8\% of events less than two isolated electrons are found, and another 10\% of events fail the invariant mass cut on the lepton pair, despite the attempt to recover bremsstrahlung photons. This indicates further room for improvement in the electron reconstruction. 

After preselection, the final states $\Plp\Plm\PQq\PAQq$ with $\PQq\PAQq \neq \PQb\PAQb$ constitute the main background. As will be seen later, the highly improved flavour tagging provides excellent separation power to suppress them.


\begin{table}
\centering
\resizebox{\columnwidth}{!}{%

\begin{tabular}{ rrrrrrrrrrr }
\hline
                                     & $\PGm\PGm\PQb\PQb$  & $\PGm\PGn\PQb\PQb\PQq\PQq$ & $\PQb\PQb\PQq\PQq\PQq\PQq$ & $\Pl\Pl\PQq\PQq\PQq\PQq$ & ($\Pl\Pl\PQb\PQb\PQb\PQb$) & $\Pl\Pl\PQq\PQq\PH$ & Other bkgs           & Total bkg      & $\Pl\Pl\PH\PH$ & ($\PGm\PGm\PQb\PQb\PQb\PQb$) \\ 
$\sigma$ [fb]                        & 21.3                & 112                        & 325                        & 14.6                     & 0.0216                     & 0.156               & $8.05\cdot 10^3$    & $8.52\cdot 10^3$   & 0.0236            & 0.0026                     \\ 
expected events                      & $4.25\cdot 10^4$          & $2.24\cdot 10^5$           & $6.51\cdot 10^5$           & $2.92\cdot 10^4$         & $43.3$                     & $311$               & $1.61\cdot 10^7$    & $1.70\cdot 10^7$   & $47.1$            & $5.20$                      \\ 
\hline 
$IsoMuons \geq  2$                   & $1.68\cdot 10^4$       & $667$                      & $6.88$                     & $3.58\cdot 10^3$         & $11.8$                     & $79.3$              & $6.57\cdot 10^4$    & $8.68\cdot 10^4$   & $15.1$            & $4.90$                      \\ 
$Paired Lep Type = 13$               & $1.68\cdot 10^4$             & $377$                      & $5.70$                      & $3.58\cdot 10^3$         & $11.8$                     & $78.2$              & $6.56\cdot 10^4$    & $8.64\cdot 10^4$   & $14.8$            & $4.90$                      \\ 
\hline 
$20.0 \leq  m_{ll}^{4C} \leq  150.0$ & $1.43\cdot 10^4$    & $330$                      & $1.50$                      & $3.23\cdot 10^3$         & $10.6$                     & $61.4$              & $5.61\cdot 10^4$    & $7.41\cdot 10^4$   & $14.4$            & $4.86$                     \\ 
$50.0 \leq  m_{H1}^{4C} \leq  200.0$ & $2.38\cdot 10^3$    & $323$                      & $1.50$                      & $3.19\cdot 10^3$         & $10.4$                     & $59.3$              & $1.30\cdot 10^4$    & $1.90\cdot 10^4$   & $14.0$            & $4.84$                     \\ 
$50.0 \leq  m_{H2}^{4C} \leq  200.0$ & $2.31\cdot 10^3$    & $301$                      & $0.88$                    & $3.04\cdot 10^3$         & $10.0$                     & $57.0$              & $1.23\cdot 10^4$    & $1.80\cdot 10^4$   & $13.7$            & $4.76$                     \\ 
$p_{t}^{miss} \leq  70.0$            & $2.29\cdot 10^3$       & $217$                      & $0.88$                    & $3.03\cdot 10^3$         & $10.0$                     & $54.3$              & $1.23\cdot 10^4$    & $1.79\cdot 10^4$   & $12.7$            & $4.72$                     \\ 
$thrust \leq  0.9$                   & $1.43\cdot 10^3$       & $213$                      & $0.88$                    & $2.96\cdot 10^3$         & $9.83$                     & $51.8$              & $7.31\cdot 10^3$    & $1.20\cdot 10^4$   & $12.7$            & $4.71$                     \\ 
\hline
\end{tabular}%
}
    \caption{Preselection cutflow in the muon channel for an integrated luminosity  of 2\,\abinv with P($\Pem,\Pep$)=($-80\%$, $+30\%$). The event categories in brackets are included in their respective parent category. 
    }
    \label{tab:mumuPreselCuts}
\end{table}

\begin{table}
\centering
\resizebox{\columnwidth}{!}{%

\begin{tabular}{ rrrrrrrrrrr }
\hline
                                     & $\Pe\Pe\PQb\PQb$     & $\Pe\PGn\PQb\PQb\PQq\PQq$  & $\PQb\PQb\PQq\PQq\PQq\PQq$ & $\Pl\Pl\PQq\PQq\PQq\PQq$ & ($\Pl\Pl\PQb\PQb\PQb\PQb$) & $\Pl\Pl\PQq\PQq\PH$   & Other bkgs & Total bkg        & $\Pl\Pl\PH\PH$ & ($\Pe\Pe\PQb\PQb\PQb\PQb$) \\ 
$\sigma$ [fb]                  & 147               & 113                                      & 325                 & 14.6                & 0.0216                & 0.156              & $7.92\cdot 10^3$    & $8.52\cdot 10^3$   & 0.0236            & 0.00277               \\ 
expected events                & $2.95\cdot 10^5$  & $2.27\cdot 10^5$                         & $6.51\cdot 10^5$    & $2.92\cdot 10^4$    & $43.3$                & $311$              & $1.58\cdot 10^7$    & $1.70\cdot 10^7$   & $47.1$            & $5.55$                \\ 
\hline 
$IsoElectrons \geq  2$            & $7.61\cdot 10^4$  & $774$                                    & $50.9$              & $1.42\cdot 10^4$    & $13.1$                & $93.1$             & $3.86\cdot 10^5$    & $4.78\cdot 10^5$   & $15.6$            & $5.04$                \\ 
$Paired Lepton Type = 11$      & $7.56\cdot 10^4$  & $406$                                    & $49.6$              & $1.42\cdot 10^4$    & $13.1$                & $92.1$             & $3.79\cdot 10^5$    & $4.70\cdot 10^5$   & $15.2$            & $5.04$                \\ 
\hline 
$20.0 \leq  m_{ll}^{4C} \leq  150.0$ & $3.36\cdot 10^4$  & $340$                                    & $2.38$              & $6.36\cdot 10^3$    & $10.9$                & $79.3$             & $1.40\cdot 10^5$    & $1.81\cdot 10^5$   & $13.5$            & $4.53$                \\ 
$50.0 \leq  m_{H1}^{4C} \leq  200.0$ & $5.58\cdot 10^3$  & $337$                                    & $2.38$              & $6.28\cdot 10^3$    & $10.6$                & $77.4$             & $3.08\cdot 10^4$    & $4.31\cdot 10^4$   & $13.2$            & $4.51$                \\ 
$50.0 \leq  m_{H2}^{4C} \leq  200.0$ & $5.36\cdot 10^3$  & $324$                                    & $1.83$              & $6.00\cdot 10^3$    & $10.3$                & $75.1$             & $2.89\cdot 10^4$    & $4.07\cdot 10^4$   & $12.9$            & $4.43$                \\ 
$p_{t}^{miss} \leq  70.0$         & $5.34\cdot 10^3$  & $208$                                    & $1.83$              & $5.99\cdot 10^3$    & $10.3$                & $72.3$             & $2.86\cdot 10^4$    & $4.02\cdot 10^4$   & $12.0$            & $4.40$                 \\ 
$thrust \leq  0.9$                & $3.02\cdot 10^3$  & $205$                                    & $1.83$              & $5.71\cdot 10^3$    & $10.1$                & $71.5$             & $1.57\cdot 10^4$    & $2.47\cdot 10^4$   & $11.9$            & $4.38$                \\ 
\hline
\end{tabular}%
}
    \caption{Preselection cutflow in the electron channel for an integrated luminosity  of 2\,\abinv with P($\Pem,\Pep$)=($-80\%$, $+30\%$). The event categories in brackets are included in their respective parent category.}
    \label{tab:eePreselCuts}
\end{table}

Following the preselection, the final selection is carried out. Separate Boosted Decision Trees (BDTs) are trained against the $\Plp\Plm\PQb\PAQb$, $\Pl\PGn\PQb\PAQb\PQq\PAQq$, $\Plp\Plm\PQq\PAQq\PQq\PAQq$ and $\Plp\Plm\PQq\PAQq\PH$ backgrounds. We use the BDT implementation of the XGBoost framework \cite{tianqi2016xgboost}. The number of weak learners and the maximum depth of a tree are chosen between $250$ and $500$, and between $3$ and $5$, respectively, depending on the available statistics. Each data set is split into a training set with 50\% and a testing set with 50\% of the full statistics. The BDTs include input observables inspired by the 2014 analysis~\cite{Durig:2016jrs}: 

\begin{itemize}
    \item All BDTs use the following set of input features: the (ordered) $\PQb$-tag values of the four jets ``bmax1'' to  ``bmax4'', the ordered invariant masses of the four jets $m_{\mathrm{jet1}}$ to $m_{\mathrm{jet4}}$; the di-lepton and di-jet invariant masses $m^{\mathrm{4C}}_{\PZ}$, $m^{\mathrm{4C}}_{\PH1}$, $m^{\mathrm{4C}}_{\PH2}$; the di-Higgs invariant mass $m_{\PH\PH}$ and the number of particle flow objects $N_{\mathrm{PFO}}$.
    \item BDT1 (against $\Plp\Plm\PQb\PAQb$) contains in addition: the thrust value and polar angle, the maximum jet momentum in the di-jet restframe when clustering the hadronic final state into two jets, the smallest angle of one of the jets to the flight direction of the $\PZ$ candidate, the lowest number of PFOs in a jet when clustering the hadronic final state into four jets and the Durham distance parameters at which the jet configuration flips from 3 to 4 and from 4 to 5 jets, respectively.
    \item BDT2 (against $\Pl\PGn\PQb\PAQb\PQq\PAQq$) contains in addition: the visible energy $E_{\mathrm{vis}}$, the smaller of the two lepton momenta, the invariant mass of the two jets with the lowest $\PQb$-tag values, the missing transverse momentum $p_{\mathrm{t}}^{\mathrm{miss}}$ and the lower MVA value of the two lepton candidates found with the BDT of the IsolatedLeptonTagger.
    \item BDT3 (against $\Plp\Plm\PQq\PAQq\PQq\PAQq$) and  BDT4 (against $\Plp\Plm\PQq\PAQq\PH$) contain in addition: $E_{\mathrm{vis}}$, $p_{\mathrm{t}}^{\mathrm{miss}}$, the $\chi^2$'s of the best $\PZ\PH\PH$ and $\PZ\PZ\PH$ fits, respectively, the absolute value and polar angle of the boson candidate with highest momentum, both for the $\PZ\PH\PH$ and the $\PZ\PZ\PH$ hypotheses, respectively.
\end{itemize}

\begin{figure}[htbp]
    \centering
    \begin{subfigure}{.33\textwidth}
        \centering
        \includegraphics[width=0.95\textwidth,page=3]{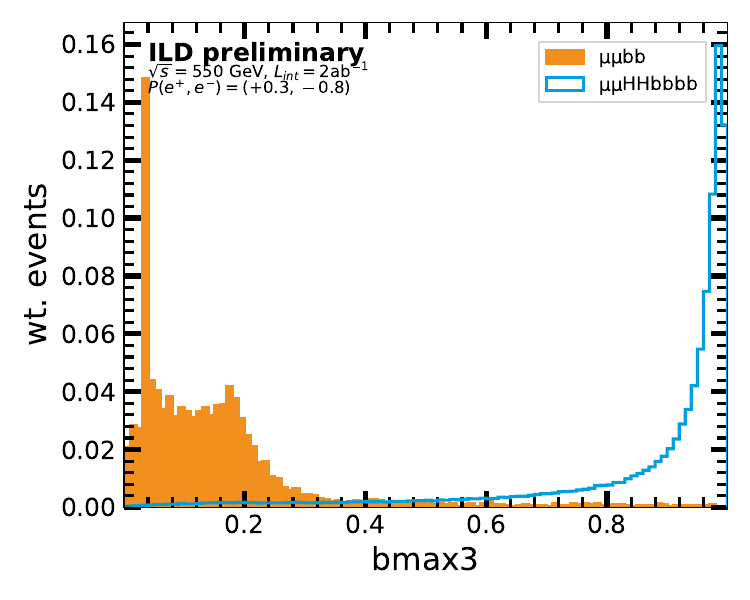}
        \caption{}
        \label{fig:bdt1mumu_inputsBmax3}    
    \end{subfigure}%
    \begin{subfigure}{.33\textwidth}
        \centering
        \includegraphics[width=0.95\textwidth,page=6]{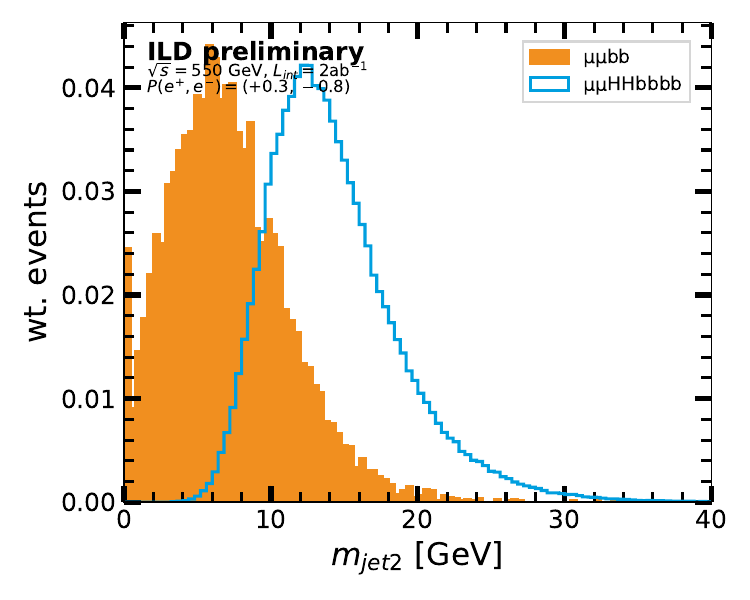}
        \caption{}
        \label{fig:bdt1mumu_inputsMjet2ndLow}
    \end{subfigure}%
    \begin{subfigure}{.33\textwidth}
        \centering
        \includegraphics[width=0.95\textwidth,page=11]{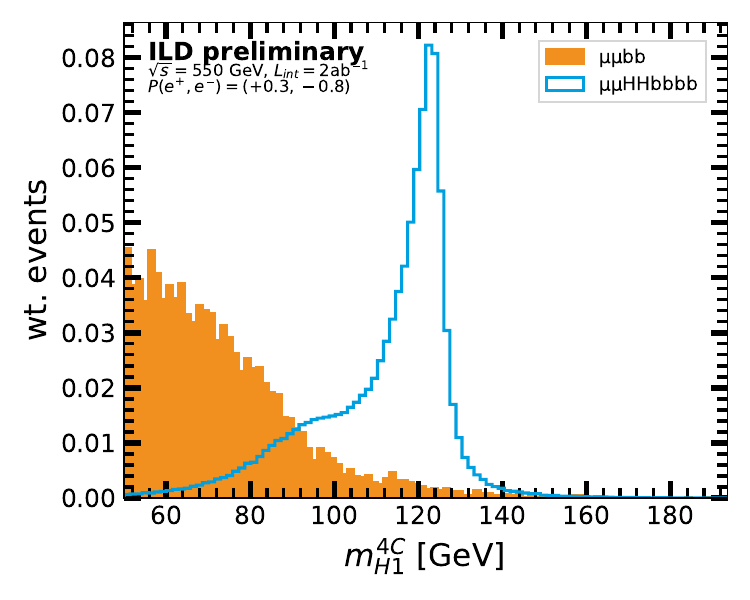}
        \caption{}
        \label{fig:bdt1mumu_inputsMH1_4C}
    \end{subfigure}\hfill
    \caption{Input variables to the first BDT which is trained to suppress $\Plp\Plm\PQb\PAQb$ backgrounds, here shown for the muon channel: (a) Third highest $\PQb$-tag value, (b) second highest jet mass and (c) lower of the two boson masses from the 4C fit, see also Figure \ref{fig:mhllqqH}.}
    \label{fig:bdt1mumu_inputs}
\end{figure}

Distributions of variables providing most separation power to suppress these are shown in Fig.~\ref{fig:bdt1mumu_inputs} for the muon channel with $\PGm\PGm\PH\PH \to \PGmp\PGmm\PQb\PAQb\PQb\PAQb$ against the $\PGmp\PGmm\PQb\PAQb$ background as an example.

Tables~\ref{tab:mumuSelCuts} and \ref{tab:eeSelCuts} show the cutflow for applying the four BDTs in the muon and electron channels, respectively. The cut values shown in the leftmost columns are optimized for significance and are obtained consecutively using the training data set and considering all backgrounds.

\begin{table}
\centering
\resizebox{\columnwidth}{!}{%

\begin{tabular}{lrrrrrrrrrr}
\hline
                               & $\PGm\PGm\PQb\PQb$ & $\PGm\PGn\PQb\PQb\PQq\PQq$ & $\PQb\PQb\PQq\PQq\PQq\PQq$ & $\Pl\Pl\PQq\PQq\PQq\PQq$ & ($\Pl\Pl\PQb\PQb\PQb\PQb$) & $\Pl\Pl\PQq\PQq\PH$ & Other bkg & Total bkg & $\Pl\Pl\PH\PH$ & ($\PGm\PGm\PQb\PQb\PQb\PQb$) \\ 
after preselection  &  $1.43\cdot 10^3$       & $213$                      & $0.884$                    & $2.96\cdot 10^3$         & $9.83$                     & $51.8$              & $7.31\cdot 10^3$    & $1.20\cdot 10^4$   & $12.7$            & $4.71$                     \\ 
\hline
BDT1 $\geq  0.9999$ & $0.96$ & $1.57$ & $0$ & $6.25$ & $5.23$ & $4.47$ & $0.0$ & $13.3$ & $4.46$ & $4.03$ \\ 
BDT2 $\geq  0.998$  & $0.96$ & $0.79$ & $0$ & $6.15$ & $5.16$ & $4.28$ & $0.0$ & $12.2$ & $4.36$ & $4.01$ \\ 
BDT3 $\geq  0.998$  & $0.72$ & $0$    & $0$ & $1.36$ & $1.34$ & $1.46$ & $0.0$ & $3.54$ & $3.72$ & $3.46$ \\ 
BDT4 $\geq  0.84$   & $0.48$ & $0$    & $0$ & $0.81$ & $0.79$ & $0.88$ & $0.0$ & $2.17$ & $3.45$ & $3.24$ \\
\hline
\end{tabular}%
}
    \caption{Main selection cutflow in the muon channel for an integrated luminosity  of 2\,\abinv with P($\Pem,\Pep$)=($-80\%$, $+30\%$). The event categories in brackets are included in their respective parent category. }
    \label{tab:mumuSelCuts}
\end{table}

\begin{table}
\centering
\resizebox{\columnwidth}{!}{%

\begin{tabular}{lrrrrrrrrrr}
\hline
                               & $\Pe\Pe\PQb\PQb$ & $\Pe\PGn\PQb\PQb\PQq\PQq$ & $\PQb\PQb\PQq\PQq\PQq\PQq$ & $\Pl\Pl\PQq\PQq\PQq\PQq$ & ($\Pl\Pl\PQb\PQb\PQb\PQb$) & $\Pl\Pl\PQq\PQq\PH$ & Other bkg  & Total bkg & $\Pl\Pl\PH\PH$ & ($\Pe\Pe\PQb\PQb\PQb\PQb$) \\ 
after preselection & $3.02\cdot 10^3$ & 205 & 1.83 & $5.71\cdot 10^3$ & 10.1 & 71.5 & $1.57\cdot 10^4$    & $9.01\cdot 10^3$ & 11.9 & 4.38 \\
\hline
BDT1 $\geq  0.9999$ & $1.19$   & $0.61$  & $0.40$  & $8.89$  & $5.23$  & $7.17$ & $0$   & $18.3$ & $4.0$  & $3.64$ \\ 
BDT2 $\geq  0.995$  & $1.19$   & $0.61$  & $0$     & $6.91$  & $5.22$  & $7.1$  & $0$   & $15.8$ & $3.98$ & $3.64$ \\ 
BDT3 $\geq  0.99$   & $0.26$   & $0.03$  & $0$     & $3.40$  & $2.38$  & $4.8$  & $0$   & $8.49$ & $3.77$ & $3.47$ \\ 
BDT4 $\geq  0.81$   & $0.01$   & $0.03$  & $0$     & $0.83$  & $0.83$  & $1.47$ & $0$   & $2.34$ & $2.93$ & $2.79$ \\ 

\hline
\end{tabular}%
}
    \caption{Main selection cutflow in the electron channel for an integrated luminosity  of 2\,\abinv with P($\Pem,\Pep$)=($-80\%$, $+30\%$). The event categories in brackets are included in their respective parent category.}
    \label{tab:eeSelCuts}
\end{table}

In the muon channel, after the cut on BDT4, the total number of remaining background events is 2.17, expectedly dominated by $\Plp\Plm\PQq\PAQq\PH $ events. Slightly more than three events of the $\PGm\PGm\PH\PH \to \PGmp\PGmm\PQb\PAQb\PQb\PAQb$ signal pass the final selection. This results in a significance of 1.39 calculated from $S/\sqrt{S+B}$, which should be compared to the significance for $\PGmp\PGmm\PH\PH \to \PGmp\PGmm\PQb\PAQb\PQb\PAQb$ of 0.82 in the 2014 analysis. The assumptions in the extrapolation discussed in Sec.~\ref{sec:results} lead to a significance of 1.14 for the $\PGmp\PGmm\PH\PH \to \PGmp\PGmm\PQb\PAQb\PQb\PAQb$ channel. 
The corresponding cutflow for the electron channel is given in Tables~\ref{tab:eePreselCuts} and~\ref{tab:eeSelCuts} for the preselection and the BDT cuts, respectively.
The selection in the electron channel yields 2.79 signal events compared to 2.34 background events, resulting in $S/\sqrt{S+B} = 1.23$. This can be compared to the significance for $\Pep\Pem\PH\PH \to \Pep\Pem\PQb\PAQb\PQb\PAQb$ of 0.84 in the 2014 analysis and of 1.15 assumed in the extrapolation.

Therefore, already the first and by far not fully optimised attempt to run the full event selection confirms (and even slightly outperforms) the assumptions made in the extrapolation for the $\PGmp\PGmm\PQb\PAQb\PQb\PAQb$ and $\Pep\Pem\PQb\PAQb\PQb\PAQb$ analyses. 

The analysis strategy outlined here can be further optimised in a number of ways. First of all, overall model performance can be further tuned by using a cross-validated hyper-parameter optimization. Second, one could train a multi-class classifier aware of different signal and background contributions. This would allow for an easier and more interpretable tuning of MVA cuts. Furthermore, additional observables like the quark-anti-quark information from the 11-class flavour tag enable to use the matrix element likelihoods in the selection, as pioneered in~\cite{Bliewert:2024}. 

\subsection{Neutrino Channel}
\label{sec:ana550:neutrino}
In the 2014 analysis, no attempt was made to separate the contributions of the $\PW$ boson fusion (WBF) production mode to the $\PGn\PAGn\PH\PH$ final-state from those of the $\PZ\PH\PH$ process. However, due to their very different interference pattern, which will be discussed in detail in Sec.~\ref{sec:results:self}, a separate cross-section measurement of both processes would be highly desirable to enhance the sensitivity to the self-coupling. Therefore, we investigated for the first time the prospects for this separation.

The increase in centre-of-mass energy is very beneficial for this enterprise, since at \SI{550}{\giga\electronvolt}, about 17\% of the $\PGn\PAGn \PH\PH$ events are expected  to come from WBF, compared to about 10\% at \SI{500}{\giga\electronvolt}.

\begin{figure}[htbp]
    \centering
    \begin{subfigure}{.5\textwidth}
    \centering
        \includegraphics[width=0.95\textwidth]{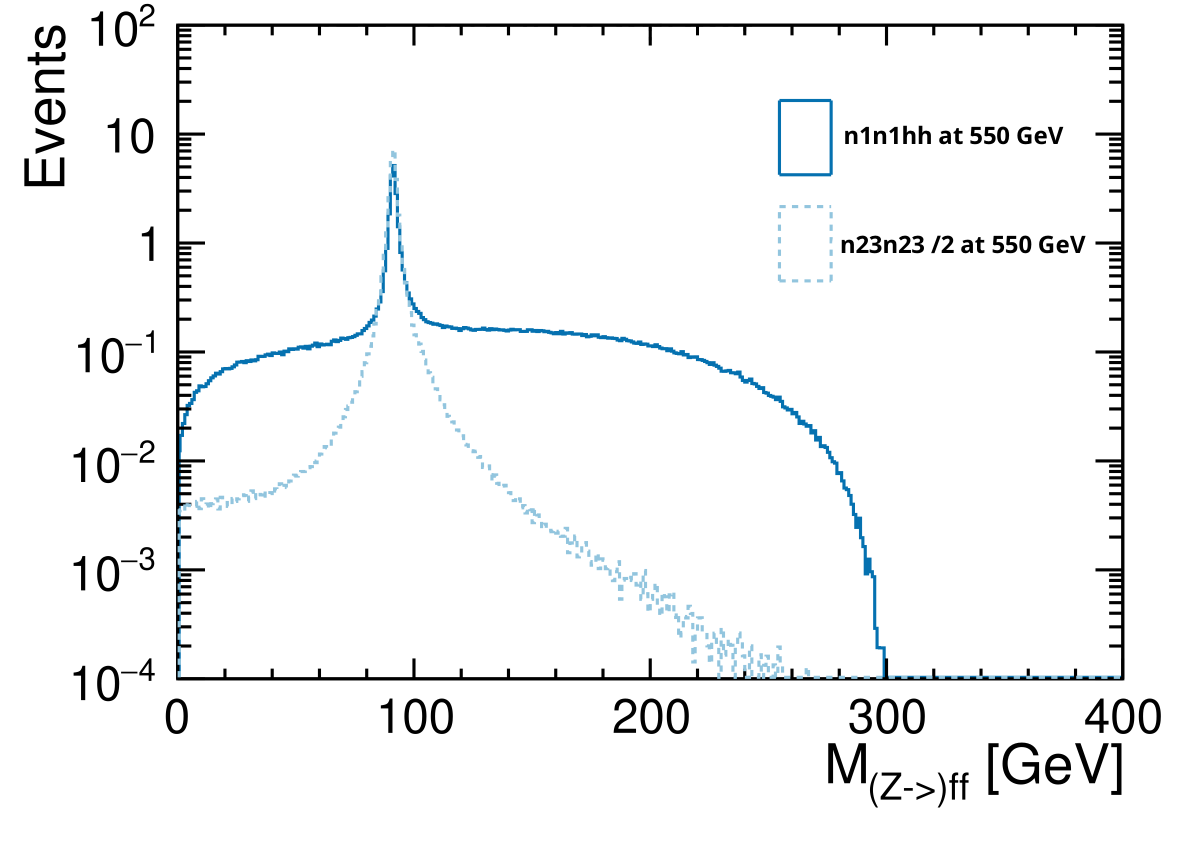}
        \caption{}
        \label{fig:wbf:total}    
    \end{subfigure}\hfill%
    \begin{subfigure}{.5\textwidth}
        \centering
        \includegraphics[width=0.95\textwidth]{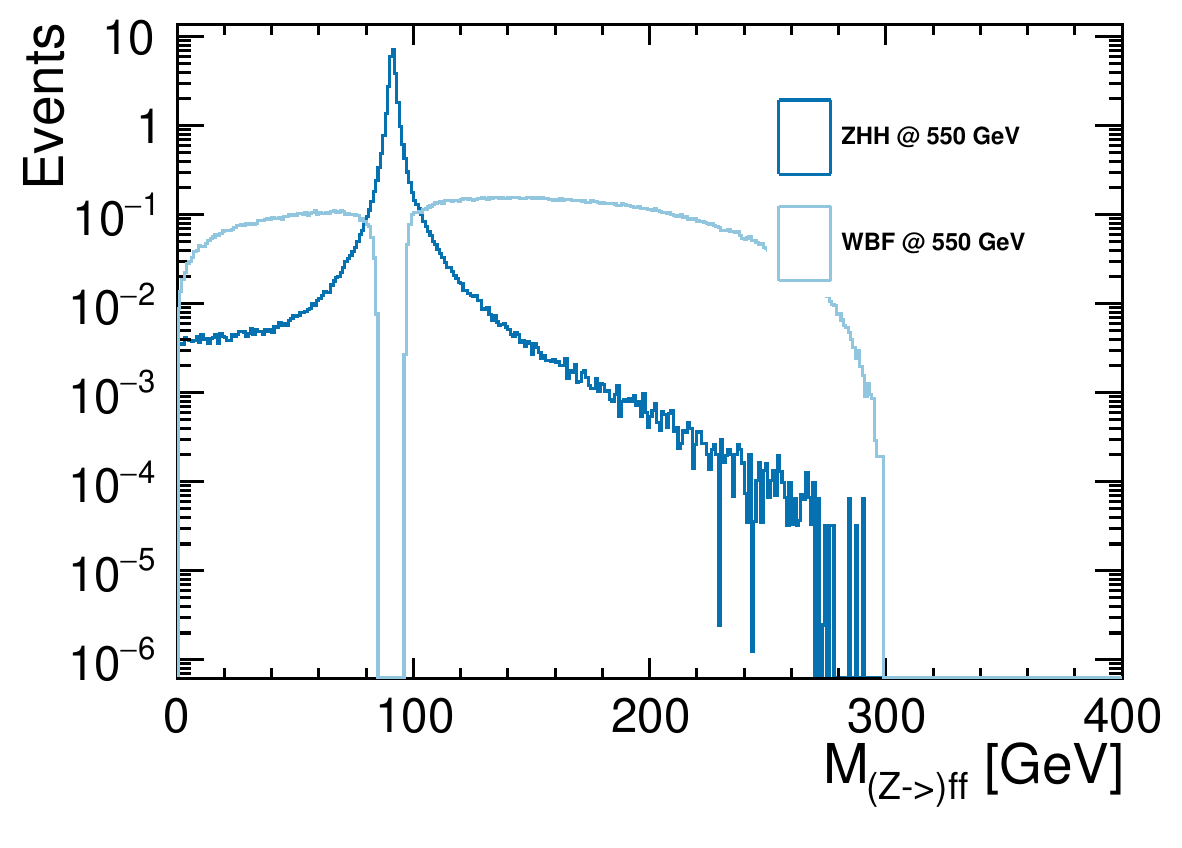}
        \caption{}
        \label{fig:wbf:split}    
    \end{subfigure}%
    \caption{Invariant mass of hard process neutrinos, $m_{\PGn\PAGn}$, in (a) for $\PGne\PAGne\PH\PH$ events and $\HepParticle{\PGn}{\!\PGm,\PGt}\Xspace\HepAntiParticle{\PGn}{\!\PGm,\PGt}\Xspace\PH\PH$ events weighted by a factor 0.5, and in (b) split into a $\PZ\PH\PH$ and a WBF contribution by subtracting the histograms in (a).}
    \label{fig:wbf}
\end{figure}

As $\PZ\PH\PH$ and WBF both contribute with diagrams to the same scattering amplitude of the $\PGn\PAGn \PH\PH$ final state, calculating them separately can always only be done to an approximation. But in the complete $\PGn\PAGn \PH\PH$ event generation, the $\PZ\PH\PH$ and WBF still have identifiable features. For $\PZ\PH\PH$ events, the neutrino pair originates from a $\PZ$-boson with a corresponding missing mass distribution peaked around the $\PZ$ mass, and for WBF events, the $m_{\PGn\PAGn}$ distribution is relatively flat across the available phase space. For illustration purposes, the $m_{\PGn\PAGn}$ distribution from pure $\PZ\PH\PH$ events can be obtained from $\PGnGm\PAGnGm \PH\PH$ and $\PGnGt\PAGnGt  \PH\PH$ events, for which  only the $s$-channel diagram is allowed, while the $\PGne\PAGne \PH\PH$ process contains both the $\PZ\PH\PH$ contribution, the $\PW$ boson fusion, and their interference. By subtracting the distributions in Fig.~\ref{fig:wbf:total}, the contribution from $\PZ\PH\PH$ can be removed leaving the contribution from WBF events including a small effect from the interference, as shown in Fig.~\ref{fig:wbf:split}. Hence with a mass cut on $m_{\PGn\PAGn}$, WBF events and $\PZ\PH\PH$ events can be separated to a good approximation at generator level. 

\begin{figure}[htbp]
    \centering
     \includegraphics[width=0.5\textwidth]{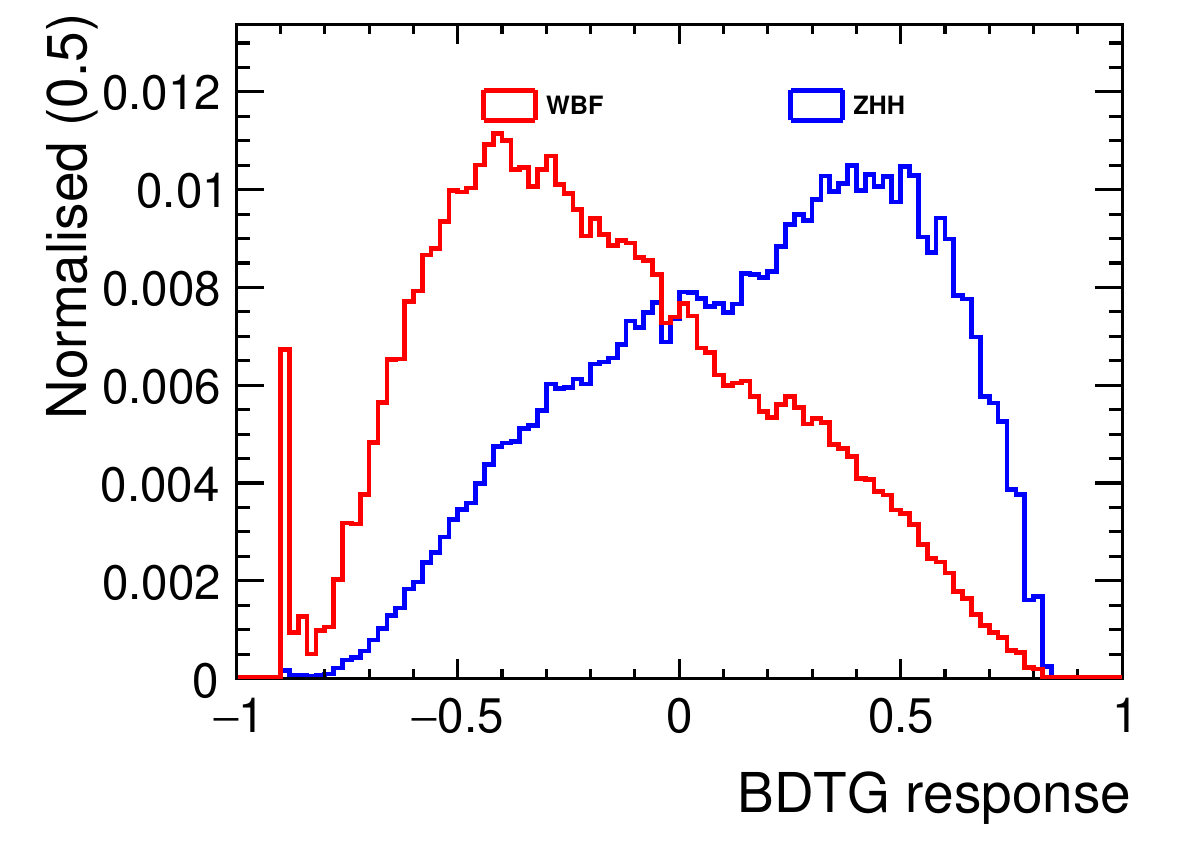}
    \caption{Normalised BDT response on $\PGn\PAGn \PH\PH$ events with a BDT trained on $\PZ\PH\PH$ events and WBF-like events based on simple splitting on true invariant mass of $\PGn\PAGn$-pair.}
    \label{fig:wbfbdt}
\end{figure}

At the reconstruction level, the separation between WBF events and $\PZ\PH\PH$ events is not as obvious, since the neutrinos can only be identified from what is missing, suffering from  detector effects as well as additional missing energy from ISR and beamstrahlung. The events can still be separated to some extent. As a proof-of-principle,  a Boosted Decision Tree (BDT) has been trained using ROOT TMVA on pure $\PZ\PH\PH$ events from $\HepParticle{\PGn}{\!\PGm,\PGt}\Xspace\HepAntiParticle{\PGn}{\!\PGm,\PGt}\Xspace \PH\PH$ events and WBF-like events from $\PGne\PAGne \PH\PH$ with a true $m_{\PGn\PAGn}$ below 79\,GeV and above 102\,GeV. This cut has been chosen by simultaneously optimising the efficiency and purity of $\PZ\PH\PH$-like events and as well as of WBF-like events, providing a large purity of $\PZ\PH\PH$-like events of around 99\% within the cut-window while the efficiency is just above 92\% and it provides a large efficiency of WBF-like events outside the cut-window of around 99\% with a purity of around 90\%, keeping more WBF-like events benefitting the training. 75\% of the events were used for training with $\sim850$k $\PZ\PH\PH$ events and $\sim650$k WBF-like events. The BDT was trained on five variables: missing $p_T$, missing invariant mass, visible energy, visible invariant mass, and thrust. Figure~\ref{fig:wbfbdt} shows the output distributions of the BDT, demonstrating that it indeed has separation power. This can be exploited for instance via a template fit, or for weighted event counting, in the upcoming re-analysis of the neutrino channel. 

\section{Current Status of Projections for Di-Higgs Production and the Trilinear Higgs Self-coupling}
\label{sec:results}
This section describes how the findings of the previous sections have been extrapolated to all final states and summarizes the resulting projections for di-Higgs cross-section measurements at a Linear Collider with centre-of-mass energies of up to 1\,TeV as well as the corresponding extraction of the trilinear self-coupling of the 125-GeV Higgs boson.
\subsection{Extrapolation}
The past and current performance of the two key features of the di-Higgs production measurement, the $\PQb$-tagging and the kinematic reconstruction, on data-sets passed through full and SGV simulation of ILD has been discussed in Sec.~\ref{sec:hlreco_extrapol}. Key observations from dedicated performance checks, re-confirmed in the full analysis of the electron and muon channels, are:
\begin{itemize}
    \item $\PQb$-tagging: at a typical working point, the $\PQb$-tagging efficiency increases by 10\% per jet at the same mis-ID rate. This is assumed to enter the efficiency for 4$\PQb$ part of the signal at the third power (as the 2014 analysis applied a cut on at least the third highest $\PQb$-tag value in an event in all channels, in the neutrino and hadronic channels also on the forth highest $\PQb$-tag value). For the non-4$\PQb$ part of the signal, mostly originating from $\PH\PH \to \PQb\PQb$+anything, the improvement is assumed to enter to the power of 1.5. Together, this is very well compatible with the improvements discussed in Sec.~\ref{sec:extrapol:flav}.
    The background processes with real $\PQb$-jets are not scaled up here, as their rejection is anyhow relying on  kinematic observables -- which receive the second, independent key improvement:
    \item kinematic reconstruction: 10\% higher event selection efficiency at the same background rejection is assumed in all channels. This is assumed to account for the effect of the improved kinematic fit including the neutrino correction (c.f.\ Sec.~\ref{sec:sldcorr}) and the inclusion of the $\PZ\PH\PH$ and $\PZ\PZ\PH$ matrix elements~\cite{Bliewert:2024} in the selection.
\end{itemize}

Furthermore, the neutrino channel has been split into the $\PZ\PH\PH$-only part and the VBF part following the procedure presented in Sec.~\ref{sec:ana550:neutrino}. This means that for the P($\Pem,\Pep$)=($-80\%$, $+30\%$) case, 75\% of the signal events in Table~\ref{tab:Nevt_Claude} are attributed to $\PZ\PH\PH$, while in the P($\Pem,\Pep$)=($+80\%$, $-30\%$) case, all events are attributed to $\PZ\PH\PH$\footnote{For P($\Pem,\Pep$)=($+80\%$, $-30\%$), the WBF cross-section is below 1\% of the total $\PGn\PAGn\PH\PH$ cross-section.}.

Table~\ref{tab:Nevt_extrapol} shows the resulting event counts corresponding to Table~\ref{tab:Nevt_Claude}, i.e.\ still for 2\,\abinv at $\sqrt{s}=500$\,GeV for each of the two beam polarisation settings as well as their combination. The combined cross-section uncertainty of 16.6\% is directly comparable to the 21.1\% of the 2014 analysis.

\begin{table}[htb]
\begin{center}
\begin{tabular}{llc|ccccc}
P($\Pem$), P($\Pep$) & channel & decay mode &  S & B &  S/$\sqrt{S+B}$ & $\delta\sigma/\sigma [\%]$ \\
\hline
$-80\%$, $+30\%$ & $\Pe\Pe\PH\PH$ & inclusive & 5.4 & 7.0 & 1.5 & 65\\
 & & $\PH\PH \to 4\PQb$ &  3.8  &   &  1.2 & 86 \\
 & $\PGm\PGm\PH\PH$ & inclusive & 6.9 & 8.9 &  1.7 & 58 \\
 & & $\PH\PH \to 4\PQb$ &  4.1  &   & 1.1 & 88 \\
 & $\PGn\PGn\PH\PH$ ($\PZ\PH\PH$ only) & inclusive & 6.1 & 6.5 &  1.8 & 55\\
 & & $\PH\PH \to 4\PQb$ &  6.0  &   & 1.7 & 60 \\
 & $\PQb\PQb\PH\PH$ & inclusive & 12.3 & 21.9 & 2.1 & 47\\
 & & $\PH\PH \to 4\PQb$ &  11.7  &   & 2.0 & 49 \\
 & $\PQq\PQq\PH\PH$ & inclusive & 18.0 & 55.0 & 2.1 & 47 \\
 & & $\PH\PH \to 4\PQb$ &  16.0  &   & 1.9 & 53\\
\hline
$-80\%$, $+30\%$ & combined & inclusive &  &  & 4.2 & 24\\
 & & $\PH\PH \to 4\PQb$ &     &   & 3.6 & 28 \\
\hline
\hline
$+80\%$, $-30\%$ &  $\Pe\Pe\PH\PH$ & inclusive & 4.0& 4.2 & 1.4 & 72\\
 & & $\PH\PH \to 4\PQb$ & 2.8 & & 1.1 & 95\\
 & $\PGm\PGm\PH\PH$ & inclusive & 5.1 & 5.3 &  1.6 & 63\\
 & & $\PH\PH \to 4\PQb$ &  2.9  &   & 1.0 & 98\\
 & $\PGn\PGn\PH\PH$ & inclusive & 5.2 & & 2.1 & 48 \\
 & & $\PH\PH \to 4\PQb$ &  5.1  &   & 2.1 & 49 \\
 & $\PQb\PQb\PH\PH$ & inclusive & 8.6 & 7.0 & 2.2 & 46 \\
 & & $\PH\PH \to 4\PQb$ &  8.2  &   & 2.1 & 48\\
 & $\PQq\PQq\PH\PH$ & inclusive & 12.0 & 16.0 &  2.3 & 44\\
 & & $\PH\PH \to 4\PQb$ &  11.4  &   & 2.2 & 46\\
\hline
$+80\%$, $-30\%$ & combined & inclusive &  &  &  4.3 & 23\\
 & & $\PH\PH \to 4\PQb$ &     &   & 4.0 & 25 \\
\hline
\hline
combined & combined & inclusive &  &  & 6.0 & 16.6\\
 & & $\PH\PH \to 4\PQb$ &     &   & 5.4 & 18.7 \\
\hline
\hline
\end{tabular}
\end{center} 
\caption{Signal and background counts for 2\,\abinv per opposite-sign polarisation dataset at 500\,GeV  and the corresponding significances and cross-section uncertainties extrapolated from the 2014 $\PH\PH\to 4\PQb$ analysis as explained in the text. Note that the actual result is, like in the 2014 analysis, the inclusive signal. The numbers for the true $\PH\PH \to 4\PQb$ events are given for comparison with the new analysis described in Sec.~\ref{sec:ana550}. The combined cross-section uncertainty of 16.6\% is directly comparable to the 21.1\% of the 2014 analysis.}
\label{tab:Nevt_extrapol}
\end{table}

Like in the original 2014 analysis, the extrapolated result in Table~\ref{tab:Nevt_extrapol} is assumed to improve further by 20\% when combining with the $\PH\PH \to \PQb\PQb\PW\PW^*$ channel. A further 8\% improvement from including $\PZ \to \PGt\PGt$ (determined already in the 2014 analysis but not included in the result at the time) is  taken into account, as well as 10\% improvement from the inclusion of all other Higgs decay modes: in particular $\PH\PH \to \PQb\PQb\PGt\PGt$, $\PH\PH \to \PW\PW^*\PW\PW^*$, $\PH\PH \to \PW\PW^*\PGt\PGt$  and $\PH\PH \to \PQb\PQb\PZ\PZ^*$, together accounting for another 30\% of the $\PH\PH$ signal. 

The second aspect of the extrapolation does not concern the analysis technique or scope, but the assumed running scenario. In particular, we considered the increase of the centre-of-mass energy from 500\,GeV to 550\,GeV as foreseen for C$^3$~\cite{Vernieri:2022fae} and LCF~\cite{LinearCollider:2025lya}, and increase of the positron polarisation to 60\%~\cite{LinearCollider:2025lya}, as well as a two-times higher integrated luminosity of 8\,\abinv as foreseen in the LCF running scenario~\cite{LinearCollider:2025lya}. The change in luminosity is a trivial scaling with the available statistics, since all relevant systematic uncertainties are negligible. The effect of the higher positron polarisation is taken as was evaluated in the 2014 analysis (c.f.\ Table~9.1 of~\cite{Durig:2016jrs}). We note that the gain from doubling of the positron polarisation is equivalent to a 10\% higher integrated luminosity (i.e.\ 8\,\abinv with 60\% positron polarization gives nearly the same result as 8.8\,\abinv with 30\% positron polaristion).

The effect of increasing the centre-of-mass energy is more subtle and has been evaluated with Whizard~\cite{Kilian:2007gr} in the same setup as used for the event generation, c.f.\ Sec.~\ref{sec:ILDMC:evtgen}. The higher energy increases the total $\PZ\PH\PH$ cross-section by 16\%, leading to a better precision on the cross-section. However, concerning the determination of the self-coupling, a large part of this increase is due to those diagrams whose contributions to the cross-section do not depend on $\lambda$. Thus the translation from cross-section uncertainty to self-coupling uncertainty (aka ``sensitivity factor'') gets slightly less favourable, but still there is a net gain also in terms of the self-coupling precision. 

The situation is drastically different for the $\PW\PW$-fusion production process (WBF). Its cross-section increases by almost a factor 2 between 500 and 550\,GeV. Thus, the higher centre-of-mass energy allows to measure the $\PW\PW$-fusion separately and to include it explicitly in the global interpretations. Despite the very modest precision on the cross-section measurement at this centre-of-mass energy, it contributes visibly to the precision on $\lambda$, since its sensitivity factor (for the SM case) is much more favourable than in the $\PZ\PH\PH$ case, as will be discussed in Sec.~\ref{sec:results:self}.

\subsection{Cross-section projections}


Table~\ref{tab:xsec-prec} summarizes the projected precisions for the cross-section measurements of di-Higgs processes at the 550\,GeV stage of the LCF, as well as the prospects for a 1\,TeV upgrade, all assuming the SM case.
The $\PZ\PH\PH$ precisions have been scaled for energy, luminosity and polarisation from Table~\ref{tab:Nevt_extrapol}, leading to precisions of 10\% (to be understood with two-digit precision) for each of the opposite-sign polarisation datasets.

\begin{table}[htb]
\begin{center}
\begin{tabular}{ll|c|l|c}
ECM [GeV] & $\mathcal{L}_{\mathrm{int}}$ [\abinv] & P($\Pem$), P($\Pep$)  & process & $\delta\sigma/\sigma [\%]$ \\
\hline
550 & 3.2 & $-80\%$, $+60\%$ & $\PZ\PH\PH$  & 10\%\\
         &           &                & $\PH\PH$ from WBF & 40\%\\ \hline 
550 & 3.2 & $+80\%$, $-60\%$ & $\PZ\PH\PH$ & 10\%\\ \hline
\hline
1000 & 3.2 & $-80\%$, $+20\%$ & $\PZ\PH\PH$ & 10\%\\
  &   &  & $\PH\PH$ from WBF  & 13\%\\ \hline 
1000   & 3.2 & $+80\%$, $-20\%$ & $\PZ\PH\PH$ & 10\%\\ \hline
\hline
\end{tabular}
\end{center} 
\caption{Projected precisions for the cross-section measurements of di-Higgs processes at the 550\,GeV stage of the LCF, as well as the prospects for a 1\,TeV upgrade, all assuming the SM case.}
\label{tab:xsec-prec}
\end{table}

For the $\PW$ boson fusion production, the technique proposed in Sec.~\ref{sec:ana550:neutrino} has been used to estimate a ``precision'' of 120\% on the WBF cross-section under the conditions of Table~\ref{tab:Nevt_extrapol}. This has been found to be quite consistent with the precision expected from scaling the previous WBF analysis at 1\,TeV~\cite{Tian:2013} down in centre-of-mass energy. Assuming the same gain from including all other $\PH\PH$ decay modes, this improves to about 85\% on the WBF cross-section measurement for 4\,\abinv at 500\,GeV. The WBF cross-section increases by about a factor of two with the increase of centre-of-mass energy from 500 to 550\,GeV, the luminosity increase gives another factor of two. Also, the WBF production profits even more than the $\PZ\PH\PH$ mode from the increased positron polarisation. Thus, with single-digit precision, we estimate an uncertainty of approximately 40\% on the WBF cross-section.

For 1\,TeV, the $\PZ\PH\PH$ extrapolation in Table~\ref{tab:Nevt_extrapol} is  scaled again for energy, luminosity and polarisation, yielding a precision of 8\%, which we round here to 10\%. Note that while this seems very competitive with the 550\,GeV result in terms of cross-section precision, it is much less sensitive to the self-coupling. For the WBF process, we quote here the result of the old 1\,TeV analysis~\cite{Tian:2013}, without any attempt to account for analysis improvements.

\begin{figure}[htbp]
    \centering
    \begin{subfigure}{.5\textwidth}
    \centering
        \includegraphics[width=0.95\textwidth]{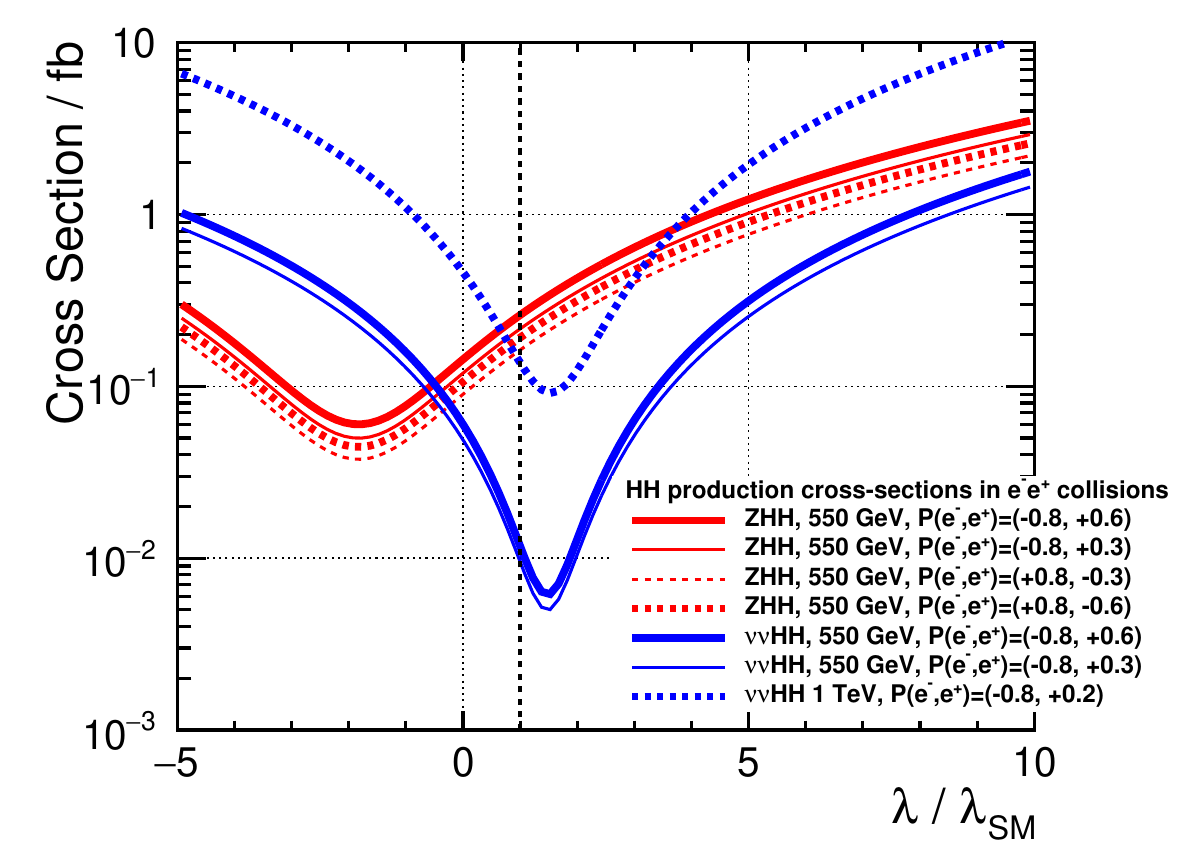}
        \caption{}
        \label{fig:xsec_vs_kala:xsec}    
    \end{subfigure}\hfill%
    \begin{subfigure}{.5\textwidth}
        \centering
        \includegraphics[width=0.95\textwidth]{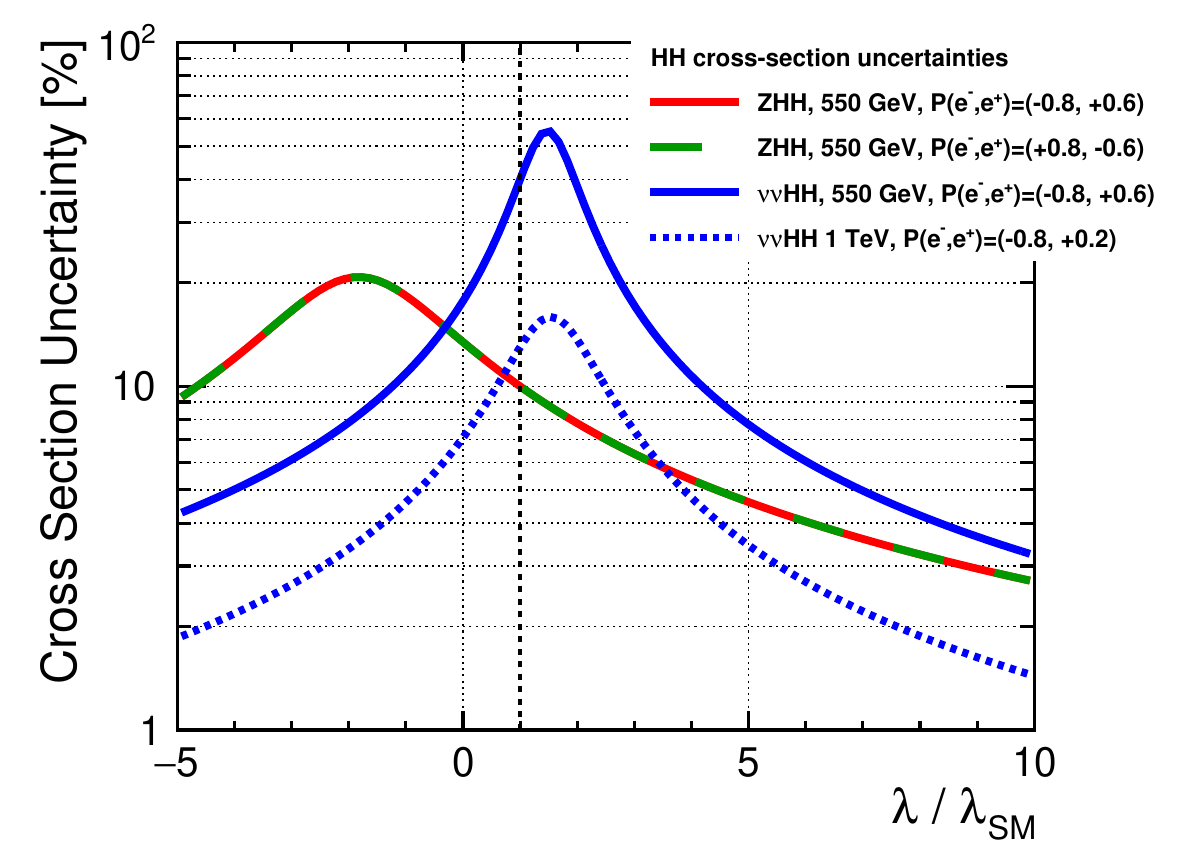}
        \caption{}
        \label{fig:xsec_vs_kala:dxsec}    
    \end{subfigure}%
    \caption{(a) Total cross-sections and (b) projected precisions on total cross-sections of di-Higgs productions as a function of $\kappa_{\lambda} = \lambda/\lambda_{\mathrm{SM}}$. The projection assumes 3.2\,\abinv per opposite-sign polarisation data set. The values SM ($\lambda/\lambda_{\mathrm{SM}} =1 $) correspond to Table~\ref{tab:xsec-prec}.}
    \label{fig:xsec_vs_kala}
\end{figure}

Figure~\ref{fig:xsec_vs_kala:xsec} shows the di-Higgs production cross-sections at 550\,GeV and 1\,TeV for various settings of the beam polarisations as a function of $\kappa_{\lambda} = \lambda/\lambda_{\mathrm{SM}}$. This already shows the effect of the different interference patterns: while the diagrams contributing to $\PZ\PH\PH$ interfere constructively, leading to a cross-section minimum for negative values of $\lambda$, the WBF diagrams interfere destructively, producing a minimum at positive values of $\lambda$, very similar as in case of gluon-gluon-fusion or VBF at the LHC. Since the cross-sections vary with $\lambda$ by more than two orders of magnitude, the actual value of $\lambda$ chosen in nature has a strong influence on the expected cross-section precision, shown in Fig.~\ref{fig:xsec_vs_kala:dxsec}. 

It is interesting to note that, within the precision of the projection, the $+-$ and $-+$ data sets give the same precision on the $\PZ\PH\PH$ cross-section, despite the about 40\% difference in rate. This is indeed an observation not only of the 2014 di-Higgs analysis, but also many other analyses targeting $s$-channel processes~\cite{LinearColliderVision:2025hlt}, which can be explained as the relevant background processes either have a similar scaling while, in the case of $t$-channel processes involving $\PW$ bosons, are even much more suppressed for the ``wrong'' polarisation setting than the signal.

\subsection{Self-coupling projections}
\label{sec:results:self}

In this section, we present a single-parameter extraction of the trilinear Higgs self-coupling $\lambda$ from the total cross-section, which has been shown to be equivalent to an extraction in leading-order dim-6 SMEFT~\cite{Barklow:2017awn}. In other extensions of the Standard Model, in particular in the presence of additional, not too heavy Higgs bosons, the interpretation will be more complex~\cite{Braathen:2025qxf, Frank:2025zmj, Arco:2025pgx, Arco:2025nii, Heinemeyer:2024nfq, Heinemeyer:2024qjp, Heinemeyer:2024hxa} and is beyond the scope of this note. We'd like to point out, however, that in particular in these cases, the additional measurements of the total and differential di-Higgs production cross-sections in $\Pep\Pem$ collisions gives qualitatively new information with respect to proton-proton collisions or $\Pep\Pem$ collisions at lower energies, and only a joint interpretation can unveil the underlying physics.  

\begin{figure}[htbp]
    \centering
    \begin{subfigure}{.5\textwidth}
    \centering
        \includegraphics[width=0.95\textwidth]{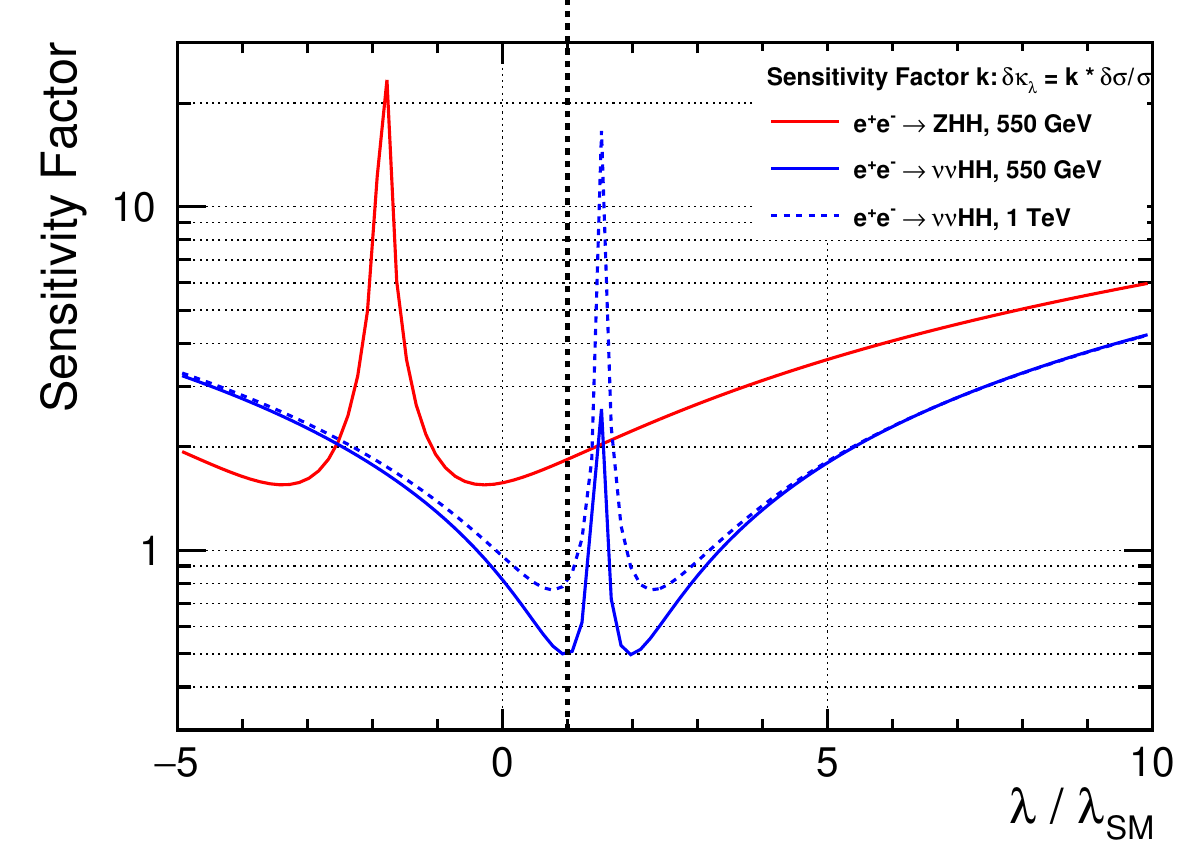}
        \caption{}
        \label{fig:kala_vs_kala:factor}    
    \end{subfigure}\hfill%
    \begin{subfigure}{.5\textwidth}
        \centering
        \includegraphics[width=0.95\textwidth]{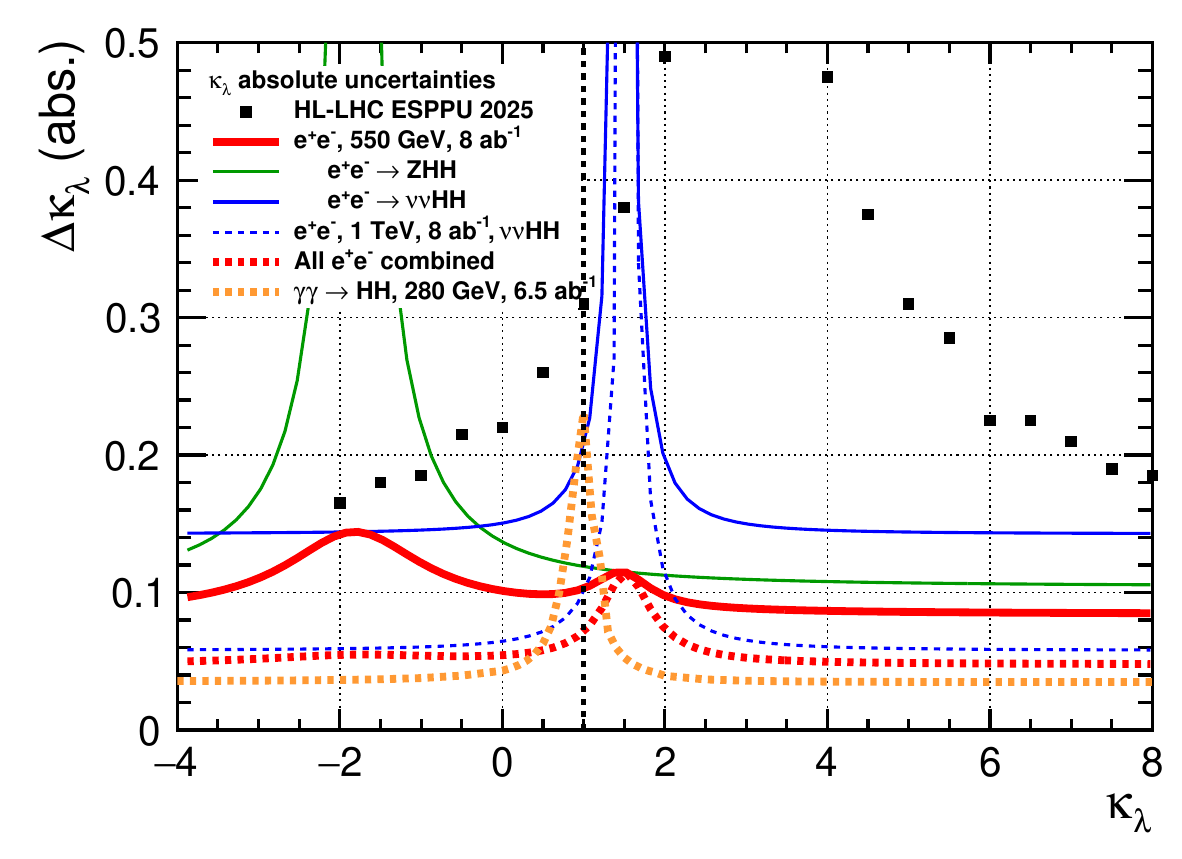}
        \caption{}
        \label{fig:kala_vs_kala:dkala}    
    \end{subfigure}%
    \caption{(a) Sensitivity factor $k$ for translating the precision on the di-Higgs cross-section into an (absolute) uncertainty on $\kappa_{\lambda}$ corresponding to a single parameter extraction. The inverse of $k$ is given by the derivative of the total cross-section with respect to $\lambda$ evaluated at the value of $\lambda$: $k^{-1}=d\sigma/d\lambda|_{\lambda}$.  (b) Absolute precision on $\kappa_{\lambda} = \lambda/\lambda_{\mathrm{SM}}$ as a function of value of $\kappa_{\lambda}$ realised in nature as expected from HL-LHC (black squares), LCF550 (full lines) and some of its upgrade options (dashed lines). The complementary  behaviour of the two production modes in $\Pep\Pem$ collisions as well as in $\Pp\Pp$ and $\PGg\PGg$ collisions is clearly visible.}
    \label{fig:dkala_vs_kala}
\end{figure}

The total di-Higgs production cross-sections are directly sensitive to the trilinear Higgs self-coupling $\lambda$. In the simplest approach possible, only the information encoded in the total cross-sections is exploited in a single-parameter extraction of $\lambda$. In this case, the inverse of the derivative of the cross-section with regard to $\lambda$ is the relevant ``sensitivity factor'' translating a cross-section uncertainty into a coupling uncertainty. If only the ``signal'' diagram with the triple-Higgs vertex were present, the sensitivity factor would always be 1/2. Due to the presence of additional ``background'' diagrams, not containing the triple-Higgs vertex, and their interference with the signal diagram, the sensitivity factor varies strongly with the value of $\lambda$ realized in nature, as shown in Fig.~\ref{fig:kala_vs_kala:factor}. In principle, additional information on $\lambda$ is encoded in the kinematic properties of the events, for instance in the di-Higgs invariant mass $m_{\PH\PH}$ distribution. In the context of the 2014 analysis, a simple weighting of events according to their $m_{\PH\PH}$ value improved the sensitivity factor by about 6\%~\cite{Durig:2016jrs}, and this improvement is also considered in Fig.~\ref{fig:kala_vs_kala:dkala}. In the future, further improvement is expected from using the full information of the matrix elements~\cite{Bliewert:2024, Durig:2016jrs}.
We stress here again that for the case of $\Pep\Pem \to \PZ\PH\PH$, it has been shown that the single-parameter extraction gives equivalent results to a full dim6-SMEFT interpretation, since all other parameters will be known much better from all other measurements at a future $\Pep\Pem$ collider~\cite{Barklow:2017awn}.

\subsection{Complementarity and synergy with other measurements}
Figure~\ref{fig:kala_vs_kala:dkala} compares the LCF550 projections with the precisions expected from other di-Higgs measurements.
At the HL-LHC, the sensitivity to the trilinear coupling degrades for $\uplambda \gtrsim 2$ since the $\Pg\Pg \to \PH\PH$ process is governed by a strong destructive interference between the top-quark box and the triangle diagram that is sensitive to $\uplambda$, which flattens the $\sigma(\Pg\Pg \to \PH\PH)$ dependence on $\uplambda$. In contrast, at an $\Pep\Pem$ collider the dominant channels $\Pep\Pem \to \PZ\PH\PH$ and $\PGn\PAGn \PH\PH$ involve interference between self-coupling and gauge-mediated diagrams with different signs and kinematic weights; at $\sqrt{s} \simeq 550$ GeV this yields a steeper dependence of the double Higgs production cross-section with respect to $\uplambda$ around $\uplambda=2$. As illustrated in Fig.~\ref{fig:kala_vs_kala:dkala}, the LCF550 projections achieve percent-level precision ($\sim 5-10\%$) in this region, substantially tighter than the HL-LHC ($\sim 50 \% $  ). The two programs are therefore highly complementary: HL-LHC anchors the measurement around $\uplambda \simeq 1$, while the LC removes the $\uplambda \simeq 2$ degeneracy and sharpens the global constraint in a combined fit.

At higher energies, the $\Pep\Pem$ measurements improve further. The old ILC result for 8\,\abinv at $\sqrt{s}=1$\,TeV is included in Fig.~\ref{fig:kala_vs_kala:dkala} without any attempt to estimate the effect of analysis improvements. For CLIC at $\sqrt{s}=3$\,TeV, the 2018 study performed in full, Geant4-based simulation of the CLICdet detector concept~\cite{Roloff:2019crr}, found a precision of $-8\%, +11\%$ for the SM value of $\lambda$. This study also showed that the coupling between two Higgs bosons and two $\PW$ bosons can be determined simultaneously with $\lambda$ as soon as differential cross-section information is exploited.

As the 2014 ILC study, the CLIC study was based on nowadays outdated tools like LCFIPlus. A re-analysis with modern techniques is therefore expected to yield a similar improvement for the 1\,TeV and 3\,TeV results as is seen in the current 550\,\GeV analysis update.

Another upgrade option for an $\Pep\Pem$ collider is to turn one of the interaction points into a $\PGg\PGg$ collider~\cite{LinearColliderVision:2025hlt, Barklow:2023ess, Telnov:2008zz, Burkhardt:2002vh, ECFADESYPhotonColliderWorkingGroup:2001ikq}. A recent preliminary study~\cite{Barklow:2025privcom} of $\PGg\PGg \to \PH\PH \to 4\PQb$ in Delphes simulation of the SiD concept as designed for the ILC indicates that for non-SM values of the self-coupling, a $\PGg\PGg$ collider at a relatively modest energy of 280\,GeV could contribute complementary measurements. 

\section{Conclusions and Outlook}
\label{sec:concl}
The ILD concept group has started a re-analysis of di-Higgs production, aiming to harvest a number of high-level reconstruction developments introduced since the last set of studies in 2014. In particular, this comprises a number of improvements to the kinematic reconstruction as well as machine-learning based flavour tagging tools. In addition, the new analysis is -- for the first time -- considering an increased centre-of-mass energy of 550\,GeV, as foreseen in the run plans of e.g.\ C$^3$ and LCF@CERN.
Since the re-analysis is not complete yet, the impact of the reconstruction developments has been estimated for the update of the European Strategy for Particle Physics based on improvement potential identified in the current  2014 analysis and performance evaluations of the new tools on signal and key background processes in full Geant4-based simulations of the ILD concept. For the 550\,GeV data set of the LCF running scenario (8\abinv, $|$P($\Pem,\Pep$)$|$=($80\%$, $60\%$)),
the resulting precision on the double Higgs-strahlungs cross-section is 10\% for each of the opposite-sign polarisation configurations, and the precision on the WBF cross-section is 40\% from the P($\Pem,\Pep$)=($-80\%$, $+60\%$) data-set alone. Together, this corresponds to a precision of 11\% on the tri-linear Higgs self-coupling for the SM case.
The assumption in this extrapolation are solidified further by the first, preliminary result of the re-analysis in the electron and muon channels only. The significances achieved in these channels indicate that the assumptions of the extrapolation are well justified.

The re-analysis of the neutrino and hadronic channels is underway and will be completed as soon as possible. Further plans comprise the inclusion of further decay modes, as well as making even more radical use of machine learning and full event reconstruction~\cite{Wang:2024eji}.

\section*{Acknowledgements}
We would like to thank the LCC generator working group and the ILD software working group for providing the simulation and reconstruction tools and producing some of the Monte Carlo samples used in this study.
This work has benefited from computing services provided by the ILC Virtual Organization, supported by the national resource providers of the EGI Federation and the Open Science GRID.
We thankfully acknowledge the support by the 
the Deutsche Forschungsgemeinschaft (DFG, German Research Foundation) under Germany's Excellence Strategy EXC 2121 ``Quantum Universe'' 390833306.


\printbibliography{}
\end{document}